\documentclass[fleqn,usenatbib]{mnras}

\usepackage{newtxtext,newtxmath}
\usepackage[T1]{fontenc}
\usepackage{ae,aecompl}
\usepackage{graphicx}
\usepackage{amsmath}
\usepackage{breqn}
\usepackage[normalem]{ulem}	
\usepackage{wasysym} 
\usepackage{cleveref}

\crefname{section}{\S}{\S\S}

\title[L-Galaxies and environmental effects]{Galaxy formation with L-GALAXIES: Modelling the environmental dependency of galaxy evolution and comparing with observations}

\author[M. Ayromlou et al.]{Mohammadreza Ayromlou,$^{1}$\thanks{E-mail: ayromlou@mpa-garching.mpg.de}
Guinevere Kauffmann,$^{1}$
Robert M. Yates,$^{1}$ 
\newauthor
Dylan Nelson,$^{1,2}$
Simon D. M. White$^{1}$
\\\\
$^{1}$Max Planck Institute for Astrophysics, Karl-Schwarzschild-Str. 1, 85741 Garching bei M{\"u}nchen, Germany\\
$^{2}$Universit{\"a}t Heidelberg, Zentrum f{\"u}r Astronomie, Institut f{\"u}r theoretische Astrophysik, Albert-Ueberle-Str. 2, 69120 Heidelberg, Germany
}

\date{}
\pubyear{2020}

\begin{document}
\label{firstpage}
\pagerange{\pageref{firstpage}--\pageref{lastpage}}
\maketitle

\begin{abstract}
We present a variation of the recently updated Munich semi-analytical galaxy formation model, \textsc{L-Galaxies}, with a new gas stripping method. Extending earlier work, we directly measure the local environmental properties of galaxies to formulate a more accurate treatment of ram-pressure stripping for all galaxies. We fully re-calibrate the modified \textsc{L-Galaxies} model using a Markov Chain Monte Carlo (MCMC) method with the stellar mass function and quenched fraction of galaxies at $0\leq z\leq2$ as constraints.
Due to this re-calibration, global galaxy population relations, including the stellar mass function, quenched fractions versus galaxy mass and HI mass function are all largely unchanged and remain consistent with observations.
By comparing to data on galaxy properties in different environments from the SDSS and HSC surveys, we demonstrate that our modified model improves the agreement with the quenched fractions and star formation rates of galaxies as a function of environment, stellar mass, and redshift. 
Overall, in the vicinity of haloes with total mass $10^{12}$ to $10^{15}\rm M_{\odot}$ at $z=0$, our new model produces higher quenched fractions and stronger environmental dependencies, better recovering observed trends with halocentric distance up to several virial radii.
By analysing the actual amount of gas stripped from galaxies in our model, we show that those in the vicinity of massive haloes lose a large fraction of their hot halo gas \textit{before} they become satellites. We demonstrate that this affects galaxy quenching both within and beyond the halo boundary. This is likely to influence the correlations between galaxies up to tens of megaparsecs.
\end{abstract}



\section{Introduction}
\label{sec: introduction}

Modern theories of galaxy formation and evolution are built on top of the Lambda Cold Dark Matter ($\rm \Lambda CDM$) cosmological model. One of the critical assumptions of $\rm \Lambda CDM$ is that cold dark matter interacts at best only very weakly with other matter except gravitationally, which has formed the basis of a number of projects that have simulated a gravity only (dark matter only) universe (e.g. \citealt{springel2005simulations,boylan2009resolving,Klypin2011Dark,Riebe2011MultiDark,Angulo2012Scaling,Skillman2014Dark,Wang2020Universal}). Above all, these simulations have shown that galaxy formation within $\rm \Lambda CDM$ can successfully reproduce the observed patterns of galaxy clustering within the cosmic web.

However, the Universe is observed in light coming from stars and gas, which make up galaxies. These are all made of baryonic matter, which interacts in much more complicated ways than weakly interacting gravity-only cold dark matter. Modelling these baryonic components and their interactions in our Universe is one of the most challenging topics in astrophysics. Due to the complex multi-physics and multi-scale nature of the problem, analytical solutions alone are impossible. This has motivated the development of a diverse set of simulation techniques from semi-analytical models (e.g. \citealt{kauffmann1993formation,de2006formation,henriques2020galaxies}) to cosmological hydrodynamical simulations (e.g. \citealt{vogelsberger2014Introducing,Schaye2015eagle,pillepich2018First}) to non-periodic zoom simulations of individual regions (e.g. \citealt{Grand2017Auriga,hopkins18,Libeskind2020Hestia}).

According to the standard hierarchical structure formation theory, baryonic matter accretes into the gravitational potential wells of dark matter structures \citep{white1978core}. It then undergoes cooling and contraction, which eventually leads to the formation of stars and galaxies. Furthermore, the evolution of baryonic matter and galaxies is influenced by a set of complex physical processes
(e.g. gas phase transitions,  stellar evolution, galaxy mergers and feedback processes), which motivates the development of a comprehensive theory of galaxy formation and evolution
(see \citealt{mo2010galaxy} for a full discussion). In addition to intrinsic physical processes, galaxy evolution is also strongly influenced by environment \citep{hubble1931velocity,boselli06}. In other words, the properties of galaxies/subhaloes are strongly correlated with their local density \citep{Yan2013Dependence}. For example, if they are within clusters \citep{Boselli2016Quenching,Pallero2019Tracing}, filaments \citep{Sarron2017filaments} and voids \citep{Tavasoli2013Challenge,Mosleh2018Stellar}. Well known examples of these phenomena are the morphology-density relation \citep{oemler1974systematic,dressler1980galaxy} and higher fraction of quenched galaxies within massive clusters \citep{kauffmann2004environmental,peng10,Davies2020Galaxy}. Environmental processes such as ram-pressure stripping \citep{Gunn_Gott1972} and tidal stripping \citep{binney1987galactic} can strip the gas out of galaxies, affecting their gas reservoirs \citep{wang2020}, star formation \citep{donnari2020,donnari2020b}, stellar properties \citep{webb2020}, morphology \citep{joshi2020}, and so on.

On the theoretical side, the treatment of many of these environmental processes in hydrodynamical simulations occurs as a result of solving the equations of gravity, (magneto)hydrodynamics, and as heating and cooling together, allowing the direct modelling of both dark and baryonic matter \citep{hernquist89,springel2010pur,Crain2015TheEagle}. Simulations using idealised setups \citep{roediger07,vijayaraghavan17,lee2020} as well as full cosmological simulations \citep{Pallero2019Tracing,yun19} reveal the strong dependency of galaxy properties on their environment. Although large-volume hydrodynamical simulations are powerful tools for understanding the physics of galaxies, they are computationally expensive and typically take millions of CPU hours for a single simulation \citep{nelson2019First,pillepich19,dubois2020}. Alternatively, one can model galaxy formation
and evolution, employing modern, computationally efficient
semi-analytical techniques.

A semi-analytical model (SAM) implements the key physical processes involved in galaxy formation and evolution on top of halo merger trees constructed by Monte Carlo methods or extracted from dark-matter-only (DMO) simulations \citep{kauffmann1993formation,kauffmann1999clustering,somerville1999semi,cole2000hierarchical,springel2001populating,de2006formation}. As there is no hydrodynamical interaction in the DMO simulations on which SAMs are implemented, baryonic environmental processes do not occur naturally in SAMs. Therefore, they need to be modelled explicitly using analytical approximations. Early SAMs assume that the entire hot gas reservoir of a galaxy is stripped once it infalls past the virial radius of a larger halo (e.g. \citealt{kauffmann1999clustering,croton2006many}). Instantaneous hot gas stripping for satellites within the virial radius remains a common approach in some recent SAMs as well \citep{lacey2016unified,lagos2018shark}. Most of modern SAMs, including the Munich model \citep[\textsc{L-Galaxies};][]{henriques2020galaxies}, have replaced this prescription by a gradual, time-evolving, gas stripping that allows satellite galaxies/subhaloes to retain a fraction of their gas over longer timescales \citep{font2008colours,tecce2010ram,guo2011dwarf,croton2016semi,stevens2016building,cora2018semi}.

Traditionally, SAMs model environmental processes only for satellite galaxies within the halo boundary (typically the virial radius) of their host haloes. That is partly because estimating ram-pressure beyond this scale is difficult without the direct measurement of the density and velocity of the local environment of galaxies. Accurate object-by-object comparisons between the \textsc{L-Galaxies} SAM (\citealt{henriques2015galaxy}, H15 hereafter) and the IllustrisTNG hydrodynamical simulations shows that environmental effects influence galaxies up to much larger halocentric distances (several virial radii) in the hydrodynamical simulation compared to the SAM \citep{Ayromlou2020Comparing}. Furthermore, several observational \citep{hansen2009galaxy,von2010star,Wetzel2012Galaxy,lu2012cfht} and theoretical \citep{balogh1999differential,bahe2012competition} studies also show that environmental effects might act beyond the halo boundary. Without accurate modelling of environmental processes for all galaxies, it is not possible to robustly reproduce the properties of galaxies which reside in different environments.

In earlier work, we devised a new method to measure, locally and accurately, ram-pressure for all galaxies in the \textsc{L-Galaxies} semi-analytical model using the particle data of the underlying DMO simulation \citep{ayromlou2019new}. Here, we employ this method and update \textsc{L-Galaxies} by extending ram-pressure stripping to all simulated galaxies, regardless of their environment. Taking advantage of the computational efficiency of \textsc{L-Galaxies}, one can quickly implement and test new ideas and theories in galaxy formation and evolution, as is done in this work. Here we probe the role of environment in galaxy evolution by investigating the idea of galaxies becoming ram-pressure stripped \textit{prior to} and after infall into massive and intermediate-mass haloes. To do so, we calibrate our updated \textsc{L-Galaxies} model against a set of observational constraints using a full MCMC approach. We then measure the actual amount of the stripped gas for every galaxy on the fly as the SAM is run. We finally contrast our model predictions with a number of other observations, specifically focusing on the role of environment, and explore the impact of our new stripping model on galaxy properties and the galaxy population as a whole.

This paper is structured as follows: In \S \ref{sec: Methodology} we explain the base L-Galaxies (2020) model and the definition of the local background environment. In \S \ref{sec: new_model_updates} we introduce a variant of the \textsc{L-Galaxies} model with a novel hot gas stripping method. The observations used in this work and the mock catalogues we make to compare our model with those observations are described in \S \ref{sec: obs_data}. Section \ref{sec: environmental_dependency} is dedicated to studying the environmental dependency of galaxy evolution in our model, comparing both with previous versions and with observations. Finally, we conclude and summarise our results in \S \ref{sec: summary}.


\section{Methodology}
\label{sec: Methodology}

\subsection{Simulations and galaxy formation model}
\label{subsec: simulations_and_model}

\subsubsection{Simulations and subhalo identification}
\label{subsubsec: simulations}

In this work, we use the particle and halo merger tree data of the Millennium and Millennium-II simulations \citep{springel2005simulations,boylan2009resolving}. Both simulations are re-scaled to the Planck cosmology \citep{planck2015_xiii} from their original older cosmology, applying the method introduced by \citet{angulo2010one} as updated by \citet{angulo2015cosmological}. The properties of the simulations are given in Table \ref{tab: simulations}. In all simulation snapshots, dark matter haloes are identified using a Friends Of Friends (FOF) algorithm \citep{Davis1985TheEvolution}. Each FOF halo has one central subhalo and its other subhaloes are categorised as satellites. 

All subhaloes are detected using the \textsc{Subfind} algorithm \citep{springel2001populating}. The algorithm sets a minimum of 20 particles for each subhalo to be included in the simulation catalogue. For every FOF halo, there is a virial radius, $R_{200}$, defined as the radius in which the matter density is 200 times the critical density of the Universe. The mass within $R_{200}$ is known as the virial mass, $M_{200}$. Although it is common to consider $R_{200}$ as the halo boundary, the FOF halo can extend beyond $R_{200}$. Therefore, satellite subhaloes of a FOF halo can exist beyond this scale as well.

\begin{table*}
	\centering
	\caption{Parameters used in the Millennium \citep{springel2005simulations} and Millennium-II \citep{boylan2009resolving} simulations in their original (WMAP1, \protect\citealt{spergel2003first}) and re-scaled cosmology (Planck1, \protect\citealt{planck2015_xiii}) based on \protect\cite{angulo2010one,angulo2015cosmological}.}
	\label{tab: simulations}
	\begin{tabular}{|*{12}{c|}} 
        \hline
        \hline
        Simulation & $\Omega_{\rm m}$ & $\Omega_{\rm b}$ & $\Omega_{\Lambda}$ & $\rm H_{0} [\rm km/s/Mpc]$ & $n_{\rm s}$ & $\sigma_{8}$ & $N_{\rm particles}$ & $m_{\rm particle}[\rm M_{\odot}]$ & $l_{\rm box}[\rm Mpc]$ & $N_{\rm snapshot}$ & $z_{\rm initial}$ \\
        \hline
        \hline
        Millennium (WMAP1) & 0.25 & 0.045 & 0.75 & 73 & 1 & 0.9 & $2160^3$  & $1.18\times 10^9$ & 684 & 64 & 127\\
        Millennium (Planck1) & 0.315 & 0.049 & 0.685 & 67.3 & 0.96 & 0.826 & $2160^3$ & $1.43\times 10^9$ & 714 & 64 & 56\\
        \hline
        Millennium-II (WMAP1) & 0.25 & 0.045 & 0.75 & 73 & 1 & 0.9 & $2160^3$ & $9.42\times 10^6$ & 137 & 68 & 127 \\
        Millennium-II (Planck1) & 0.315 & 0.049 & 0.685 & 67.3 & 0.96 & 0.826 & $2160^3$ & $1.14\times 10^7$ & 143 & 68 & 56\\
        \hline
        \hline

    \end{tabular}
\end{table*}

\subsubsection{\textsc{L-Galaxies} semi-analytical model of galaxy formation}
\label{subsubsec: L-Galaxies}

\textsc{L-Galaxies} is a semi-analytical galaxy formation model that uses a set of equations to model baryonic physics on top of dark matter halo merger trees. More recent versions of the model typically run on top of the halo merger trees from the Millennium and Millennium-II simulations \citep{springel2005simulations,croton2006many,de2006formation,bertone2007recycling,boylan2009resolving,guo2011dwarf,yates2013modelling,henriques2015galaxy,henriques2020galaxies}. In this paper, we fully calibrate a variant of the latest version of the model \citep[][H20 hereafter]{henriques2020galaxies}, in which galaxy discs are resolved into a set of concentric rings and a novel gas stripping method is implemented to improve the model's predictions for the environmental dependence of galaxy properties. Following previous versions, we also run the model on top of the Millennium and Millennium-II simulations. A full description of the H20 model is given in the supplementary material of H20 \footnote{An updated version of this Supplementary Material, including a full description of our local environment estimation method and our stripping recipes, is appended to the online version of this paper.}.

In \textsc{L-Galaxies}, baryonic matter bound to each galaxy/subhalo is divided into seven main components: hot gas, cold gas (partitioned into HI and $\rm H_2$), stellar disc, bulge stars, halo stars, the supermassive black holes, and ejected material. The model initiates baryonic physics for every subhalo by seeding it with the expected fraction of diffuse hot (non-star-forming) gas at its formation time. This gas radiates energy and cools to form cold gas, containing neutral and molecular hydrogen components. Once the $\rm H_2$ surface mass density exceeds a certain limit, stars are born. Both the cold gas and stellar components in the galactic disc are followed in twelve concentric rings and all the relevant physical processes happen within these rings.

Galaxy mergers play a key role in starbursts and the formation and growth of supermassive black holes at the centres of galaxies. Feedback processes such as stellar and black hole feedback quench star formation for low-mass and massive galaxies, respectively. The energy released by supernova feedback heats the cold gas and pushes it out of the galactic disc into the hot gas component. The remaining energy is able to drive the material of the hot gas component out of the subhalo into a reservoir of ejected material where it is no longer available for cooling. The timescale for the reincorporation of the ejected gas is assumed to be proportional to $1/M_{200}$ (see Fig. 1 in the supplementary material). As a result, gas returns to massive haloes quickly, while a fraction of the gas ejected out of low-mass haloes may never return.

Another important set of physical processes that can quench star formation is environmental effects such as tidal and ram-pressure stripping. Early versions of \textsc{L-Galaxies} completely stripped hot gas out of satellite galaxies once they fell within $R_{200}$. Modern versions, since \cite{guo2011dwarf}, have gradual hot gas stripping, and therefore satellite galaxies are able to retain a fraction of their hot gas. The latest version of the model \citep{henriques2020galaxies} implements tidal stripping for satellite galaxies within the halo boundary, $R_{200}$, and limits ram-pressure stripping to satellites within $R_{200}$ of massive clusters with $M_{200}>M_{\rm r.p.}$, where $\log_{10}(M_{\rm r.p.}/{\rm M_{\odot}})=14.7$. This ram-pressure stripping threshold, $M_{\rm r.p.}$, was a free parameter in the H15 and H20 model calibrations, and the resulting value was found necessary in order to avoid having too many low-mass, red galaxies. We note that although this results in good agreement with the observed quenched fraction, the approach is merely a numerical fix and is not physical. In this work, we completely remove this mass threshold and extend ram-pressure stripping to all galaxies in the simulations. The implementation of the new gas stripping method is fully explained in \cite{ayromlou2019new} and also briefly in \S \ref{sec: new_model_updates} of this paper.

In \textsc{L-Galaxies} there are three types of galaxies: (i) central galaxies, also known as type 0 galaxies, which reside in the central subhalo of each FOF halo. (ii) Satellite galaxies with an identified subhalo, i.e. detected by the subhalo finder algorithm (here \textsc{Subfind}), which comprise all other subhalos in a FOF besides the central. These are called type 1 galaxies. (iii) Satellite galaxies without an identified subhalo, i.e. their subhaloes have been tidally disrupted below the simulation resolution and are no longer detectable. These are called orphan, or type 2, galaxies. The model tracks the position and velocity of type 2 galaxies by following the most bound particle of their former subhalo. This way, \textsc{L-Galaxies} continues the evolution of orphan galaxies even when their subhaloes are no longer detectable. In the rest of this paper, we use the phrases "subhalo" and "galaxy" interchangeably with the exception of "orphan galaxies".

\subsection{Local Background Environment Measurements}
\label{subsec: LBE}

In order to model ram-pressure, we must quantify the local environmental properties of galaxies. To do so, we follow the method introduced by \cite{ayromlou2019new} which measures the Local Background Environment (LBE) of each galaxy directly from the particle data of the simulation. This background is defined within a spherical shell surrounding the galaxy and its subhalo. Following \citet{ayromlou2019new}, we choose the background shell's radii to exclude the galaxy and its subhalo, and keep the LBE sufficiently local. We choose the inner radius, $r_{\rm in} = 1.25\,r_{\rm subhalo}$ and the outer radius, $r_{\rm out} = 2\,r_{\rm subhalo}$, where $r_{\rm subhalo}$ is the subhalo radius and is defined as the distance between the most bound and the most distant subhalo particles. In order to have proper statistics, we set a minimum of $n_{\rm min} = 30$ for the number of particles within the shell, which results in a statistical error smaller than $1/\sqrt{n_{\rm min}}\sim 20\%$. If there are fewer than $n_{\rm min}$ particles within the shell, we allow the outer radius to extend as needed. The shell density, $\rho_{\rm shell}$, is the total mass within the shell divided by its volume. The mean velocity of the shell, $\vec{v}_{\rm shell}$, is calculated by averaging over the velocities of all the shell particles.

We choose the inner radius of the background shell larger than the subhalo size to remove all the detected (via \textsc{Subfind}) bound subhalo particles. However, the background shell might still contain subhalo-associated particles which are not detected as a part of the subhalo by the subhalo finder algorithm. These particles contaminate the background shell and cause misleading values for its density and velocity, and in \cite{ayromlou2019new} we devised a Gaussian mixture method to remove this contribution. In this paper we use the decontaminated, pure LBE properties to measure ram-pressure for every simulation galaxy.

Our final step is to convert the LBE properties taken from DMO simulations to local estimates of the properties of the gas, needed to estimate ram-pressure. For the LBE velocity, we assume that gas and dark matter follow each other and, therefore, $\vec{v}_{\rm LBE,gas}=\vec{v}_{\rm LBE}$. To derive the gas density we take slightly different approaches for central and satellite galaxies. For central galaxies, we multiply the total LBE density, $\rho_{\rm LBE}$, by the cosmic baryon fraction $\Omega_{\rm b}$, while for satellites we multiply it by $f_{\rm hotgas}$, which is the hot gas fraction of the central subahalo of their parent host FOF halo. The details of the method are explained in \S 3 of \cite{ayromlou2019new}, while \textcolor{blue}{Ayromlou et al.} (\textcolor{blue}{in prep}) extends this method to measure the local gaseous environment of galaxies in the IllustrisTNG and Eagle hydrodynamical simulations.


\section{New model updates}
\label{sec: new_model_updates}

\subsection{Hot gas stripping}
\label{sec: rps and tidal}

Haloes accrete hot gas (and dark matter), a process known as cosmic infall. In \textsc{L-Galaxies}, the amount of this gas is set to be proportional to the amount of infalling dark matter. The hot gas radius, $R_{\rm hotgas}$, of every central subhalo is set equal to the virial radius ($R_{200}$) of its host halo. Satellite subhaloes are assumed to have no gas infall, and their hot gas radius is determined based both on their hot gas radius at infall time and on the environmental effects they experience.

If a subhalo moves fast enough in a dense environment, it can lose gas due to stripping processes such as tidal and ram-pressure stripping. In the H20 version of \textsc{L-Galaxies}, these processes only occur to satellites within the halo $R_{200}$ (tidal stripping) and only in satellites within $R_{200}$ of massive clusters with $\log_{10}\,(M_{200}/{\rm M_{\odot}})>14.7$ (ram-pressure stripping). We follow the method introduced by \cite{ayromlou2019new} to calculate ram-pressure stripping more self-consistently, removing these artificial boundaries in position and halo mass. We extend stripping processes so that they happen to all galaxies, regardless of whether they are a central or satellite. Below we explain our implementation of ram-pressure and tidal stripping of hot gas in the new variant (see \citealt{ayromlou2019new} for a more extensive derivation). The changes are also summarised in Table \ref{tab: model_updates}.

\begin{table*}
	\centering
	\caption{Summary of model updates regarding gas stripping as described in \S \ref{sec: new_model_updates}.}
	\label{tab: model_updates}
    \begin{tabular}{ |l|l|l| }
        \hline
        \hline
        \multicolumn{3}{|c|}{\textbf{Ram-pressure stripping}} \\
        \\ \hline
        \hline
        \textbf{Model ->} & \textbf{H20} & \textbf{This work}\\
        \hline
        Formula & Eq. \ref{eq: R_rp} & Eq. \ref{eq: R_rp}: $R_{\rm rp} = \left(\frac{GM_{\rm g}M_{\rm hotgas}}{4\pi R_{\rm g} R_{\rm hotgas} \rho_{\rm LBE,hotgas} v_{\rm gal,LBE}^2} \right)^{1/2}$\\
        \hline
        Implemented for & Satellite subhaloes within $R_{200}$ of massive clusters & All satellite and central galaxies \\
        \hline
        $\rho_{\rm LBE,hotgas}$ (Local density) & Isothermal $\rho\propto r^{-2}$ & Local density measured directly from the particle data of simulations \\
        \hline
        $v_{\rm gal,LBE}$ (galaxy's velocity & Virial velocity of the satellite's host halo & The velocity of galaxy relative to its local environment, measured\\
        relative to its environment) & & from the particle data of simulations\\
        \hline
        $R_{\rm g}$ (known radius to & Satellite galaxies: $R_{200}$ at infall time& Satellite galaxies: Half mass radius $R_{\rm halfmass}$\\
         estimate gravity) & Central galaxies: N/A & Central galaxies: $R_{200}$\\
        \hline
        $M_{\rm g}$ (known mass to & Satellite galaxies: $M_{200}$ at infall time& Satellite galaxies: Total mass within $R_{\rm halfmass}$\\
         estimate gravity) & Central galaxies: N/A & Central galaxies: $M_{200}$\\
        \hline
        \hline
        \multicolumn{3}{|c|}{\textbf{Tidal Stripping}} \\
        \\ \hline
        \hline
        Formula & Eq. \ref{eq: R_tidal} & Eq. \ref{eq: R_tidal}: $R_{\rm tidal} = \frac{M_{\rm DM}}{M_{\rm DM,infall}}R_{\rm hot,infall}$\\
        \hline
        Implemented for & Satellite galaxies within $R_{200}$ of haloes & All FOF satellite galaxies \\
        \hline
        \hline
        \multicolumn{3}{|c|}{\textbf{The stripping radius and the fate of the stripped gas}} \\
        \\ \hline
        \hline
        Satellite within $R_{200}$ & $R_{\rm strip} = {\rm min}\,(R_{\rm rp},R_{\rm tidal})$ &  Same as H20\\
        & The stripped gas is added to the hot gas of its host halo &  \\
        & After stripping: $R_{\rm hotgas} = {\rm min}\,(R_{\rm hotgas,old},R_{\rm strip})$ & \\
        \hline
        Satellite beyond $R_{200}$ & N/A &  $R_{\rm strip} = {\rm min}\,(R_{\rm rp},R_{\rm tidal})$\\
        & & The stripped gas stays outside its host halo and is allowed to fall inside \\
        & & $R_{200}$ if the galaxy itself falls inside $R_{200}$ of its host halo.\\
        & & After stripping: $R_{\rm hotgas} = {\rm min}\,(R_{\rm hotgas,old},R_{\rm strip})$\\
        \hline
        Central galaxy & N/A & $R_{\rm strip} = R_{\rm rp}$\\
        & & If the galaxy falls into $R_{200}$ of a halo later, the stripped gas is\\
        & &  also allowed to fall into that halo.\\
        & & After stripping: $R_{\rm hotgas} = R_{200}$\\
        \hline
        \hline
    \end{tabular}
\end{table*}

\subsubsection{Ram-pressure stripping}
\label{subsubsec: rps}

If the ram-pressure force from a subhalo's environment becomes stronger than the subhalo's self-gravity on its gas, the gas beyond a given scale, called the stripping radius $R_{\rm rp}$, will be stripped out of the subhalo. The stripping radius is defined as the radius where these two forces are equal. We calculate the ram-pressure using the formula derived by \cite{Gunn_Gott1972}
\begin{equation}
\label{eq: rp_rp}
P_{\rm rp} = \rho_{\rm LBE,gas} \, v_{\rm gal,LBE}^2 \,,
\end{equation}
where $\rho_{\rm LBE,gas}$ and $v_{\rm gal,LBE}$ are the gas density of the galaxy's LBE, and the velocity of the galaxy relative to its local environment, respectively. Both are measured directly using the simulation particle data as described in \S \ref{subsec: LBE}. In contrast, \textsc{L-Galaxies} models prior to this work have adopted $\rho_{\rm LBE,gas}\propto r^{-2}$ and $v_{\rm gal,LBE}$ equal to the virial velocity ($V_{200}$) of the satellite's host halo, with no ram-pressure force for central galaxies nor satellites beyond $R_{200}$.

For the gravitational restoring force we adopt the approach suggested by \cite{mccarthy2007ram} as extended in \cite{ayromlou2019new}. This approach was implemented in L-Galaxies by \cite{guo2011dwarf}. However, it was limited to satellite galaxies within the halo $R_{200}$. The gravitational restoring force per unit area is
\begin{equation}
\label{eq: rp_grav}
F_{\rm g}(r) = g_{\rm max}(r) \,\rho^{\rm proj}_{\rm hotgas}(r) \,,
\end{equation}
where $g_{\rm max}(r)$ is the maximum gravitational acceleration of the subhalo on its hot gas and $\rho^{\rm proj}_{\rm hotgas}(r)$ is the subhalo's projected hot gas density. Considering a spherically symmetric isothermal profile with $\rho_{\rm hotgas}\propto r^{-2}$ for the hot gas, $g_{\rm max}(r)$ equals
\begin{equation}
g_{\rm max}(r) = \frac{GM_{\rm subhalo}(r)}{2r^2},
\end{equation}
where $M_{\rm subhalo}(r)$ is the subhalo mass within the radius r. Taking a similar isothermal $\rho_{\rm subhalo} \propto r^{-2}$ profile for the subhalo, we can calculate the subhalo mass at any given scale as
\begin{equation}
\label{eq: M_subhalo_rp}
M_{\rm subhalo}(r) = M_{\rm g}\frac{r}{R_{\rm g}},
\end{equation}
where $M_{\rm g}$ is the mass within $R_{\rm g}$ and both are known variables which we use to estimate $M_{\rm subhalo}(r)$. In earlier \textsc{L-Galaxies} versions, $R_{\rm g}$ and $M_{\rm g}$ were taken to be $R_{200}$ and $M_{200}$ at infall time for satellite galaxies. Although these are reasonable estimates, they represent the satellite subhalo at an earlier time and ignore its evolution. Therefore, for satellite galaxies we take $R_{\rm g}$ and $M_{\rm g}$ to be the half mass radius, $R_{\rm halfmass}$, and the total mass within $R_{\rm halfmass}$ radius ($M_{\rm halfmass}$) at every redshift, thus representing the subhalo's current properties rather than its properties at infall. For central galaxies, we take $R_{\rm g}=R_{200}$ and $M_{\rm g}=M_{200}$.

Assuming an isothermal ($\rho_{\rm hotgas}\propto r^{-2}$) profile for the subhalo's hot gas, with the boundary condition of having total mass of $M_{\rm hotgas}$ within the hot gas radius, $R_{\rm hotgas}$, the 2D projected hot gas density is given as
\begin{equation}
\rho^{\rm proj}_{\rm hotgas}(r) = \frac{M_{\rm hotgas}}{2 \pi R_{\rm hotgas}r}.
\end{equation}
Finally, the ram-pressure radius, the scale on which the ram-pressure equals the gravitational restoring force per unit area, is calculated to be
\begin{equation}
\label{eq: R_rp}
R_{\rm rp} = \left(\frac{GM_{\rm g}M_{\rm hotgas}}{4\pi R_{\rm g} R_{\rm hotgas} \rho_{\rm LBE,hotgas} v_{\rm gal,LBE}^2} \right)^{1/2}.
\end{equation}
We apply the above formula to all the subhaloes uniformly.

\subsubsection{Tidal stripping}
\label{subsubsec: tidal stripping}

Unlike ram-pressure that strips gas from subhaloes, tidal stripping is able to strip both dark and baryonic matter out of subhaloes and galaxies. As the total mass change of satellite subhaloes is known from our DMO simulations, we use it to estimate the amount of the stripped gas. We assume that the fraction of gas lost is the same as the fraction of dark matter lost
\begin{equation}
\label{eq: tidal_stripping}
\frac{M_{\rm hot}(R_{\rm tidal})}{M_{\rm hot,infall}} = \frac{M_{\rm DM}}{M_{\rm DM,infall}},
\end{equation}
where $M_{\rm DM,infall}$ and $M_{\rm hot,infall}$ are the satellite's virial mass and hot gas mass at infall. In addition, $M_{\rm DM}$ and $M_{\rm hot}(R_{\rm tidal})$ are the satellite's mass and hot gas mass after tidal stripping. The only unknown variable in Eq. \ref{eq: tidal_stripping} is $M_{\rm hot}(R_{\rm tidal})$. Assuming an isothermal profile ($\rho\propto r^{-2})$, we calculate the tidal stripping radius beyond which the gas is stripped
\begin{equation}
\label{eq: R_tidal}
R_{\rm tidal} = \frac{M_{\rm DM}}{M_{\rm DM,infall}}R_{\rm hot,infall} \,,
\end{equation}
where $R_{\rm hot,infall}$ is the satellite's hot gas radius at infall time.

As mentioned in \S \ref{subsubsec: L-Galaxies}, orphan galaxies do not have an identified subhalo, and therefore their subhalo mass is set to zero. Eq. \ref{eq: tidal_stripping} then implies that they will lose all their hot gas as well. As a result, orphan galaxies are entirely empty of hot gas due to tidal stripping. In some studies, analytic formulae of the evolution of  the subhalo mass over time are used to extend the estimation of the subhalo mass beyond the point where it is no longer detectable in the simulation, and in this case orphan galaxies retain a nonzero fraction of their hot gas \citep[e.g. see][]{Luo2016Resolution}. However, we do not adopt any such scheme in this work.

\begin{figure*}
    \includegraphics[width=0.7\textwidth]{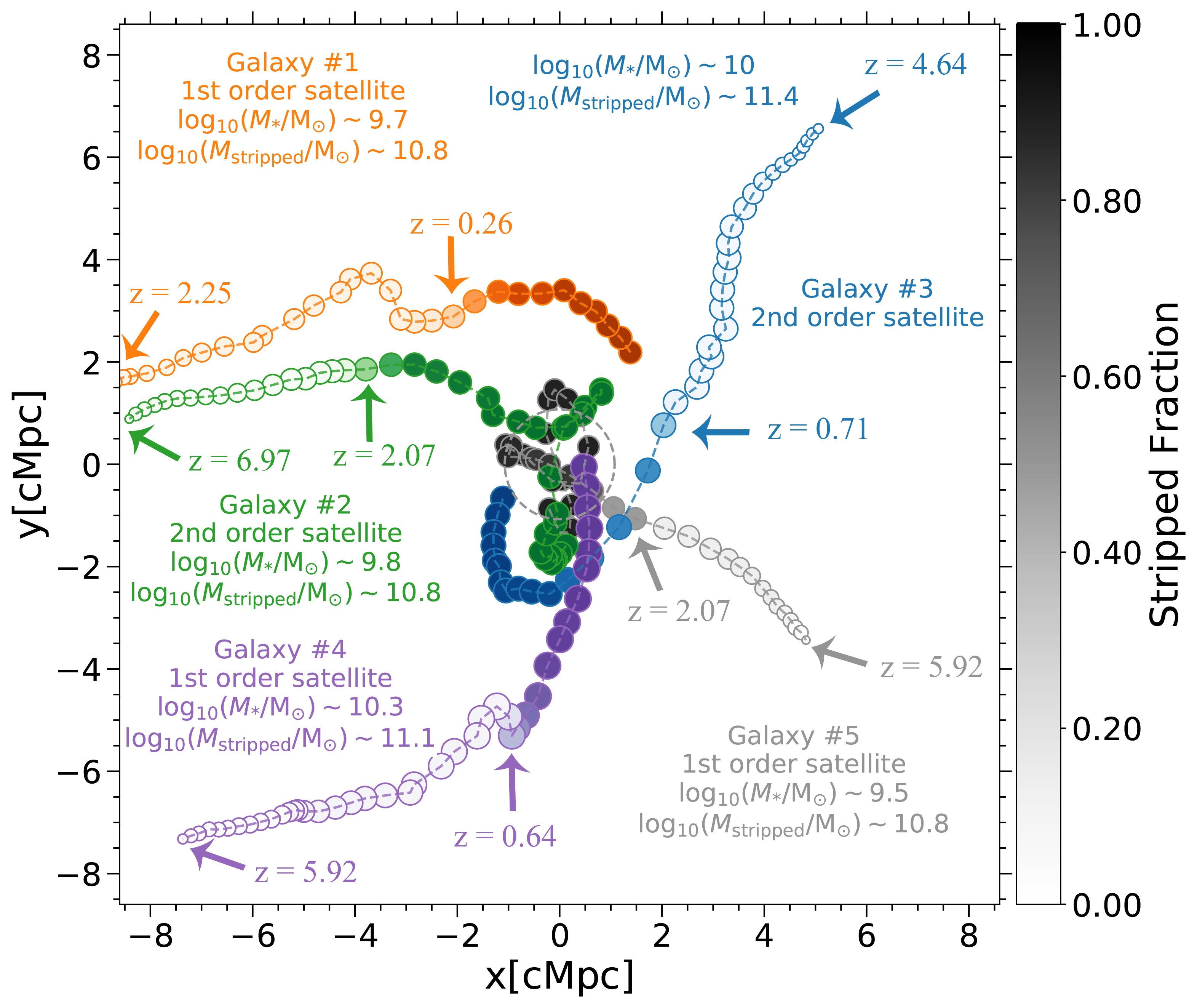}
    \includegraphics[width=0.4\textwidth]{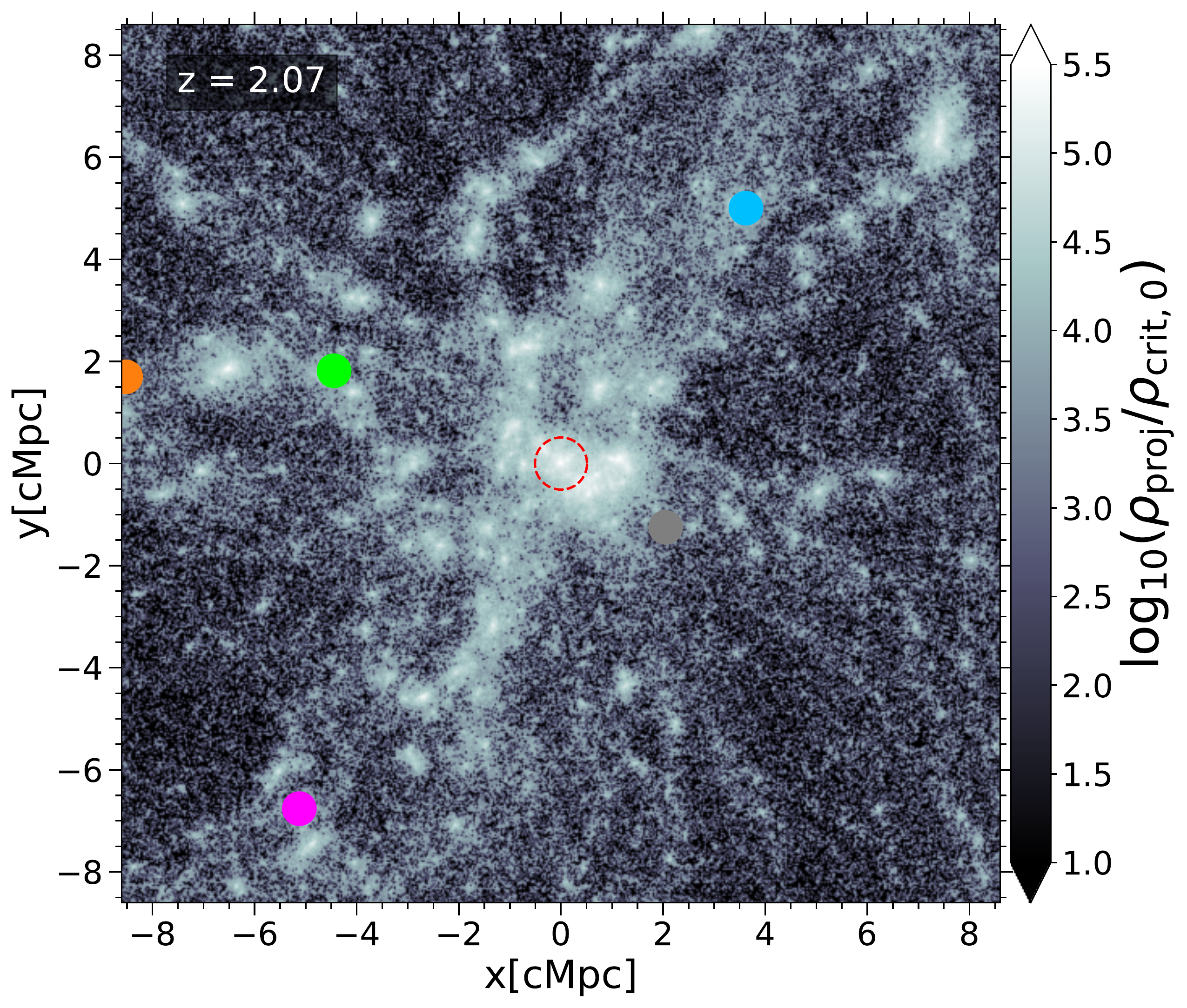}
    \includegraphics[width=0.4\textwidth]{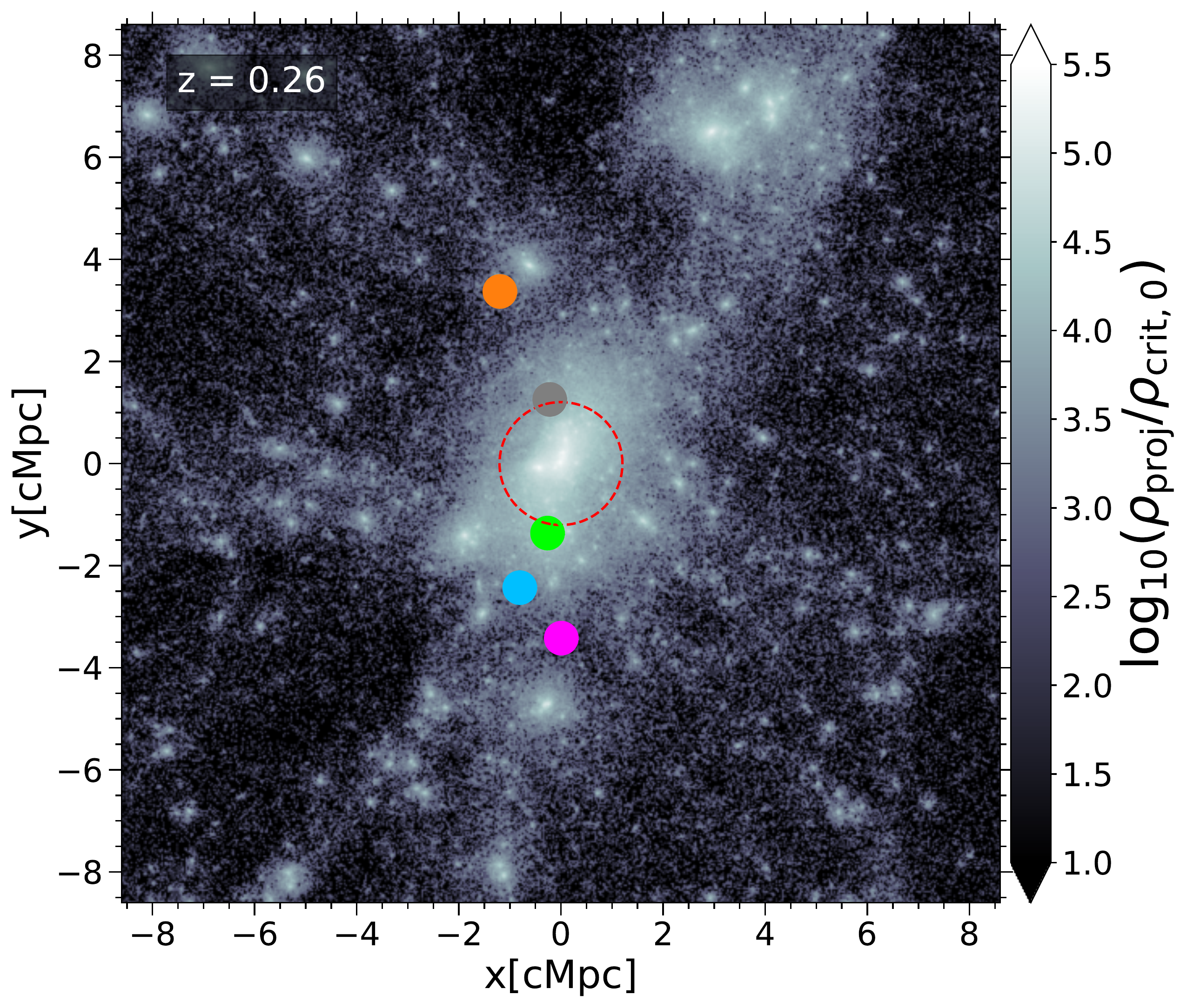}
    \includegraphics[width=0.33\textwidth]{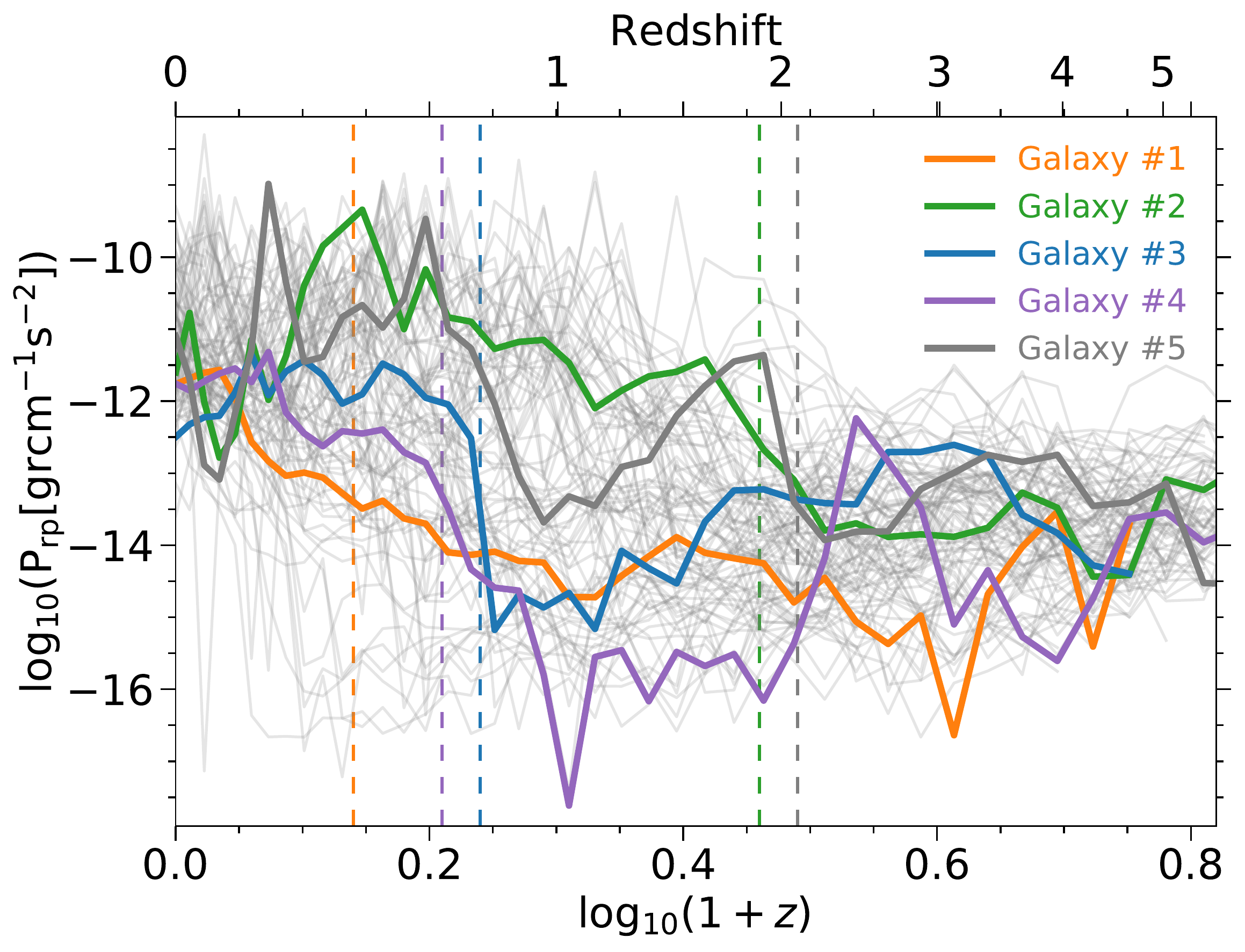}
    \includegraphics[width=0.33\textwidth]{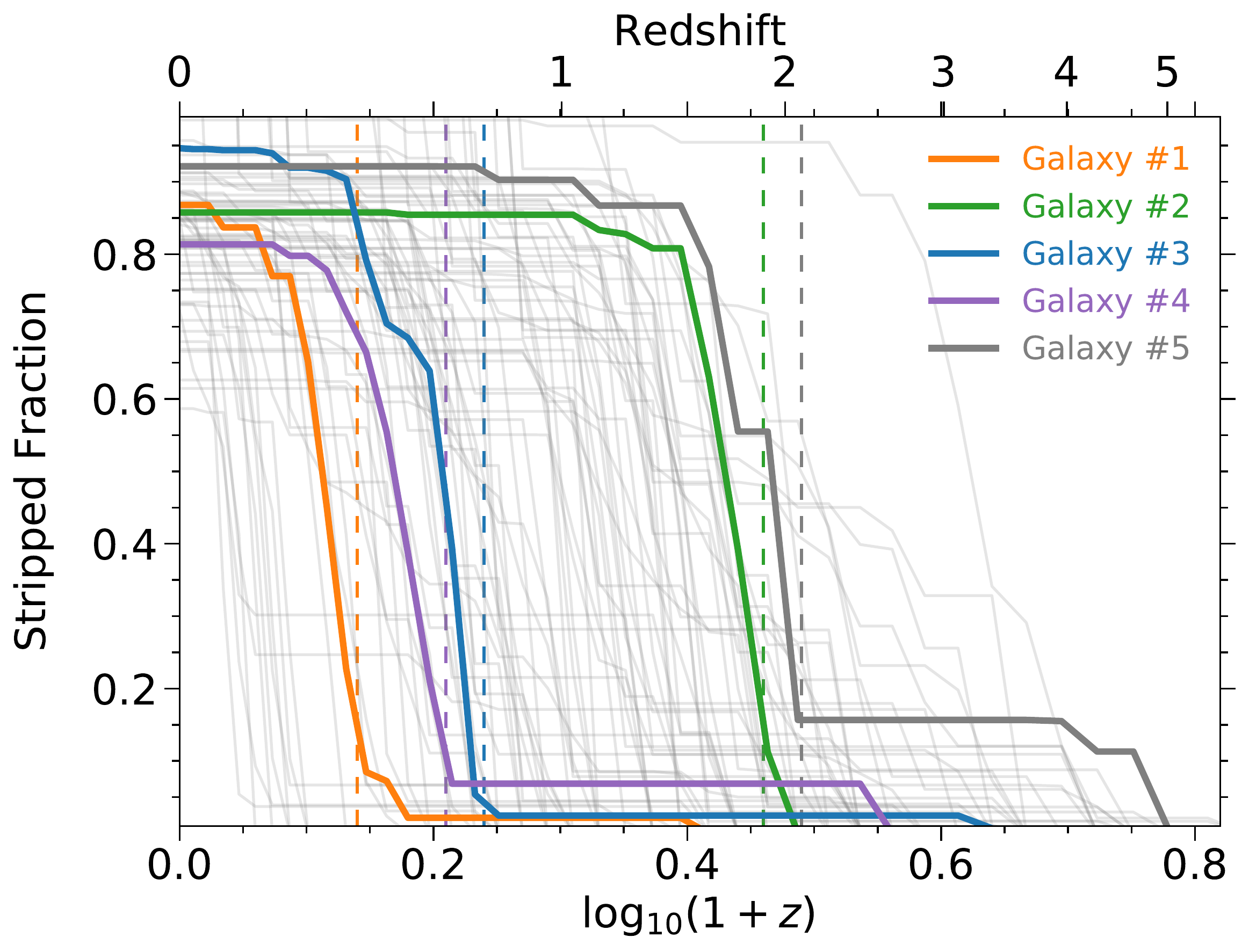}
    \includegraphics[width=0.33\textwidth]{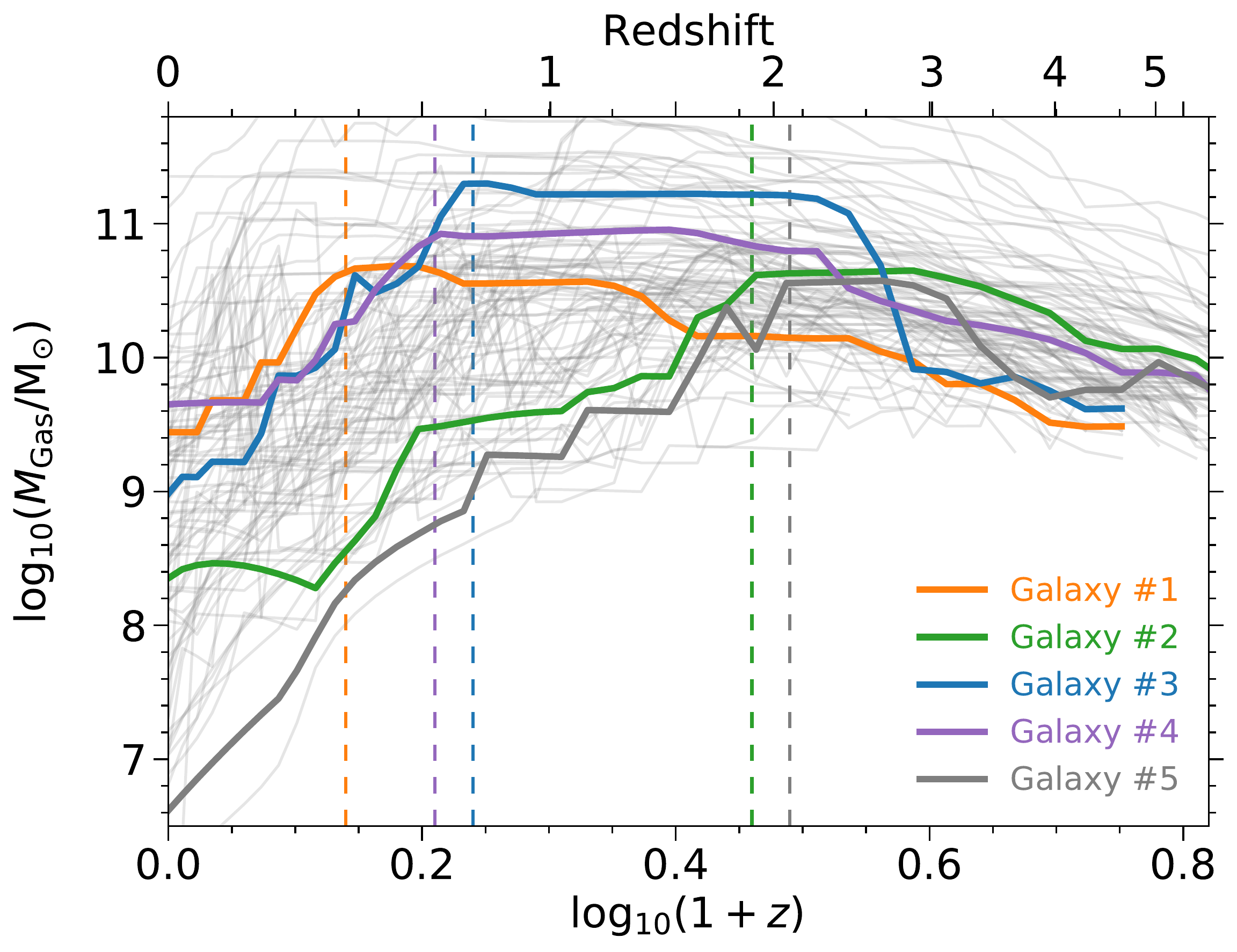}
    \caption{Illustration of gas stripping in our model. Top panel: Five sample galaxies in the vicinity of a massive cluster with $\log_{10}(M_{200}/{\rm M_{\odot}})\sim 14.5$. Each colour corresponds to a galaxy, and different circles of the same colour demonstrate the evolution of that galaxy at different redshifts. The size of each circle represents, qualitatively, the galaxy's stellar mass. Shown by the colourbar, the transparency of colours denotes the fraction of the gas stripped (Eq. \ref{eq: f_strippedhot}). Here, 1st order satellites are satellites of the central cluster at $z=0$ which have not been satellites of other haloes previously, while 2nd order satellites were previously satellites of other haloes at higher redshifts. The stellar mass and total stripped mass of each galaxy are quoted at $z=0$. Middle panel: Matter distribution at two redshifts of the simulation, depicting the positions of the five galaxies in the cosmic web. All coordinate units are comoving. Bottom panel: ram-pressure (left), stripped fraction (middle) and gas mass (right) versus redshift. Each curve corresponds to a galaxy that is currently a satellite of the central cluster. The dashed vertical lines correspond to the redshift where our five chosen galaxies start getting stripped significantly.}
\label{Fig: Stripping_Schematic1}
\end{figure*}

\subsubsection{Stripping implementation and the fate of the stripped gas}
\label{subsubsec: stripping_implementation}

Satellite galaxies are subject to both tidal and ram-pressure stripping. Therefore, we take the stripping radius to be the smaller of the tidal stripping radius and the ram-pressure stripping radius. For central galaxies, the stripping radius is equal to the ram-pressure stripping radius, as we do not consider them subject to tidal stripping. After stripping, we assume the density profile of the remaining gas remains isothermal with $\rho \propto r^{-2}$. We note that ram-pressure stripping is the dominant effect for most galaxies.

In addition to hot gas stripping we also strip the ejected reservoirs of galaxies. The fraction of gas stripped from the ejected reservoir of a galaxy is assumed to be the same as the fraction of its stripped hot gas. Stripping of material that has been expelled from the galaxy by feedback processes and is assumed to lie at least temporarily in this ejecta reservoir plays a critical role in the evolution of galaxies (see \S \ref{subsec: total_stripped_gas}).

For every galaxy, if $R_{\rm strip} < R_{\rm hotgas}$, all the gas beyond $R_{\rm strip}$ is removed. For satellite galaxies within the halo boundary, $R_{200}$, the stripped gas goes directly to the hot gas component of their host halo. For central galaxies and satellites beyond the halo $R_{200}$, the stripped gas does not go to any halo immediately, but we keep track of it through time. If the galaxy subsequently falls into $R_{200}$ of a halo, then the stripped gas is added to the host halo's hot gas with the condition that the host halo's baryon fraction does not exceed the cosmic value. After stripping, the new hot gas radius of a satellite will be the minimum of its former hot gas radius and $R_{\rm strip}$. On the other hand, as long as a galaxy is categorised as a central galaxy, since it accretes hot gas from its environment, its hot gas radius is set to its FOF $R_{200}$.

\subsubsection{Tracking gas stripping}
\label{subsubsec: tracking_stripping}

We measure the amount of gas lost due to stripping in different conditions and separate it from other intrinsic physical processes. In order to evaluate the impact and predictions of our method quantitatively, we define the "cumulative stripped fraction", or simply the "stripped fraction" of a subhalo at a given redshift as
\begin{equation}
\label{eq: f_strippedhot}
f_{\rm stripped}(z) = M_{\rm stripped}(z) \,/\, (f_{\rm b} \,M_{\rm 200,max}) \,,
\end{equation}
where $M_{\rm 200,max}$ is the maximum $M_{200}$ of the main progenitor of the subhalo through time. $M_{\rm stripped}(z)$ is the total mass which has been stripped from the subhalo since its formation time until redshift $z$. This includes gas stripped from the hot gas component and the ejected reservoir including material stripped while the subhalo was a central, unless we explicitly decompose the two. This is a cumulative quantity that we track for every galaxy across time.

As an overview of our new model, Fig. \ref{Fig: Stripping_Schematic1} illustrates a massive cluster with $\log_{10}(M_{200}/{\rm M_{\odot}})\sim 14.5$ and five galaxies that end up as its satellites by $z=0$. Each galaxy is shown by one colour, and in the top panel, the transparency of that colour shows the cumulative stripped fraction of the galaxy (Eq. \ref{eq: f_strippedhot}) at each simulation snapshot. The formation redshifts, and the redshifts at which each galaxy starts to be significantly stripped, are indicated by arrows. It is clear that stripping processes start long before galaxies fall within the virial radius ($R_{200}$) of their present-day halo (dashed circle at the centre). We find that, on average, a satellite galaxy loses more than 70-80\% of its hot gas due to stripping prior to its infall (see \S \ref{subsec: total_stripped_gas}). Galaxy \#2 (green) and \#3 (blue) are pre-infall satellites, i.e. they are already satellites of other haloes at $z>0$, prior to their infall into the central cluster. These two galaxies have lost a considerable fraction of their hot gas through pre-processing in their initial hosts.

On the other hand, the three other galaxies (\#1, \#4 and \#5) are categorised as central galaxies until they fall directly into their $z=0$ host halo. These galaxies have lost their hot gas via ram-pressure stripping when moving fast through the dense environment beyond the halo boundary of their present-day host halo. The middle panels of Fig. \ref{Fig: Stripping_Schematic1} show the position of the sample galaxies in the cosmic web at two different redshifts, $z=2$ and $z=0.26$. They illustrate the times when the green (\#2) and purple (\#4) galaxies (left panel) and the orange galaxy (\#1) are strongly stripped. For example, galaxy \#1 moves through the overdensity around the central host halo, but extending far beyond its virial radius, at which there is no abrupt change in density. All five galaxies pass through dense environments at some time in their history, losing a large fraction of their gas \footnote{A video showing the stripping process for our sample galaxies is given in the supplementary material of this paper.}. It is clear that the matter distribution of the proto-cluster is very extended, but there is no significant change in the density of the central region of the halo.

Finally, the bottom panels of Fig. \ref{Fig: Stripping_Schematic1} depict the ram-pressure (left), the cumulative stripped fraction (middle), and the gas mass (right) of all the cluster's $z=0$ surviving satellite subhaloes as a function of redshift. The previously highlighted galaxies are illustrated as thick curves, while the rest of the galaxies are shown with thin grey lines. The vertical lines correspond to the redshifts where the five sample galaxies start becoming significantly stripped. There is a strong correlation between high values of ram-pressure and the times when galaxies lose gas. Overall, Fig. \ref{Fig: Stripping_Schematic1} depicts the importance of modelling stripping beyond the virial radius, showing that galaxies can even lose their entire hot gas reservoir while still centrals, without ever infalling into a more massive host halo.

\subsection{Gas infall into haloes}
The \textsc{L-Galaxies} infall recipe in H20 is that the gas accretion rate onto haloes is equal to the accretion rate of total matter times the cosmic baryon fraction. It is possible that the baryon fraction exceeds the cosmic value (see Fig. 8 of \citealt{Ayromlou2020Comparing}). There are two reasons for this discrepancy, as described by \cite{Yates+17}. The first and more dominant effect is due to the implementation of environmental effects such as tidal and ram-pressure stripping. Satellite galaxies in the infall regions lose dark matter due to tidal effects in the DMO simulation on which \textsc{L-Galaxies} is run. The lack of environmental stripping effects for gas beyond $R_{200}$ can give these satellites a baryon fractions greater than the cosmic value when they fall into $R_{200}$ of their host halo. This artificially increases the baryon fraction of their host, given that its infall has already topped-up the halo's baryon budget to the cosmic baryon fraction. In this work we fully resolves this issue by having gas stripping for satellites beyond the halo $R_{200}$ (see \S \ref{subsubsec: stripping_implementation}).

The second reason is that total halo mass can decrease with time because of changes in morphology or halo concentration or simply through numerical fluctuations (see \citealt{DeLucia2004Substructures} for subhalo mass fluctuations). At the same time, in \textsc{L-Galaxies} the halo baryonic mass within $R_{200}$ is kept unchanged by construction, causing an increase in the  baryon fraction. Following the prescription of \cite{Yates+17} we correct the input halo merger trees to prevent the $M_{200}$ from decreasing with time. This accounts for any artificial decrease in $M_{200}$ measured when determining $R_{200}$ based on the assumption of spherical symmetry for haloes.

\subsection{Model calibration}
\label{subsec: model_calibration}

\begin{table*}
	\centering
	\caption{Free parameters used in the MCMC model calibration in this work and previous models \protect\citep{Guo2013Galaxy,henriques2015galaxy,henriques2020galaxies}. Those values labelled as "F.P." are fixed parameters, which are not taken from the MCMC best fit. A full model description is given in the Supplementary Material appended to the online version of this paper.}
	\label{tab: MCMC_free_params}
	\begin{tabular}{|*{7}{c|}}
		\hline \hline
		Model Parameter & Equation in \textsc{L-Galaxies} & G13 & H15 & H20 & This Work & Units\\
		\hline \hline
		$\alpha_{\rm SF}$ (Star formation efficiency) & $\Sigma_{\rm SFR} = \alpha_{\rm SF}\, \Sigma_{\rm H_2}/t_{\rm dyn}$ & 0.011 & 0.025 & 0.06 & 0.073\\\\
		$\alpha_{\rm SF,burst}$ (Star formation burst efficiency) & $M_{\star,\rm burst} = \alpha_{\rm SF,burst}\left(\frac{M_1}{M_2}\right)^{\beta_{\rm SF.burst}}M_{\rm cold}$ & 0.56 & 0.60 & 0.50 & 0.116\\\\
		$\beta_{\rm SF,burst}$ (Star formation slope) & same as above & 0.70 & 1.9 & 0.38 & 0.674\\
		\hline
		$k_{\rm AGN}$ (Radio feedback efficiency) & $\Dot{M}_{\rm BH} = k_{\rm AGN}\left(\frac{M_{\rm hot}}{10^{11}\rm M_{\odot}}\right)\,\left(\frac{M_{\rm BH}}{10^8 \rm M_{\odot}} \right)$ & N/A. & $5.3\times 10^{-3}$ & $2.5\times 10^{-3}$ & $8.5\times 10^{-3}$ & ${\rm M_{\odot}}/\rm yr$\\\\
		$f_{\rm BH}$ (Black hole growth efficiency) & $\Delta M_{\rm BH,Q} = \frac{f_{\rm BH}(M_{\rm sat}/M_{\rm cen})\, M_{\rm cold}}{1+(V_{\rm BH}/V_{\rm 200c})^2}$ & 0.03 & 0.041 & 0.066 & 0.011\\\\
		$V_{\rm BH}$ (Quasar growth scale) & same as above & 280 & 750 & 700 & 1068 & km/s\\
		\hline
		$\epsilon_{\rm reheat}$ (Mass-loading efficiency) & $\epsilon_{\rm disc} = \epsilon_{\rm reheat}\left[0.5+\left(\frac{V_{\rm max}}{V_{\rm reheat}}\right)^{-\beta_{\rm reheat}}\right]$ & 4.0 & 2.6 & 5.6 & 9.7\\\\
		$V_{\rm reheat}$ (Mass-loading scale) & same as above & 80 & 480 & 110 & 119 & km/s\\\\
		$\beta_{\rm reheat}$ (Mass-loading slope) & same as above & 3.2 & 0.72 & 2.9 & 2.9\\\\
		$\eta_{\rm eject}$ (Supernova ejection efficiency) & $\epsilon_{\rm halo} = \eta_{\rm eject}\left[0.5+\left(\frac{V_{\rm max}}{V_{\rm eject}}\right)^{-\beta_{\rm eject}}\right]$ & 0.18 & 0.62 & 5.5 & 9.56\\\\
		$V_{\rm eject}$ (Supernova ejection scale) & same as above & 90 & 100 & 220 & 172 & km/s\\\\
		$\beta_{\rm eject}$ (Supernova ejection slope) & same as above & 3.2 & 0.80 & 2.0 & 1.88\\\\
		$\gamma_{\rm reinc}$ (Ejecta reincorporation) & $t_{\rm reinc} = \gamma_{\rm reinc}\frac{10^{10}\rm M_{\odot}}{M_{\rm 200c}}$ & N/A. & $3.0 \times 10^{10}$ & $1.2\times 10^{10}$ & $6.6\times 10^{9}$ & 1/yr\\
		\hline
		$M_{\rm rp}$ (Ram-pressure threshold) & N/A & 0.0 & $1.2\times 10^{4}$ & $5.1\times 10^{4}$ & N/A & $10^{10}{\rm M_{\odot}}$\\\\
		$R_{\rm merger}$ (Major-merger threshold) & N/A & 0.30 & 0.1 (F.P.) & 0.1 (F.P.) & 0.1 (F.P.)\\\\
		$\alpha_{\rm friction}$ (Dynamical friction) & $\alpha_{\rm friction}\frac{V_{\rm 200c}r_{\rm sat}^2}{GM_{\rm sat}\ln(1+M_{\rm 200c}/M_{\rm sat})}$ & 2.0 & 2.5 & 1.8 & 0.312\\\\
		$f_{\rm z,hot,TypeII}$ & N/A & N/A. & N/A. & 0.3 (F.P.) & 0.3 (F.P.)\\\\
		$f_{\rm z,hot,TypeIa}$ & N/A & N/A. & N/A. & 0.3 (F.P.) & 0.3 (F.P.)\\\\
		$v_{\rm inflow}$ & $v_{\rm inflow}=r/t_{v}$ & N/A. & N/A. & 1000 (F.P.) & 1000 (F.P.) & $\rm km/s/Mpc$\\
		\hline
		\hline
	\end{tabular}
\end{table*}

\begin{figure*}
    \includegraphics[width=1\textwidth]{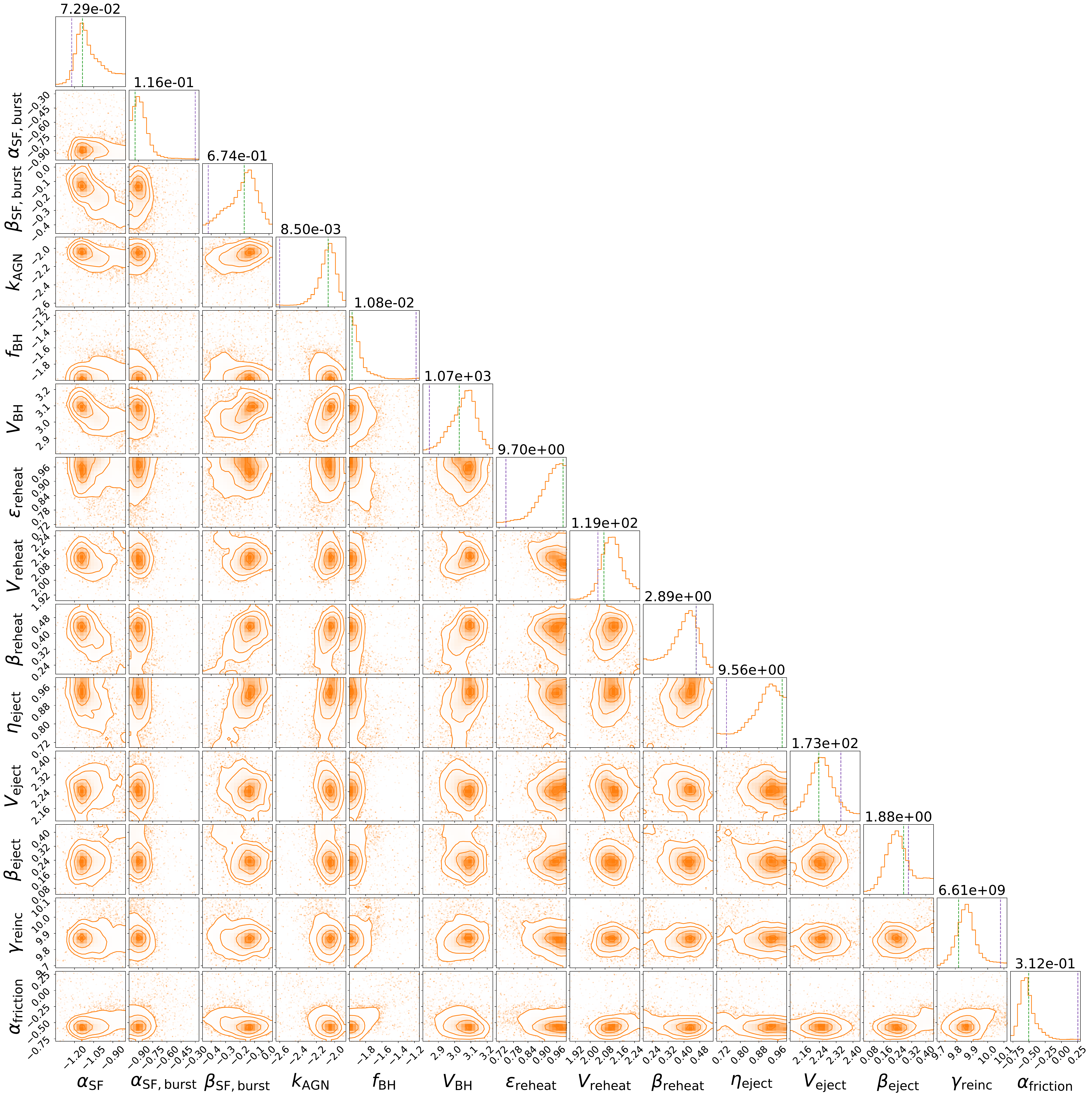}
    \caption{Histograms of the weighted accepted MCMC steps (after burn-in) for all the free parameters used in this model. The green dashed lines correspond to our best fit for each parameter and the purple dashed lines are the best fit in H20. The x-axes and y-axes in all panels are reported in logarithmic units. The value above each 1D histograms corresponds to the best fit (in linear units) of that parameter in this work. This figure is generated using $\sim 23000$ MCMC steps.}
\label{Fig: Corner_MCMC}
\end{figure*}

\begin{figure*}
    \includegraphics[width=0.33\textwidth]{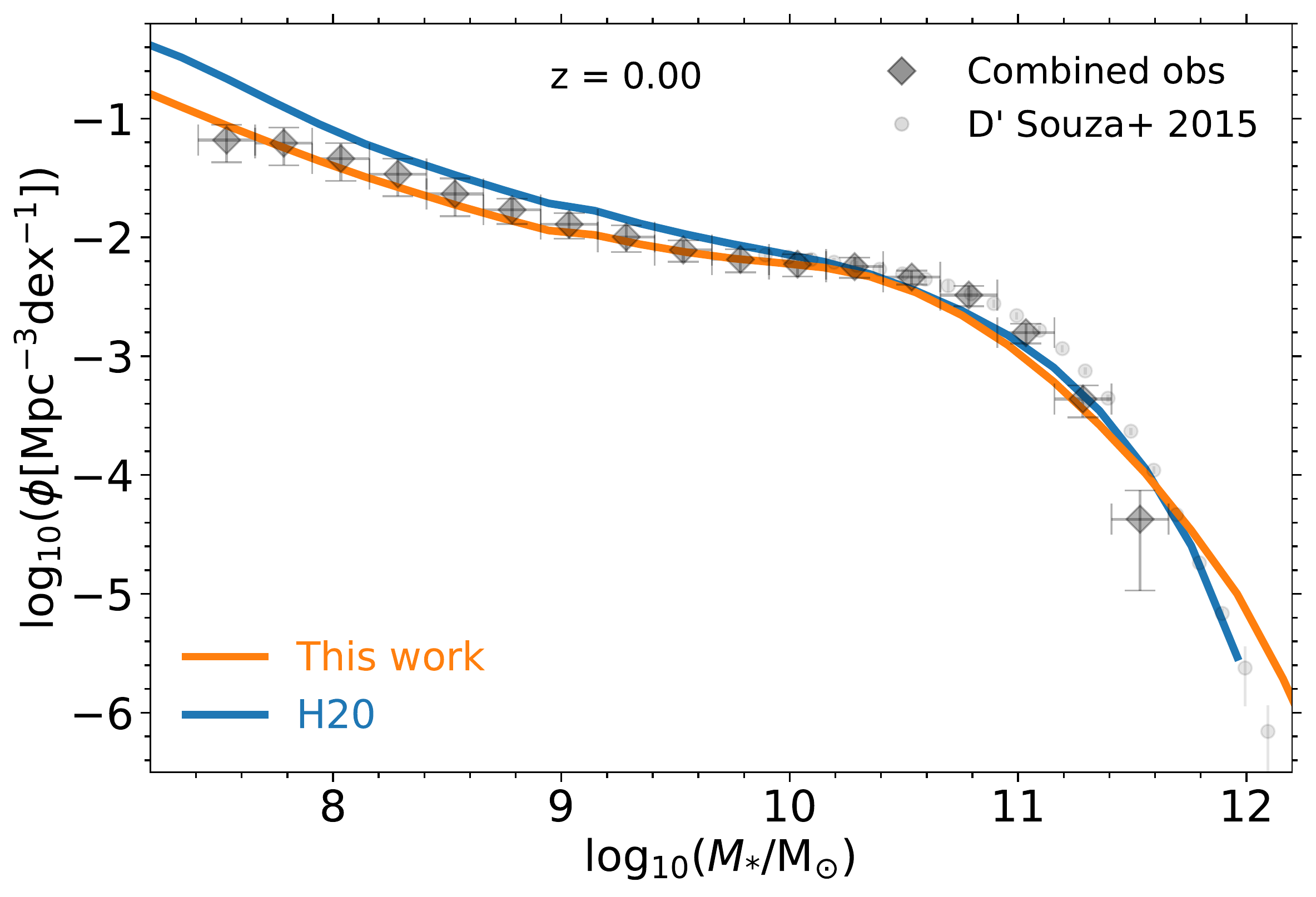}
    \includegraphics[width=0.33\textwidth]{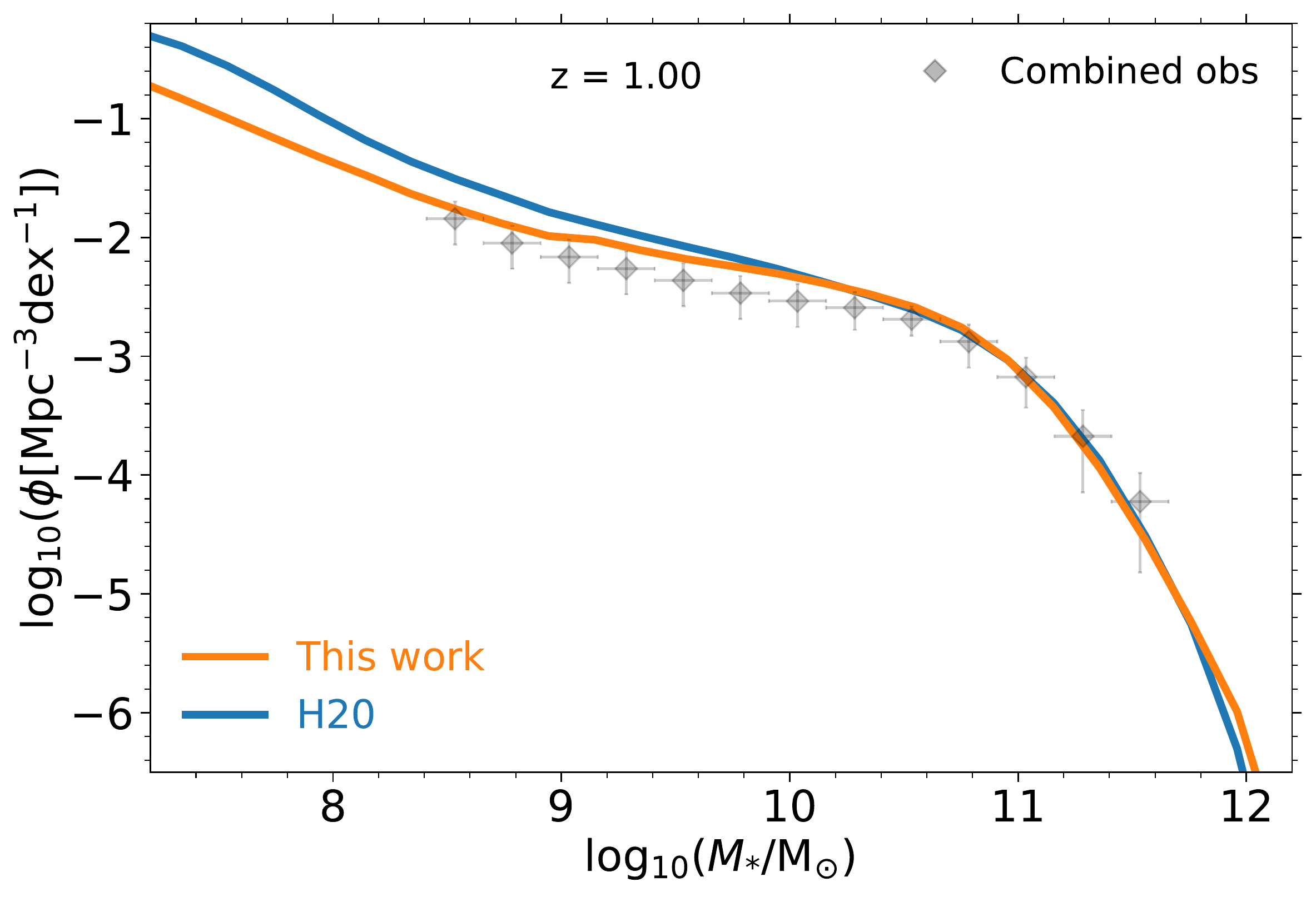}
    \includegraphics[width=0.33\textwidth]{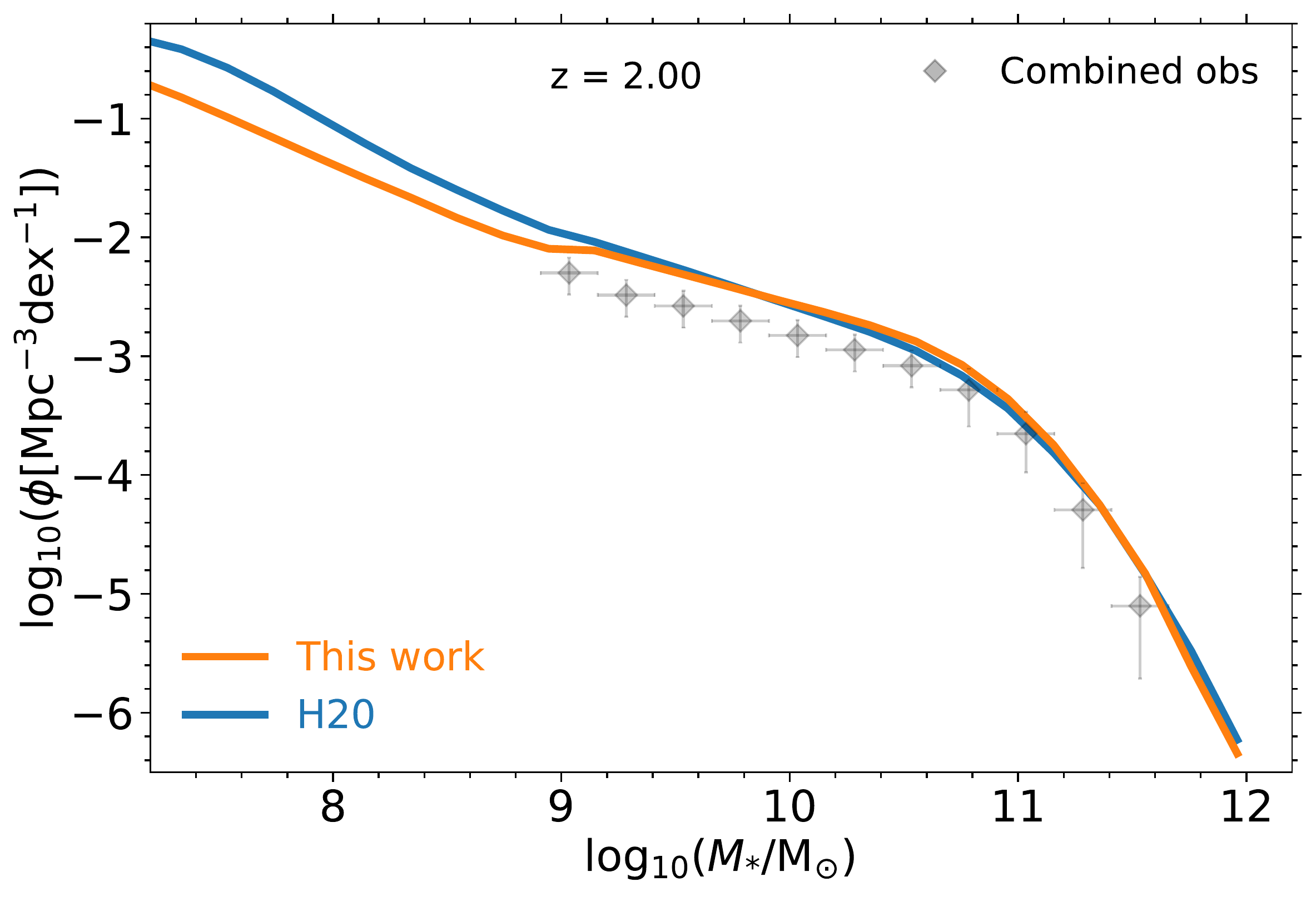}
    \includegraphics[width=0.33\textwidth]{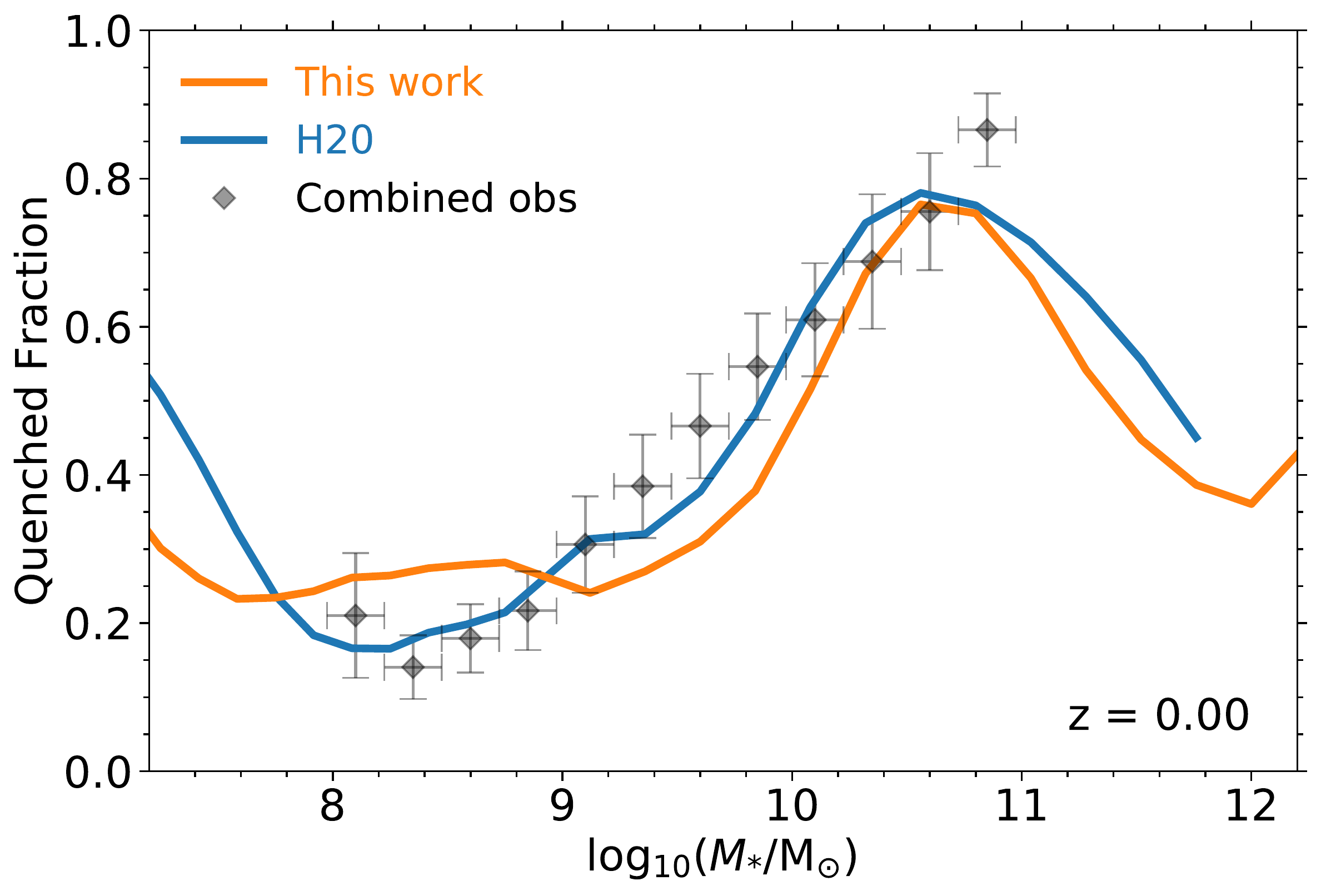}
    \includegraphics[width=0.33\textwidth]{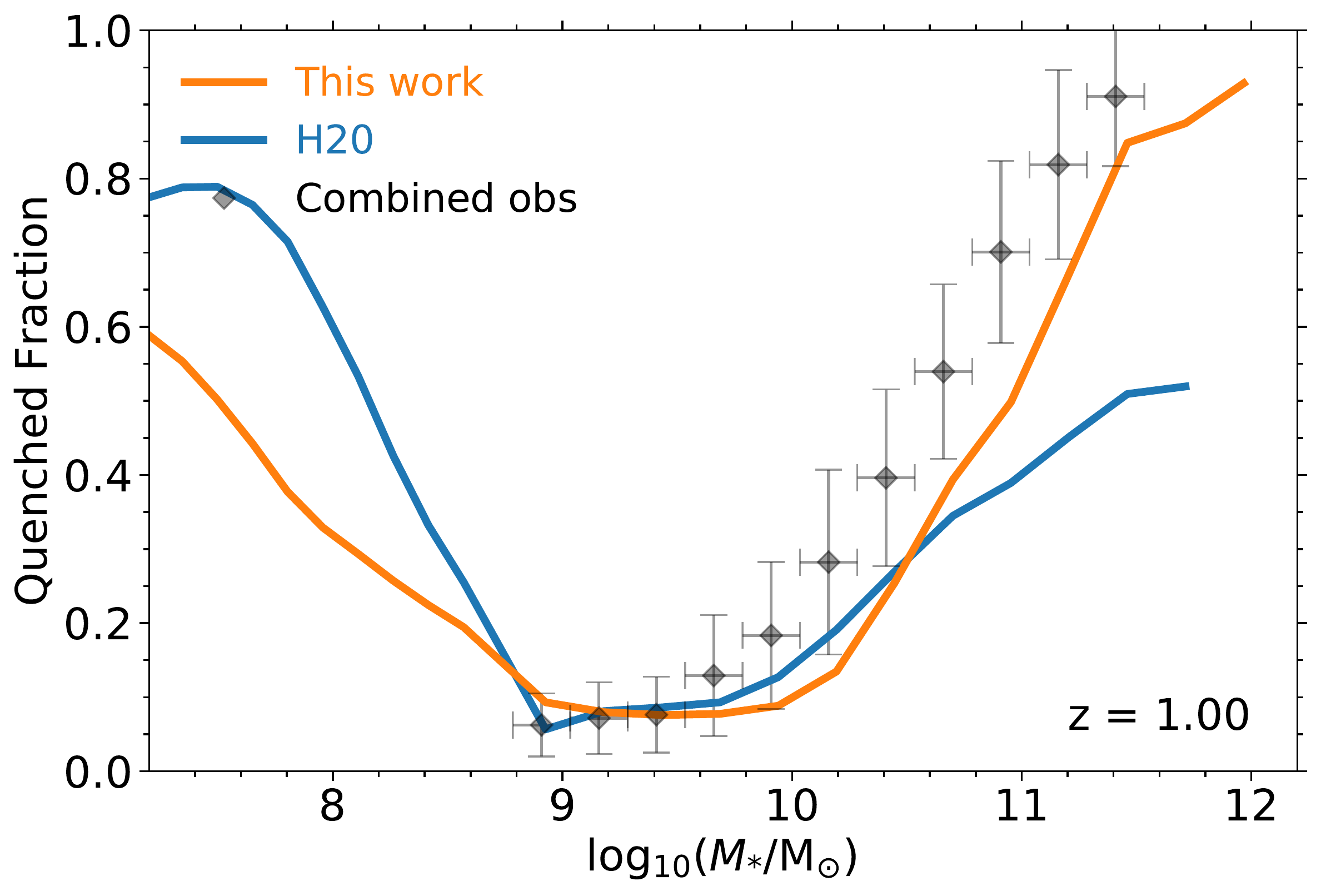}
    \includegraphics[width=0.33\textwidth]{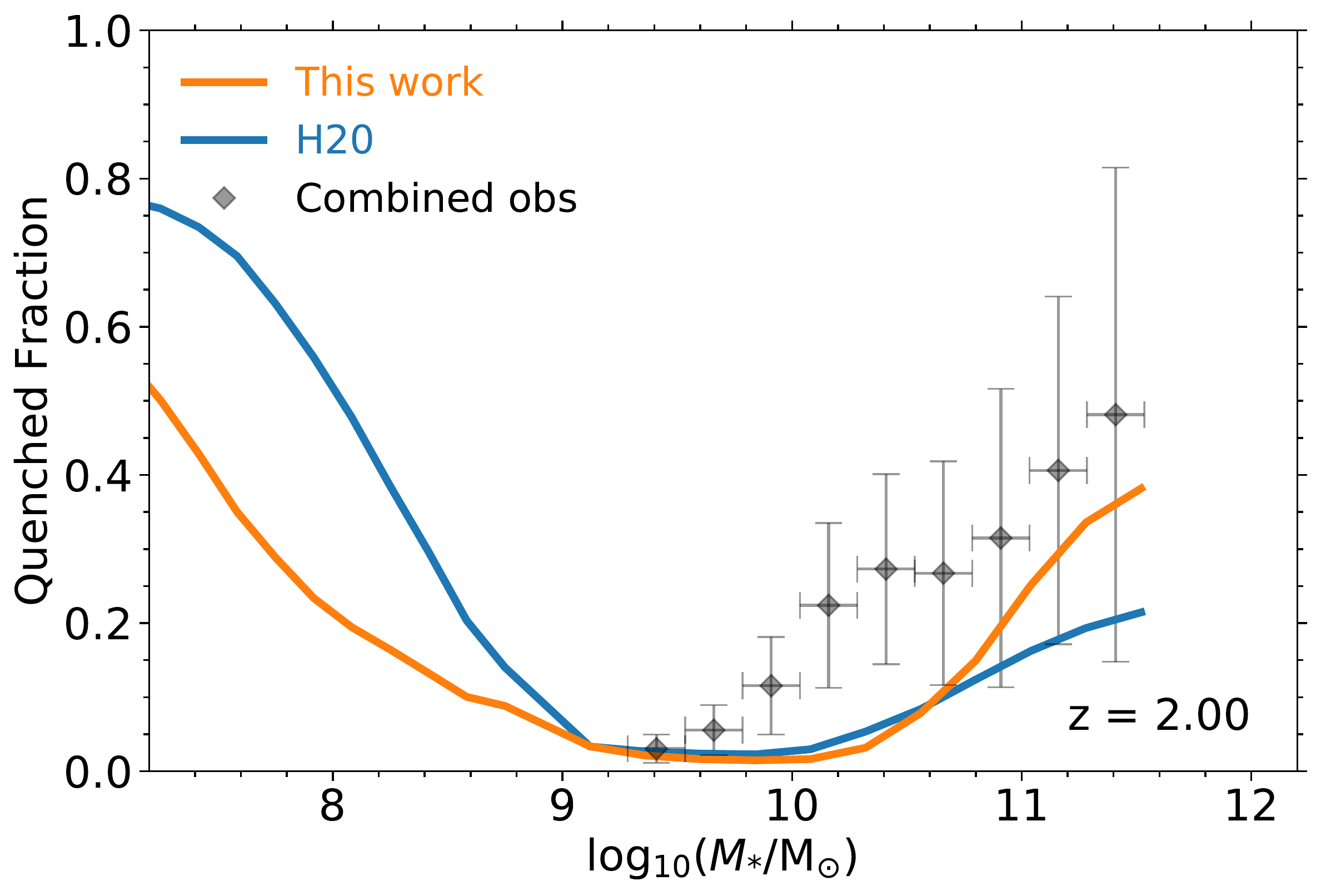}
    \caption{Model results for the quantities used as observational constraints in the MCMC in comparison with observations and with the H20 model. The top panels show the stellar mass functions and the bottom panels show the fraction of quenched galaxies. The combined observations used in the MCMC calibration are taken from H15 (see their Appendix 2 for more details) and include \protect\cite{baldry2008galaxy,baldry2012galaxy,li2009distribution} for SMF at $z=0$, \protect\cite{marchesini2009evolution,marchesini2010most,ilbert2010galaxy,ilbert2013mass,dominguez2011evolution,muzzin2013evolution,tomczak2014galaxy} for the SMF at $z>0$, \protect\cite{Bell2003Optical,Baldry2004Quantifying} for the quenched fraction at $z=0$ and \protect\cite{muzzin2013evolution,ilbert2013mass,tomczak2014galaxy} for the quenched fraction at $z>0$.}
\label{Fig: Obs_constraints_MCMC}
\end{figure*}

Like all galaxy formation models and simulations, \textsc{L-Galaxies} has a number of parameters (e.g. the star formation efficiency) that need to be fit. Therefore, we calibrate our new model against a set of observational constraints using the Markov Chain Monte Carlo (MCMC) approach developed by \cite{henriques2009monte} and used by previous \textsc{L-Galaxies} versions \citep[e.g.][]{Guo2013Galaxy,henriques2020galaxies}. To speed up the model calibration, the MCMC approach runs the model on a representative set of halo merger trees \citep[see][]{Henriques2013Simulations} instead of on the full simulations, a scheme we also adopt. In addition, we extend the previous MCMC method to also incorporate our new treatment of environmental effects.

We use six independent observational constraints: the stellar mass function and the fraction of quenched galaxies, each at $z=0,1,2$. Unlike H20, we do not use the HI gas mass function at $z=0$ as an observational constraint. However, our final model output for the HI mass function is in relatively good agreement with observations and H20 (bottom left panel of Fig. \ref{Fig: coldgas_comb_plot}). Choosing proper observational constraints and weighting them to converge to an acceptable fit can be a tricky task. For instance, weighting all observational constraint equally, would result in a rather bad fit for the stellar mass function at $z=0$, especially for galaxies with $10<\log_{10}(M_{\star}/{\rm M_{\odot}})<11$. Ultimately, exploring different weightings for different datasets, we find it best to give the highest weight to the observational constraints at $z=0$. Furthermore, at $z=0$ itself, we give the stellar mass function a higher weight than the quenched fraction. To properly fit the $z=0$ stellar mass function for $M^{\star}$ galaxies, we give an additional weight to the stellar mass function at $10<\log_{10}(M_{\star}/{\rm M_{\odot}})<11$. At $z>0$, observational constraints are weighted equally.

We run the MCMC for several tens of thousands of steps, i.e. we execute our model with different free parameters tens of thousands of times. During the calibration, we use \textsc{L-Galaxies} run on the Millennium simulation for galaxies with $\log_{10}(M_{\star}/{\rm M_{\odot}})>9$ and \textsc{L-Galaxies} run on Millennium-II for lower stellar masses. This stellar mass transition value is chosen following H20 and also by
monitoring the approximate stellar mass where the two runs converge
for a few smaller runs of the model \citep[see also][]{guo2011dwarf,henriques2020galaxies}. We note that our model has one fewer free parameter than H20, as we no longer limit ram-pressure stripping to satellites of massive clusters. Table \ref{tab: MCMC_free_params} compares our best fit parameters to previous models. The relevant equation for each free parameter is also given in that table. The description of the equations can be found in the supplementary material of this paper. In addition, histograms of the PDFs of accepted MCMC steps for all free parameters are shown in a corner plot in Fig. \ref{Fig: Corner_MCMC}. 2D contour plots show the marginalised (and normalised) 2D posterior distributions for all the possible free parameter pairs of our model. Furthermore, 1D histograms depict the individual constraints on each parameter. The vertical dashed lines show the best fit from our model (green) and from H20 (purple), and the value quoted indicates the best fit value (in linear units).

Our best fits to the observational constraints are shown in Fig. \ref{Fig: Obs_constraints_MCMC} in comparison with observations and the H20 model. The top panels of Fig. \ref{Fig: Obs_constraints_MCMC} show that our fits (orange lines) to the stellar mass functions at different redshifts are slightly better than H20, in particular for low-mass galaxies. In addition, the bottom panels show that we have a higher fraction of massive quenched galaxies at higher redshifts, in better agreement with observational data. This is primarily due to the higher AGN feedback efficiency parameter (see Table \ref{tab: MCMC_free_params}). Overall, our model fits the targeted observational constraints successfully.

We emphasise that we \textit{do not} include any environment-dependent quantities in the observational constraints used in our model calibration. As a result, we retain all environmental dependencies and correlations related to galaxy evolution as \textit{predictions}, rather than the result of fits to observations. Our observational constraints, i.e. the stellar mass function and quenched fraction, are global values and do not distinguish between field galaxies and those in dense environments. In the rest of this work, we study our model's predictions regarding the properties of galaxies in different environments.


\section{Observational data and Mock catalogues}

\label{sec: obs_data}
\subsection{Observational data}

\label{subsec: obs_data}
\subsubsection{The fraction of quenched galaxies at $z=0$}
\label{subsubsec: methods_sdss_z0}

For consistency and to ensure the most robust comparisons between our model and empirical data, we carry out a new analysis of observations which constrain galaxy quenched fractions as a function of environment. Our galaxy sample at $z=0$ is taken from the Sloan Digital Sky Survey Data Release 7 (SDSS-DR7, \citealt{Abazajian2009SDSS}), with stellar masses and star formation rates calculated using the methodologies described in  \cite{Kauffmann2003Stellar,Brinchmann2004Physical,Salim2007UV}. We cross-match the SDSS-DR7 catalogue with the group catalogue from \cite{Yang2005halo-based,Yang2007Galaxy} to identify the most massive galaxy of each group, the so-called central galaxy. We note that there are a lot of complexities associated with group-finding in observational data \citep[e.g. see][]{Bravo2020From,Tinker2020Self-calibrating}. We estimate the halo $M_{200}$ and $R_{200}$ using our galaxy formation model by fitting the central galaxy stellar to halo mass relation as described in \S \ref{subsubsec: halomass_mock_simulation} and Appendix \ref{app: halomass_stellarmass_fit}.

For each halo, we find galaxies with line-of-sight velocity separation from the central galaxy in the halo $|v_{\rm gal,LOS}-v_{\rm halo,LOS}|\, \leq \pm 10V_{\rm 200,halo}$ which lie within a 2D projected halocentric distance of $10R_{200}$. We note that, for most of our analysis we only take galaxies with $|v_{\rm gal,LOS}-v_{\rm halo,LOS}|\, \leq \pm 2V_{\rm 200,halo}$ (e.g. see Figs. \ref{Fig: quenchedFrac_dis_proj_z0},\ref{Fig: quenchedFrac_M*_proj_z0}). In this work, we limit our SDSS galaxy sample to galaxies with $\log_{10}(M_{\star}/{\rm M_{\odot}})\geq9.5$ and our observational halo sample to haloes with $\log_{10}(M_{200}/{\rm M_{\odot}})\geq12$. The redshift interval of our host haloes is taken as $0.01\leq z\leq0.04$. Finally, we adopt the definition that galaxies with $\log_{10} (\rm SSFR/yr^{-1})<-11$ are quenched.

A similar analysis has been carried out by \cite{Wetzel2012Galaxy,Wetzel2014Galaxy}, where they use the SDSS data and a modified \cite{Yang2007Galaxy} halo catalogue to study the properties of galaxies in different environments. In contrast, our new analysis is based on a different halo catalogue, halo mass derivation method, stellar and halo mass bins, and so on. Furthermore, the analysis in this paper is designed so that the results can be compared directly and self-consistent with our SAM model predictions.

\subsubsection{The fraction of quenched galaxies at $z>0$}
\label{subsubsec: methods_z>0}

At $z>0$, we compare our results with the fraction of quenched galaxies inferred from the Hyper Suprime-Cam Subaru Strategic Program (HSC-SSP) survey data of optical broadband photometry from \cite{Pintos-Castro2019Evolution}. Their analysis is based on the HSC-SSP project \citep{Aihara2018HSC}. In their study, galaxy clusters are taken from the SpARCS \citep{Muzzin2009Spectroscopic,Wilson2009Spectroscopic} fields, using a modified cluster red sequence algorithm by \cite{Gladders2000New,Gladders2005Red-Sequence,Muzzin2008Evolution} that employs photometric redshifts.

\subsection{Mock catalogues}
\label{subsec: mock_catalogues}

In order to make a fair comparison with observations, we create mock catalogues both from our own model and from that of H20. The details are explained in this subsection.

\subsubsection{Projection in 2D}
\label{subsubsec: 2d_projection}

In order to compare with the SDSS data at $z=0$, to locate galaxies in the vicinity of each halo in our model, we first transform the galaxy distribution into redshift/velocity space. This is done using their positions and peculiar velocities along the z-axis of the simulation domain, accounting for expansion due to the Hubble flow. For each simulated halo we locate nearby galaxies with $|v_{\rm gal,LOS}-v_{\rm halo,LOS}|\, \leq \pm 2V_{\rm 200,halo}$ within a 10 Mpc projected halocentric distance.

For our comparison with the HSC data \citep{Pintos-Castro2019Evolution} at $z>0$, we adopt the projection made by \cite{Pintos-Castro2019Evolution} within a redshift slice of
\begin{equation}
    \Delta z = z_{\rm halo}\pm 0.05\times (1+z_{\rm halo}).
\end{equation}
The physical distance corresponding to $\Delta z$ could exceed our simulation box size for some redshifts. Therefore for simplicity we project the entire simulation box along the z-axis of the simulation volume ($l_{\rm box}\sim 700\, \rm Mpc$ for the Millennium simulation in the Planck cosmology).

\subsubsection{Deriving halo mass and radius}
\label{subsubsec: halomass_mock_simulation}

For each central galaxy in both models and observations we calculate a halo mass, $M_{200}$, from its stellar mass with the fitting formula below (motivated by \citealt{Yang2003Costraining,Moster2010Constraints}):
\begin{equation}
\label{eq: Mstellar_to_M200_conversion}
\log_{10} (M_{200}/{\rm M_{\odot}}) = \alpha_1  \log_{10} (M_{\star}/{\rm M_{\odot}}) + \beta_1,
\end{equation}
where $\alpha_1$ and $\beta_1$ are free parameters which we derive using our final, calibrated model output (see Figs. \ref{Fig: append_SMHM},\ref{Fig: append_M200_mock} in appendix \ref{app: halomass_stellarmass_fit}): $\alpha_1 = 1.65$, $\beta_1 = -5.16$ for $\log_{10}(M_{\star}/{\rm M_{\odot}})\geq10.5$, and $\alpha_1 = 0.80$, $\beta_1 = 3.70$ for $\log_{10}(M_{\star}/{\rm M_{\odot}})<10.5$.
The above equation is valid for galaxies with $\log_{10}(M_{\star}/{\rm M_{\odot}})>9.5$. The stellar mass at which $\alpha_1$ and $\beta_1$ change is $\log_{10}(M_{\star}/{\rm M_{\odot}})=10.5$ which roughly corresponds to haloes with $\log_{10}(M_{200}/{\rm M_{\odot}})\sim 12.2$. For each halo, $R_{200}$ is calculated from $M_{200}$ through
\begin{equation}
    \label{eq: R200_M200_conversion}
    R_{200} = \left( \frac{3M_{200}}{4\pi \times 200\rho_{\rm crit}} \right)^{1/3}.
\end{equation}
We note that independent values of halo mass and radius are given in the \cite{Yang2005halo-based,Yang2007Galaxy} catalogue as well. There were calculated from abundance matching using the observed stellar mass function and an analytical halo mass function. In contrast, here we use the median stellar mass to halo mass ratio at each stellar mass bin in our model, which does not invoke the same assumptions as abundance matching. In practice, the differences between their values and ours are quite small.

For our direct comparisons with the SDSS observations, wherever the halo mass, $M_{200}$ and the halo radius, $R_{200}$, are needed, we derive $M_{200}$ using Eq. \ref{eq: Mstellar_to_M200_conversion}. The corresponding $R_{200}$ is then derived using Eq. \ref{eq: R200_M200_conversion}. We do this for the data \textit{as well as} for the models. We note that such a derivation of $M_{200}$ and $R_{200}$ leads to considerable scatter in halo mass and radius compared with the actual values measured directly from the simulation. This is explained in more detail in Appendix \ref{app: halomass_stellarmass_fit} (see Figs. \ref{Fig: append_SMHM}, \ref{Fig: append_M200_mock}). 

For our comparison with the HSC data \citep{Pintos-Castro2019Evolution} at $z>0$, we use the direct simulation halo masses rather than making a mock catalogue. This is because there are many complexities in the technique that work employs to derive halo masses, that are beyond the level we wish to emulate in this paper. We note that this makes our high redshift comparisons less reliable (see \S \ref{sec: environmental_dependency}).

\subsubsection{Homogenization of the halo mass distributions between models and observations}
\label{subsubsec: Homogenization_M200}

In general, the halo mass distribution within a given halo mass bin (e.g. $14<\log_{10}(M_{200}/{\rm M_{\odot}})<15$) can differ between the observations and the models. This bias could affect our results and conclusions. Therefore, we re-sample the distribution of simulated haloes to match the observed distributions.

Assume that we want to compare the quantity "$Q$" (e.g. the fraction of quenched galaxies) between a model and an observation, as a function of $M_{200}$. We divide each halo mass bin (e.g. $14<\log_{10}(M_{200}/{\rm M_{\odot}})<15$) in the model into smaller sub-bins and calculate the quantity $Q$ within those sub-bins. The final value of the quantity $Q$ within the larger bin is
\begin{equation}
    Q = \displaystyle\sum_{\rm i=1}^{\rm n} Q_{\rm i}W_{\rm i},
\end{equation}
where $Q_{\rm i}$ is the value (mean, median, etc.) of $Q$ within the ith sub-bin and "n" is the number of sub-bins. In addition, $W_{\rm i}$ is the weight of the ith sub-bin, defined as the ratio of the fraction of haloes in the ith sub-bin of the observational sample to the fraction of haloes in the ith sub-bin of the model. This homogenization is only done for our comparison with the SDSS data at $z=0$.


\section{Results}
\label{sec: environmental_dependency}

\subsection{Gas stripping through time}
\label{subsec: total_stripped_gas}

\begin{figure}
    \includegraphics[width=0.95\columnwidth]{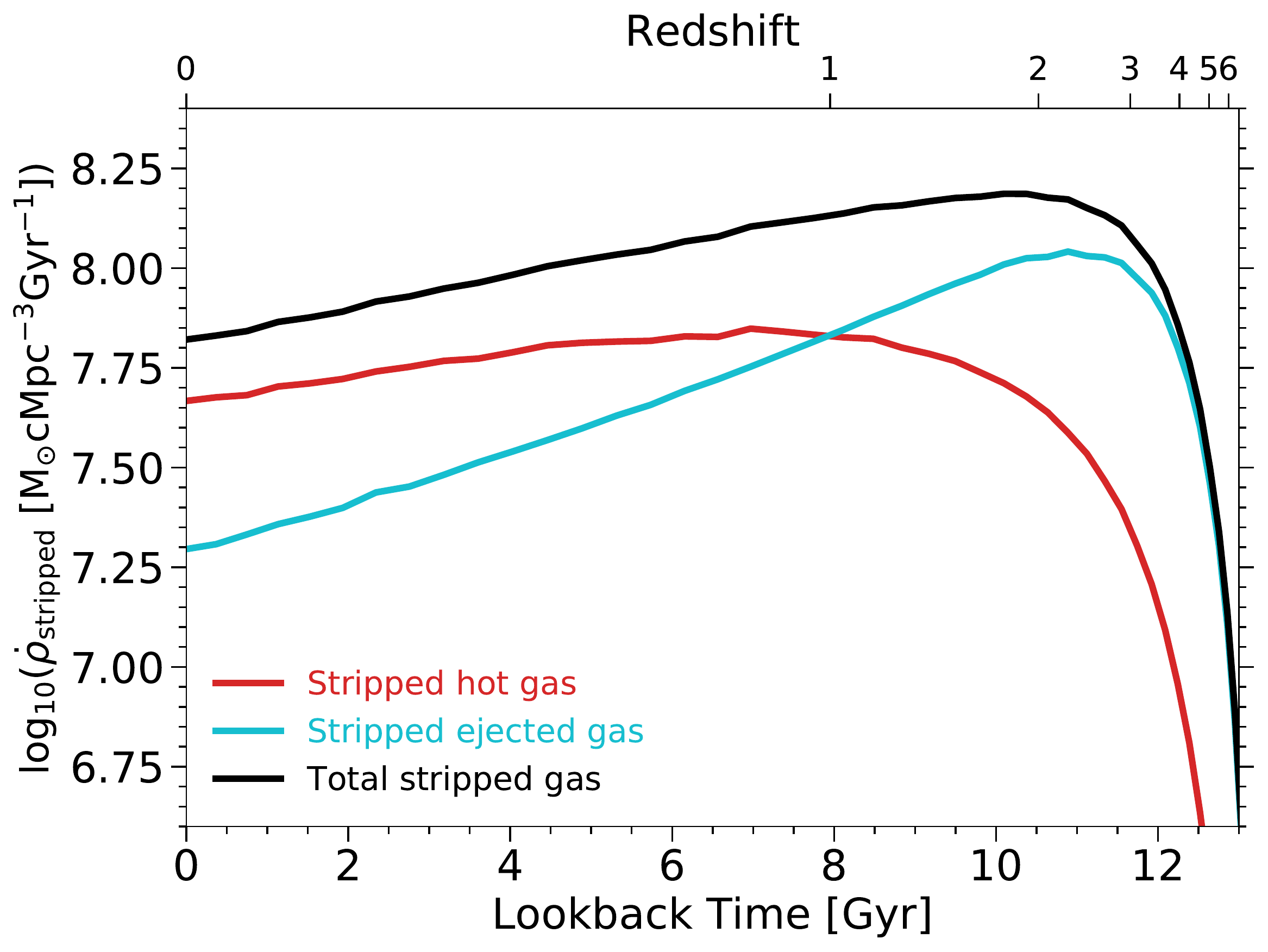}
    \caption{Time evolution of the stripping rate density as defined in Eq. \ref{eq: rho_stripped}. The red and cyan lines correspond to the stripping rate density of the hot gas (red) and the ejected reservoir (cyan), respectively. Summing the two, the black line is the total gas mass stripping rate density.}
\label{Fig: Stripped_density_vs_redshift}
\end{figure}

\begin{figure*}
    \includegraphics[width=0.33\textwidth]{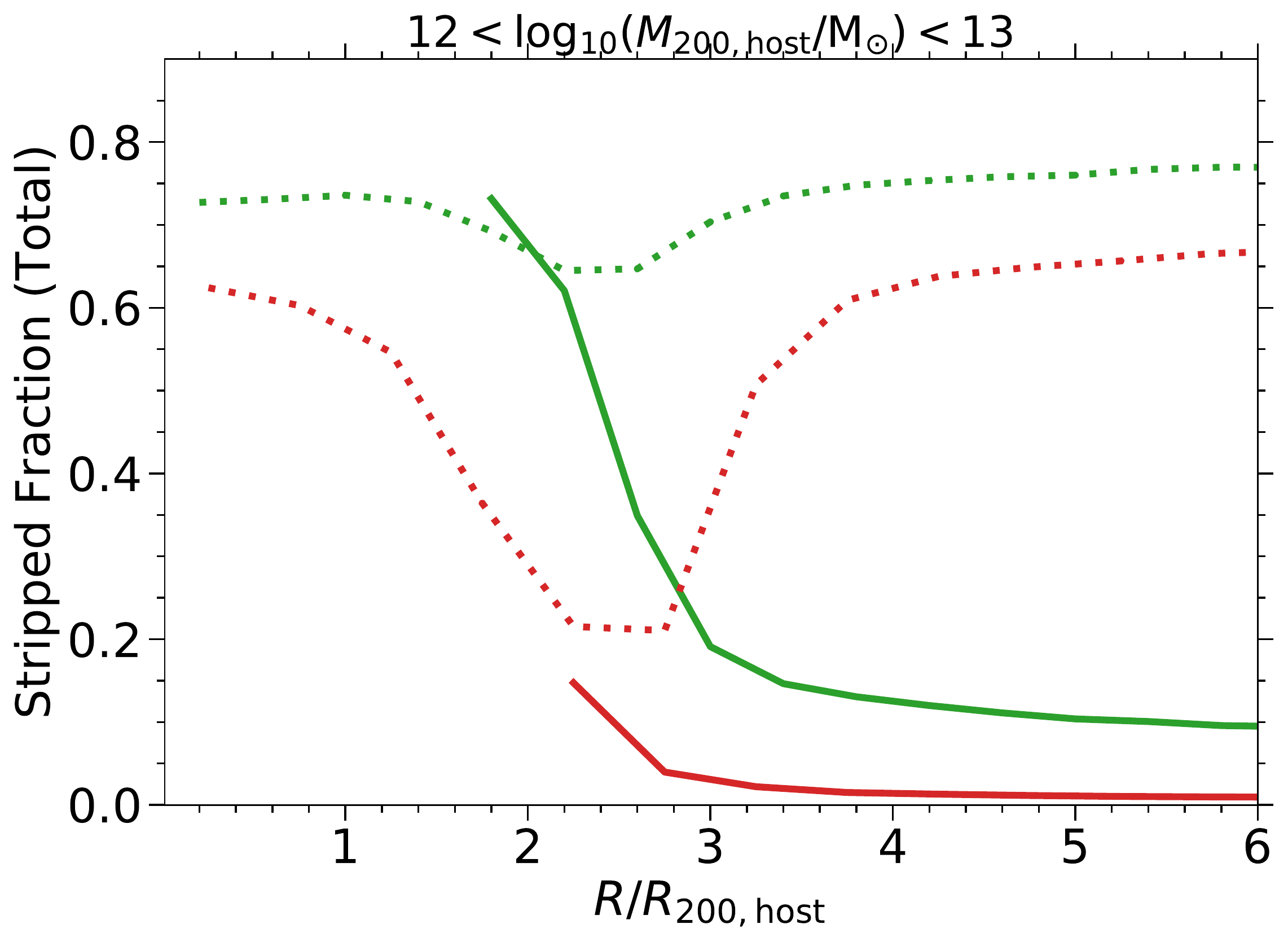}
    \includegraphics[width=0.33\textwidth]{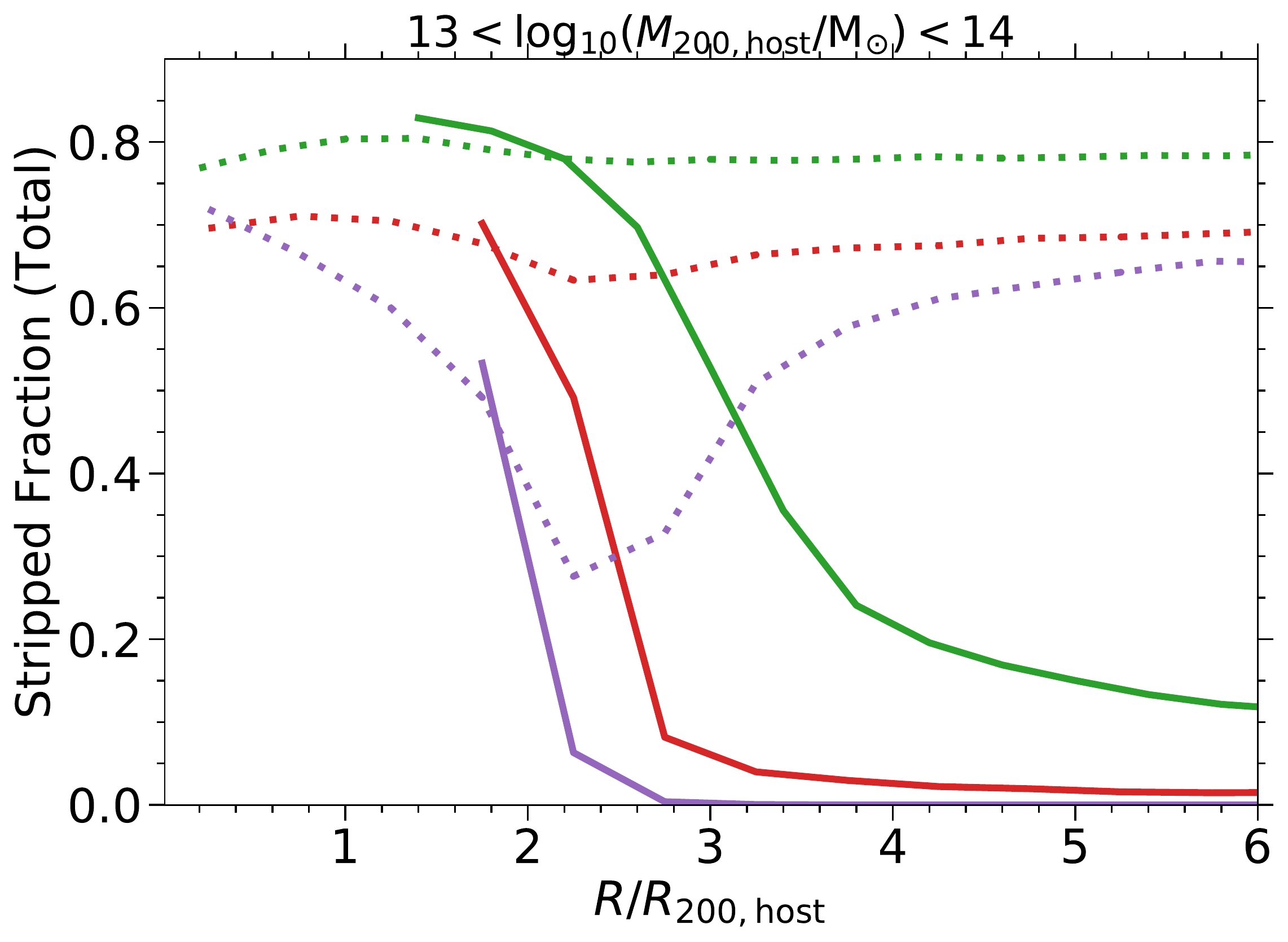}
    \includegraphics[width=0.33\textwidth]{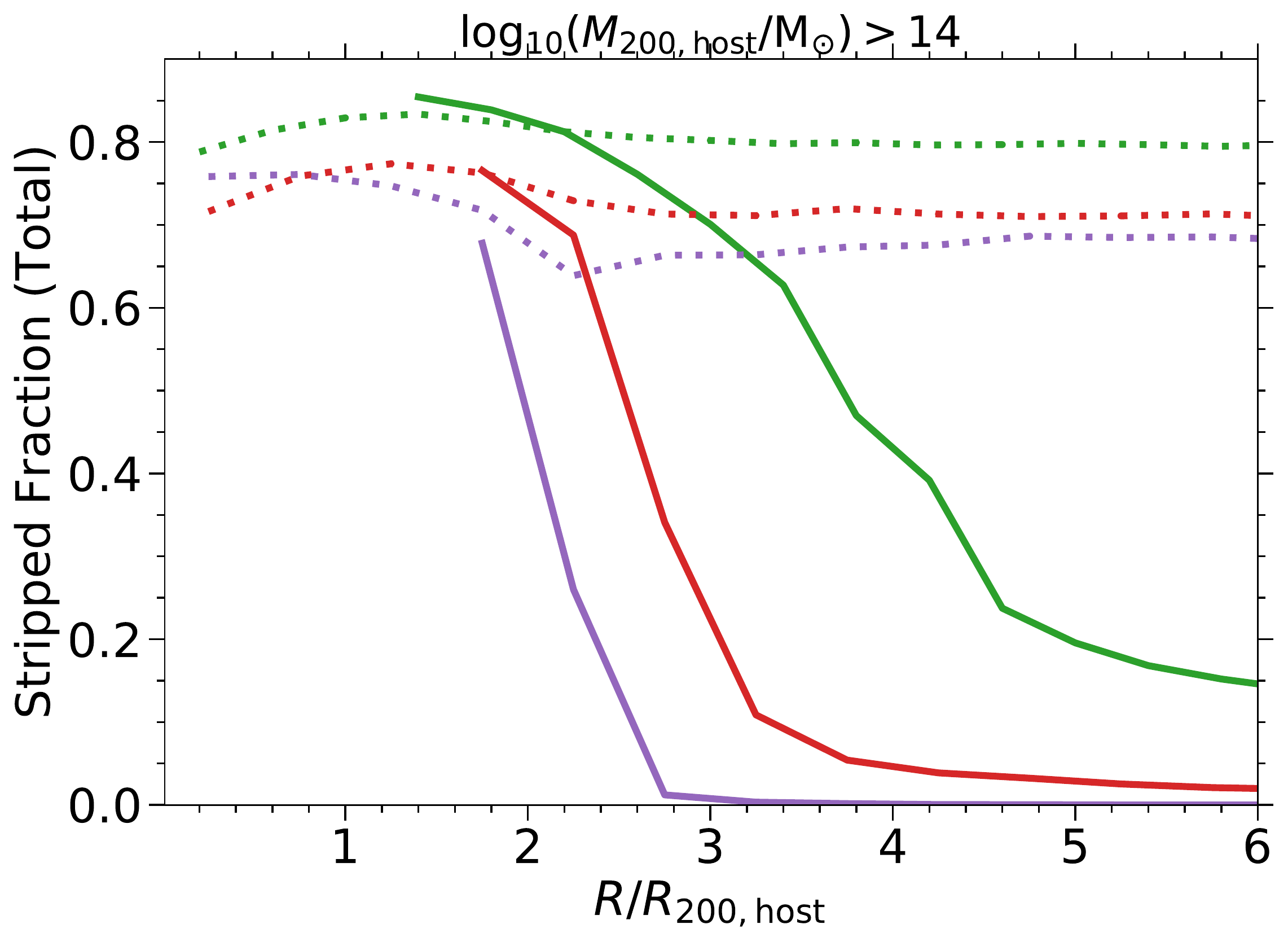}

    \includegraphics[width=0.33\textwidth]{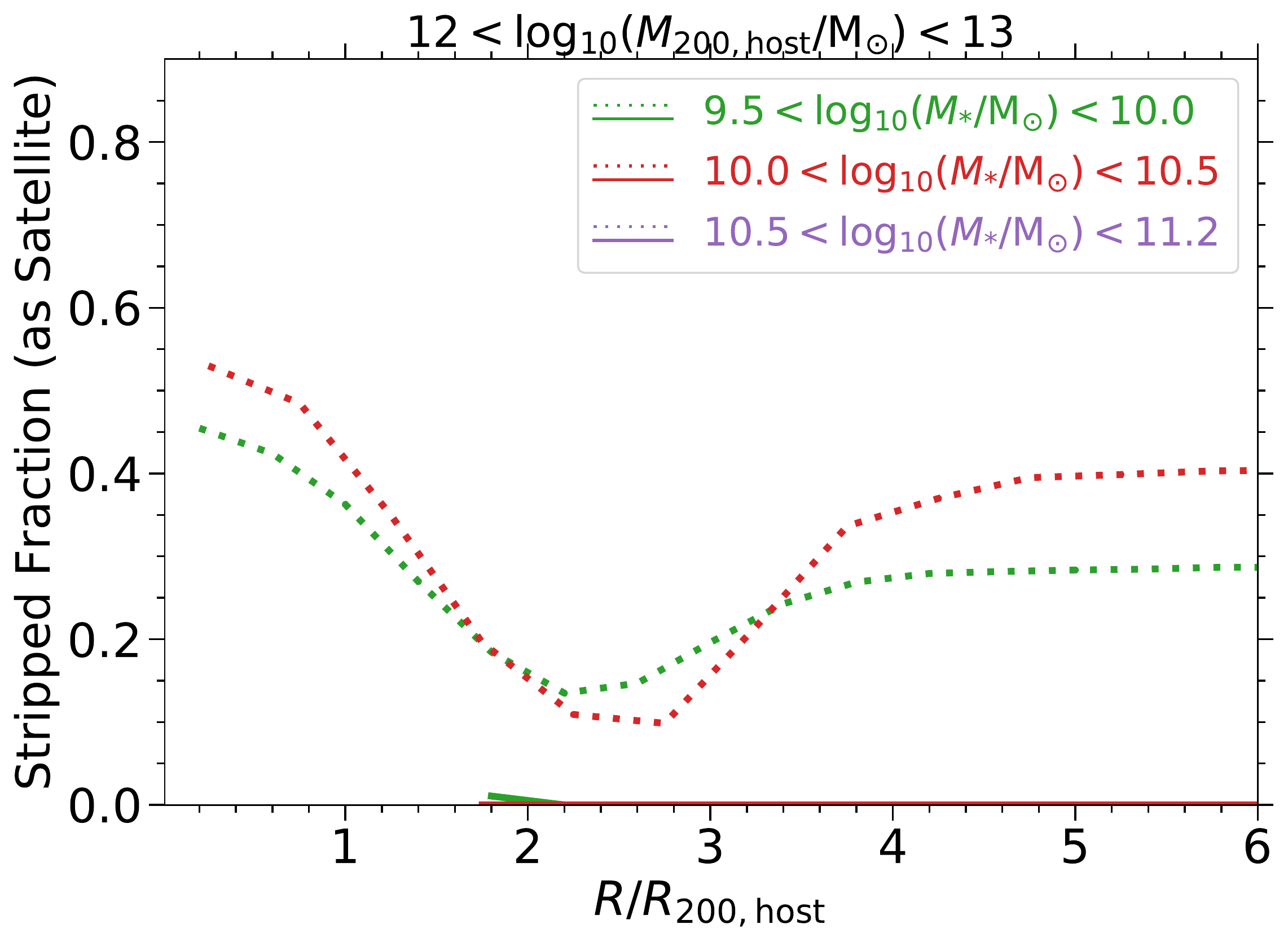}
    \includegraphics[width=0.33\textwidth]{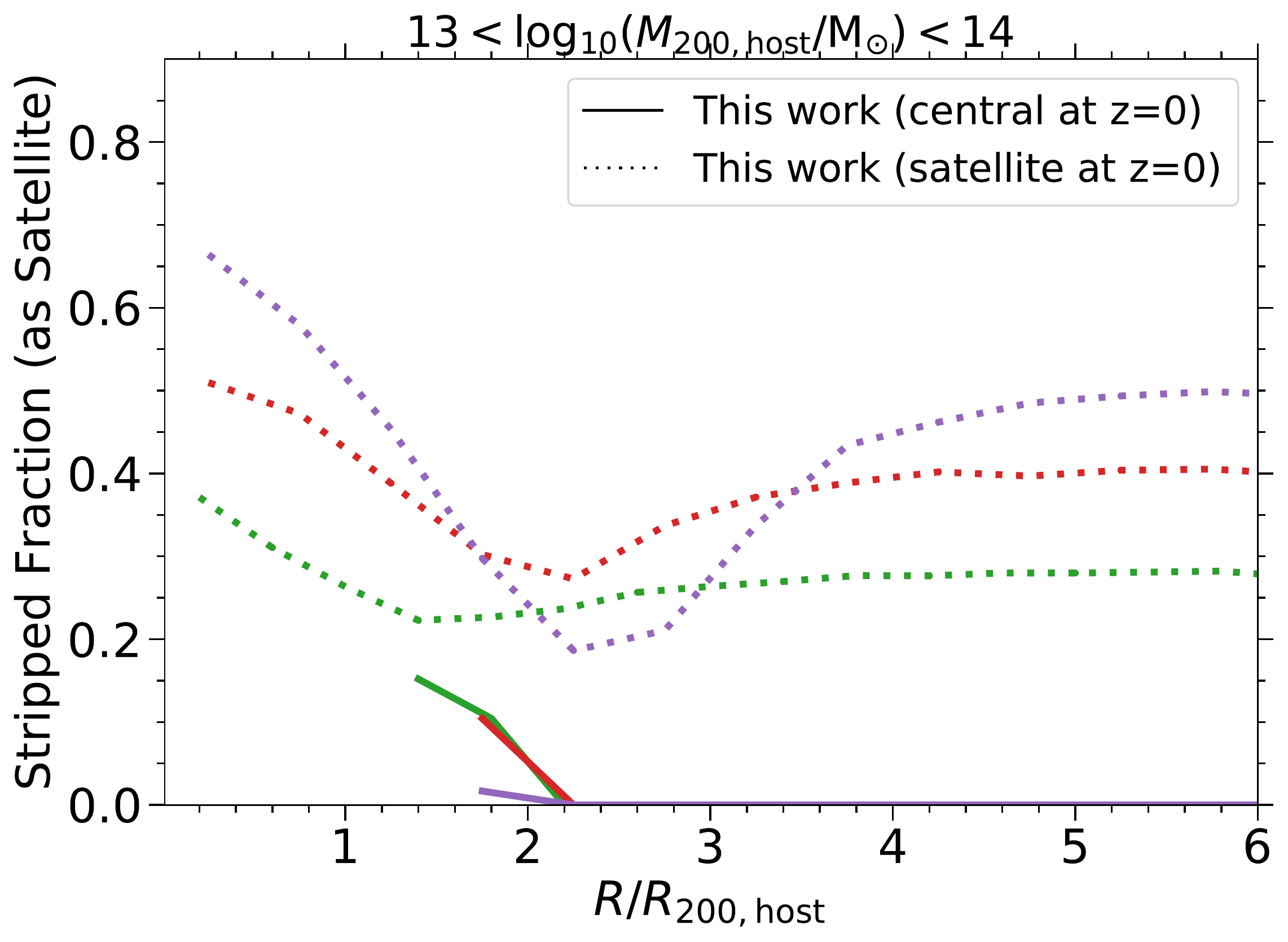}
    \includegraphics[width=0.33\textwidth]{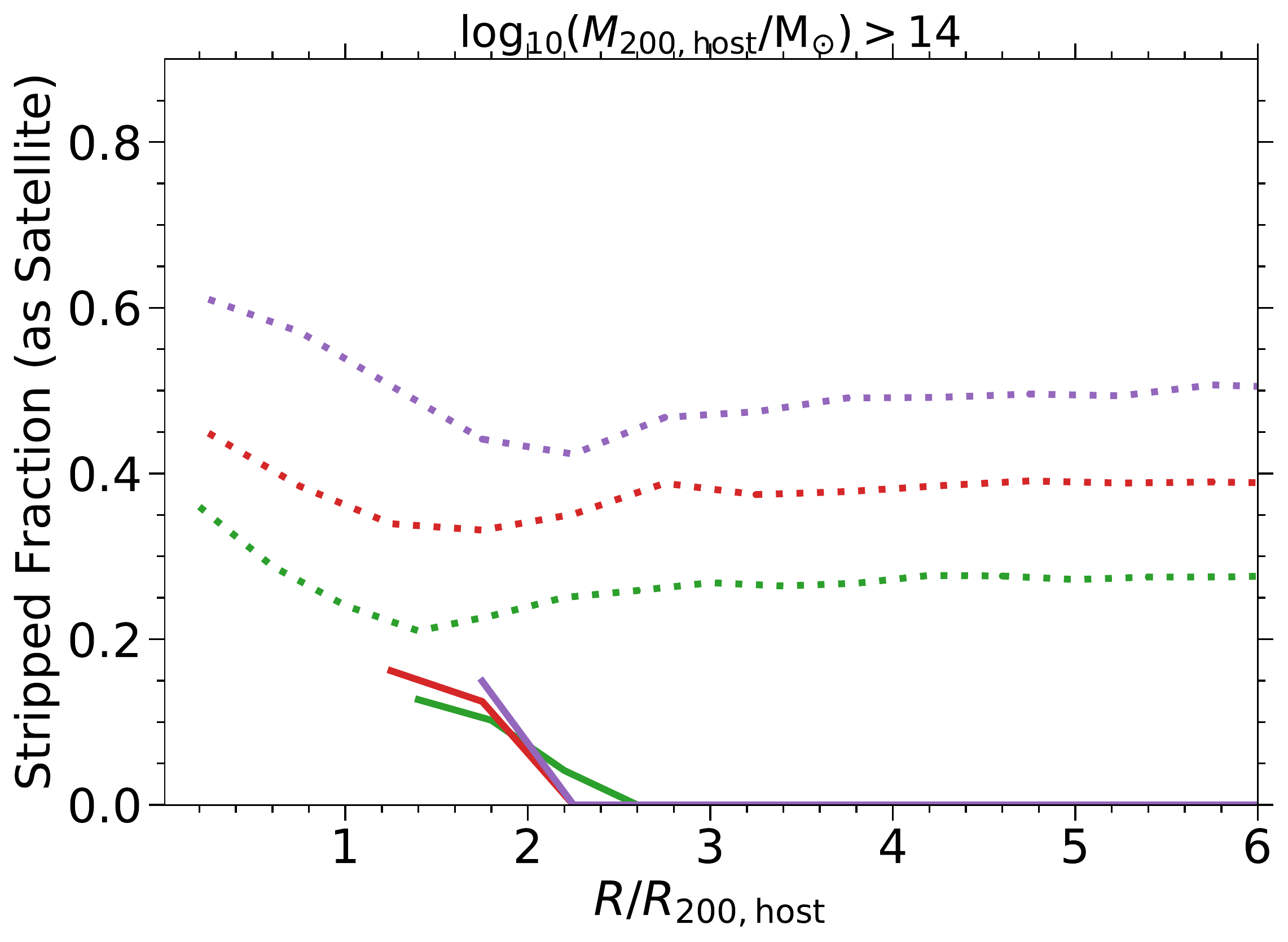}
    
    \includegraphics[width=0.33\textwidth]{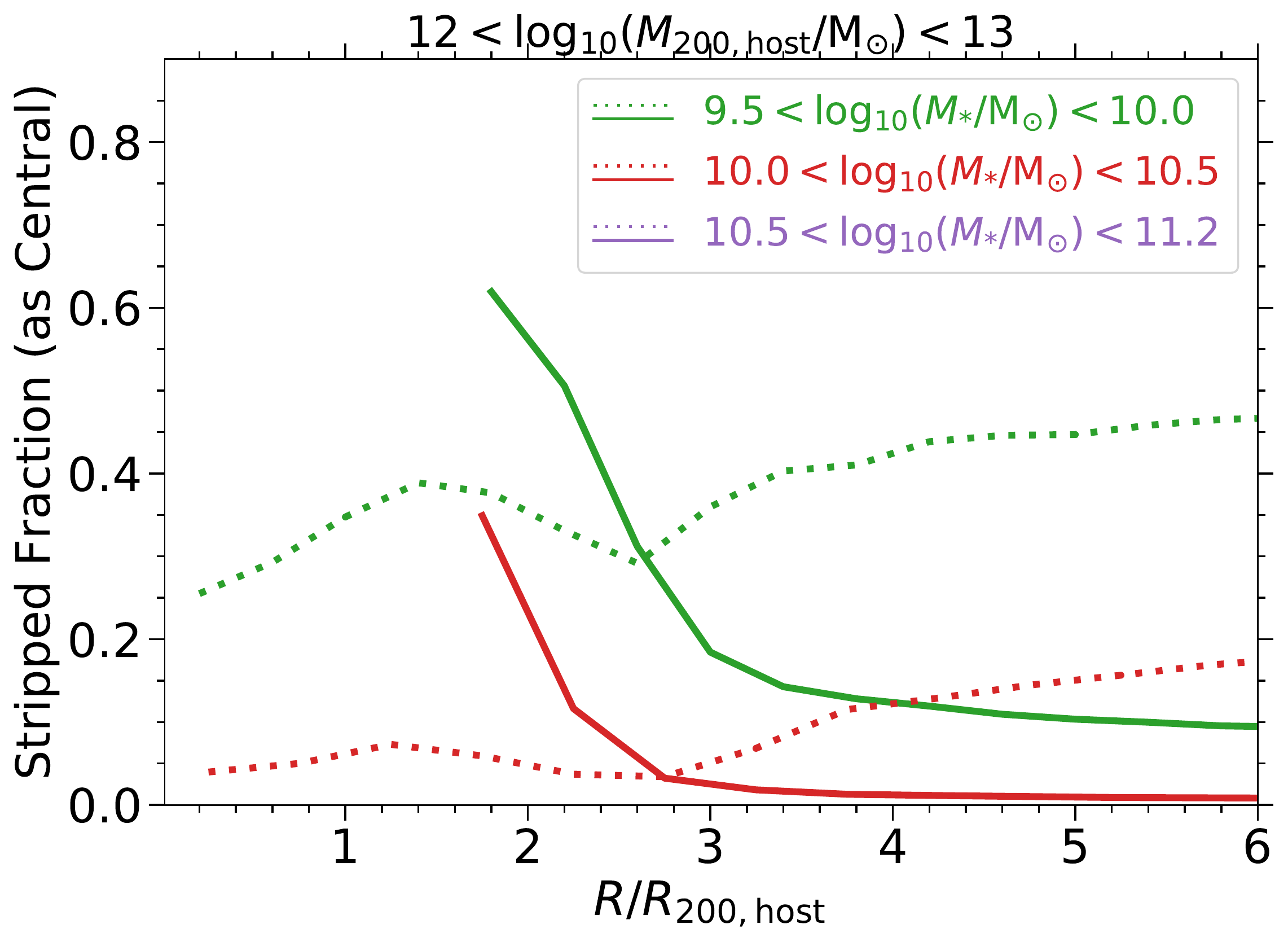}
    \includegraphics[width=0.33\textwidth]{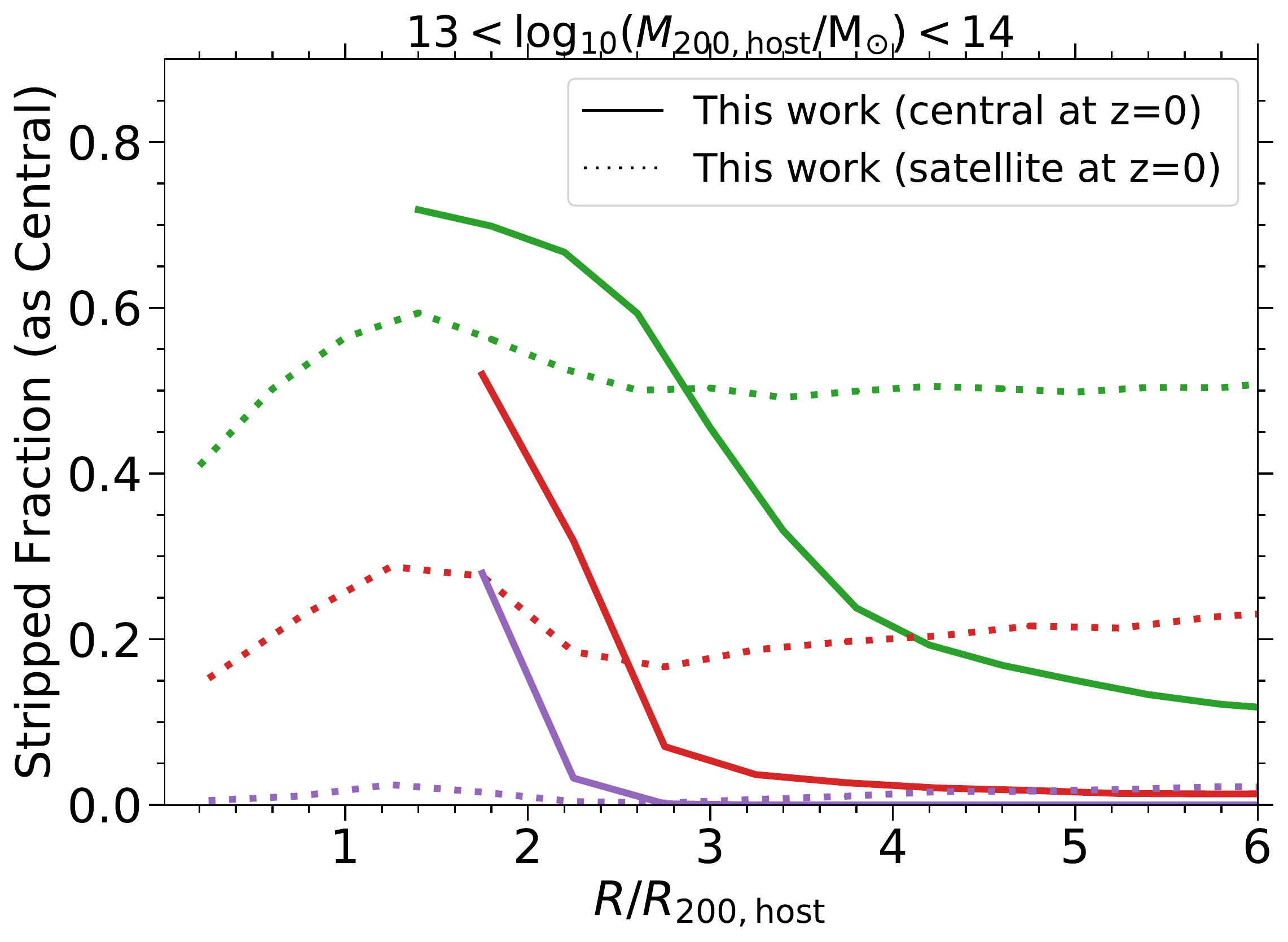}
    \includegraphics[width=0.33\textwidth]{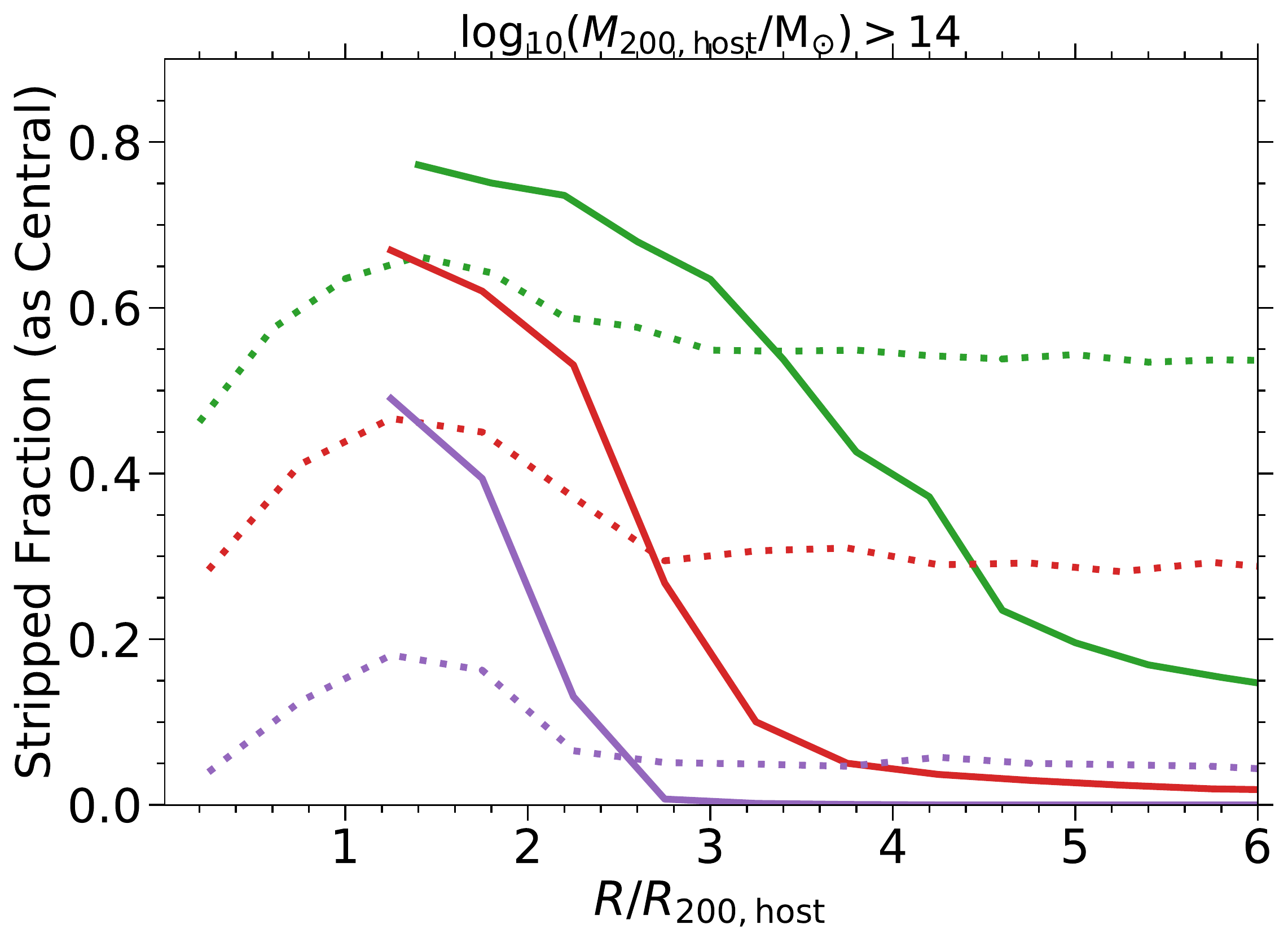}
    
    \caption{The median cumulative stripped fraction of gas (Eq. \ref{eq: f_strippedhot}) in galaxies as a function of halocentric distance at $z=0$. Different columns show results for different central host halo masses. In all the panels, different colours correspond to different stellar mass ranges (indicated in the legends) and different line styles correspond to central galaxies (solid) versus satellites (dotted). The top row shows the total stripped fraction, which is divided into two categories as shown in the middle and bottom rows: the cumulative gas stripped from galaxies when they were satellites (middle row) versus when they were centrals (bottom row). Galaxies experience significant stripping, both when they are satellites, and when they are centrals (see text for details).}
\label{Fig: totGasStripped_dis_z0}
\end{figure*}

As a first, global exploration of the role of gas stripping, we define the stripping rate density, $\dot{\rho}_{\rm stripped}(t)$, as the stripped mass in the entire simulation per unit time (gigayear), normalised to the simulation volume:
\begin{equation}
\label{eq: rho_stripped}
    \dot{\rho}_{\rm stripped}(t) = \frac{1}{V_{\rm simulation}}\displaystyle\sum^{\rm N_{\rm gal}}_{i=1} \dot{m}_{\rm stripped,i}(t)\,,
\end{equation}
where $\dot{m}_{\rm stripped,i}(t)$ is the gas stripping rate of the i$^{\rm th}$ galaxy per unit time, and the sum is over all galaxies ($\rm N_{\rm gal}$). In addition, $V_{\rm simulation}$ is the comoving simulation volume.

Fig. \ref{Fig: Stripped_density_vs_redshift} shows $\dot{\rho}_{\rm stripped}(t)$ as a function of lookback time (lower x-axis) as well as redshift (upped x-axis). Considering the redshift evolution of $\dot{\rho}_{\rm stripped}$, most of the stripped gas at $z>1.2$ is due to the stripping of the ejected material (cyan line) rather than the hot halo gas (red line), demonstrating a strong correlation between feedback processes and gas stripping at high redshifts. At $z<1.2$, stripping of the hot gas is the dominant stripping process.

In \textsc{L-Galaxies}, the outflows contributing to the ejected reservoir are caused by supernova feedback and are present mostly for low-mass galaxies. We note that AGN feedback in our model prevents the hot gas from cooling but does not eject it outside the halo. Therefore, it does not contribute to the material in the ejected reservoir. Some hydrodynamical simulations (e.g. IllustrisTNG) have strong outflows caused by black hole feedback in massive galaxies. As discussed in \cite{Ayromlou2020Comparing}, the presence of ejective black hole feedback results in a significant enhancement in ram-pressure stripping for massive galaxies. We will consider the possibility of ejective black hole feedback and study its influence on gas stripping in \textsc{L-Galaxies} in future work.

The total stripping rate density (black line) does not decrease monotonically with redshift, but it has a maximum at $z\sim 2$. The same is true for the ejected stripped density rate (blue line) with a maximum at $z\sim 2$, implying a close relationship to the peak of cosmic star formation rate density which also occurs at $z\sim 2$ (see Fig. \ref{Fig: cosmic_SFR_density}). Due to the maximal total star formation rate at $z\sim 2$ there is an abundance of powerful stellar feedback activity in \textsc{L-Galaxies}, and the largest fraction of ejected material at this cosmic epoch. This results in the suppression of the hot gas density in galaxies and, as a result, the suppression of the gravitational force between the galaxies and their hot halo gas (Eq. \ref{eq: rp_grav}). Therefore, the hot and ejected gas of galaxies are more easily stripped.

\begin{figure*}
    \centering
    \includegraphics[width=0.33\textwidth]{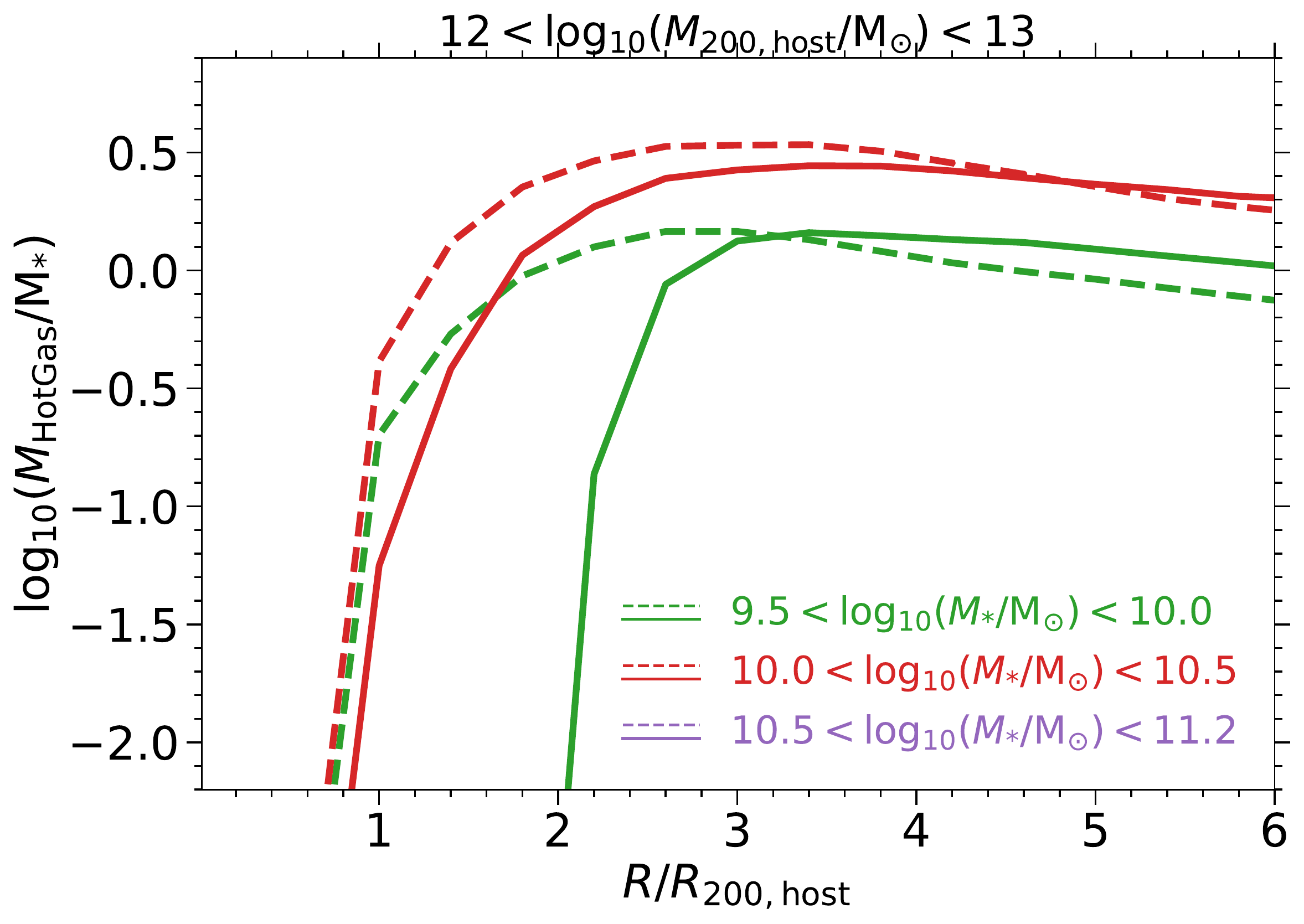}
    \includegraphics[width=0.33\textwidth]{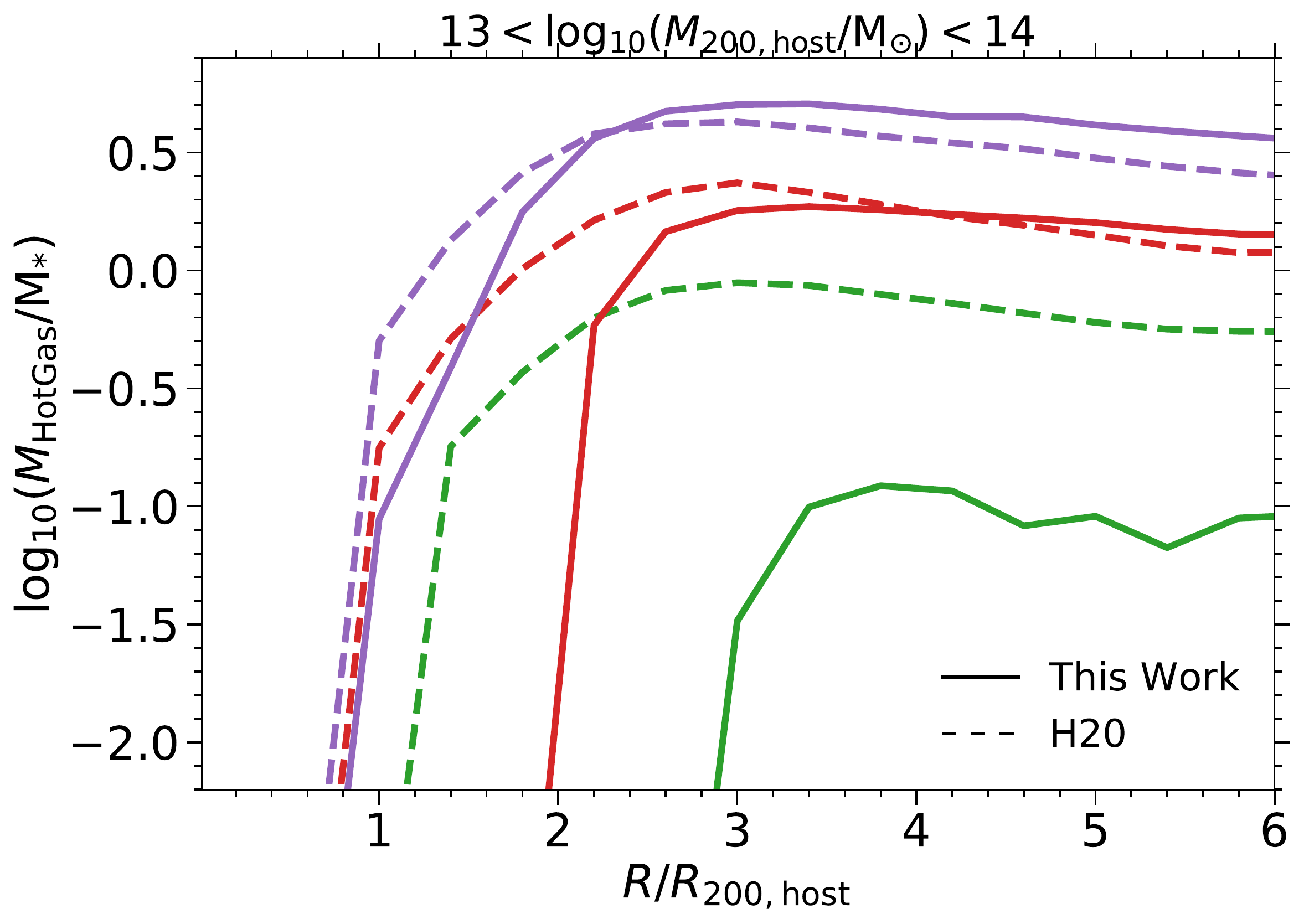}
    \includegraphics[width=0.33\textwidth]{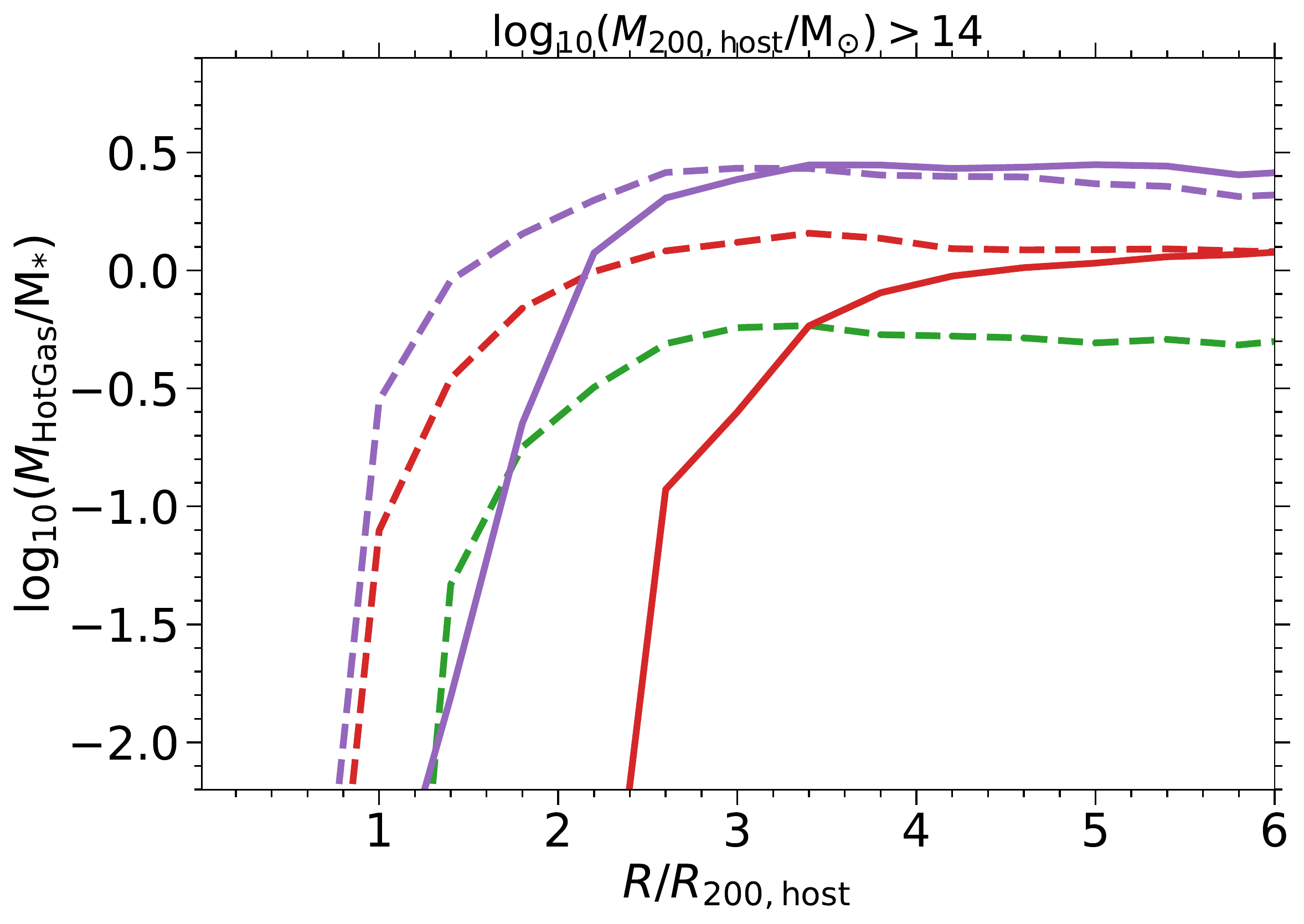}
    \includegraphics[width=0.4\textwidth]{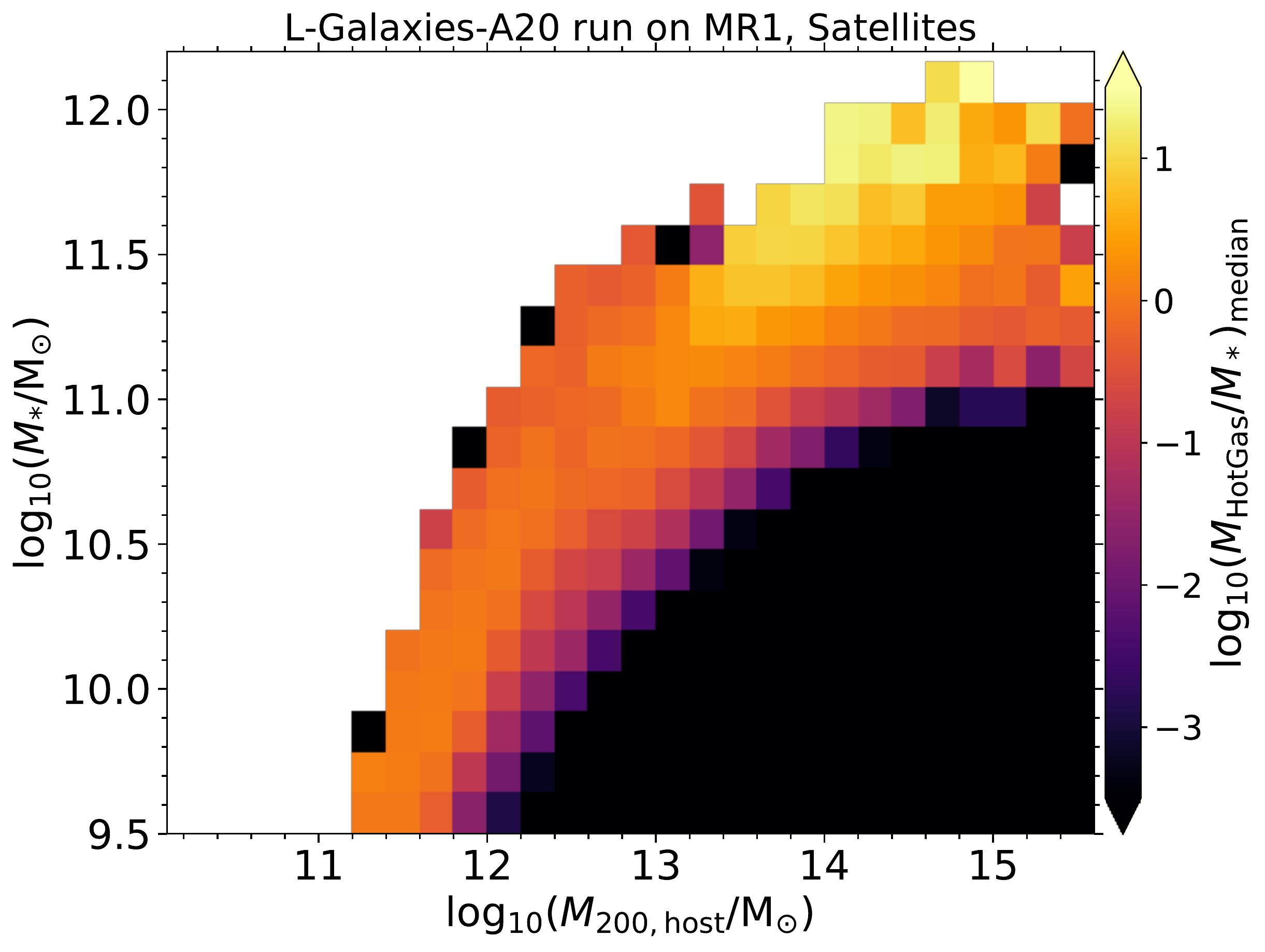}
    \includegraphics[width=0.4\textwidth]{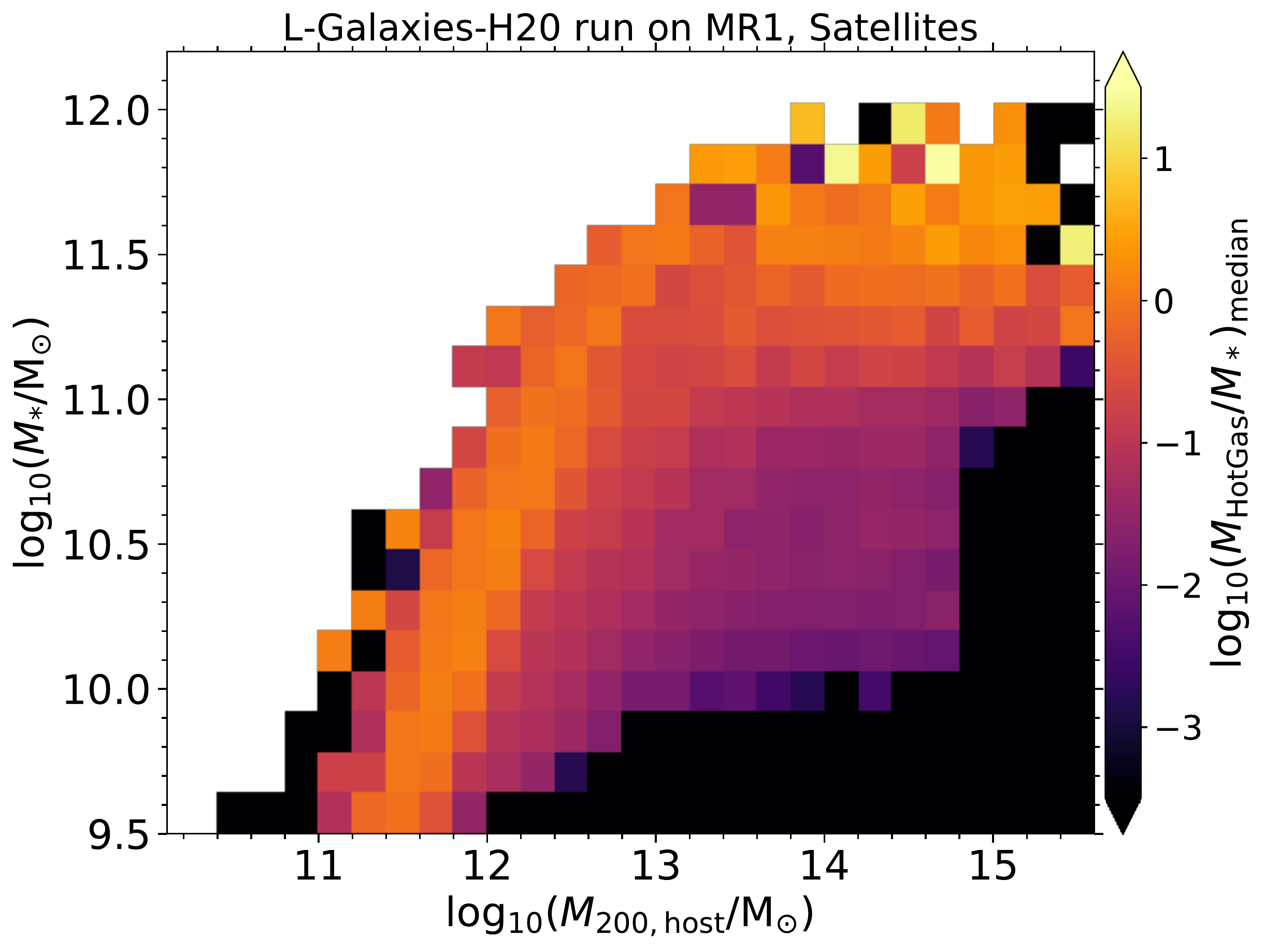}
    \caption{Top panel: Hot gas to stellar mass ratio as a function of halocentric distance at $z=0$, as a function of host mass (three columns), stellar mass (line colors), and model (line styles). Bottom panel: 2D histograms of the hot gas to stellar mass ratio of satellite galaxies as a function of their host halo mass (x-axis) and stellar mass (y-axis), contrasting this work (left panel) and H20 (right panel).}
\label{Fig: hotgas_comb_plot}
\end{figure*}

Fig. \ref{Fig: totGasStripped_dis_z0} shows the median cumulative stripped fraction (Eq. \ref{eq: f_strippedhot}) up to $z=0$, versus halocentric distance, for galaxies in the vicinity of clusters (right row), groups (middle row) and lower mass haloes (left row). We stack all galaxies in the vicinity of haloes as a function of distance, and in each panel the curves show median values at each distance bin. The solid and dotted lines correspond to central and satellite galaxies, respectively.

The top panel of Fig. \ref{Fig: totGasStripped_dis_z0} shows the total stripped fraction of galaxies by $z=0$. We also divide the stripped fraction into two categories based on whether they were central or satellite galaxies when stripping occurred. In particular, the stripped fractions in the middle and bottom rows are calculated only when a galaxy is either a satellite, or a central, respectively. Therefore, the top panel is the sum over the middle and bottom panels. In the rest of this subsection, we use Fig. \ref{Fig: totGasStripped_dis_z0} as a reference to describe the model predictions for gas stripping in different types of galaxies.

\subsubsection{Gas stripped out of satellite galaxies}

The middle panel of Fig. \ref{Fig: totGasStripped_dis_z0} shows that $z=0$ satellite galaxies  (dotted lines) lose a fraction of their gas after infall. There is a trend with distance for galaxies near FOF haloes: galaxies closer to the halo centre have lost more gas. That is a direct influence of environmental processes, as galaxies closer to the halo centre reside in denser environments and are more strongly stripped. There is a minimum amount of stripping at $R/R_{200}\sim1-3$. This is where the most distant satellites of the FOF haloes reside, i.e. the characteristic scale at which galaxies change type from central to satellite according to the halo finder algorithm. Among all satellites, the most distant are the least influenced by environmental effects. Beyond this scale, satellite galaxies belong to other FOF haloes and are mostly closer to the centre of their respective hosts. As a result, they are more strongly stripped. The resulting stripped fraction increases with distance beyond this characteristic scale, until reaching a constant, mass-dependent, global value.

Galaxies that are centrals at $z=0$ could have been satellites at some point in their history. These galaxies are mainly flyby or splashback systems, i.e. former satellites of other central haloes. Flyby galaxies enter a halo, traverse it, and then leave without becoming bound to the halo, while splashback galaxies remain bound to that halo and can reach apocenter before possibly returning. The solid lines in the middle row of Fig. \ref{Fig: totGasStripped_dis_z0} represent the gas stripped out of these two kinds of galaxies. They are the central galaxy of their own halos at $z=0$, and the stripped gas shown with solid lines is calculated when they were satellites of other haloes at $z>0$. There is a characteristic scale at $R/R_{200}\sim 1.5-2$ where these galaxies show any signs of having lost gas, which corresponds to the splashback radius of the nearby central halo at $z=0$. 

\subsubsection{Gas stripped out of central galaxies}
\label{subsubsection: Gas_stripping_out_of_centrals}

While gas stripping out of satellite galaxies is expected, central galaxies are less often imagined to experience significant stripping. As illustrated in the bottom panel of Fig. \ref{Fig: totGasStripped_dis_z0}, our model shows that galaxies can in fact lose a non-negligible fraction of their hot gas due to stripping while they are centrals. The stripped fraction decreases with halocentric distance for galaxies that are centrals at $z=0$ (solid lines), and can reach 80\% in the median for low-mass central galaxies in the vicinity of clusters (bottom right panel, solid lines). Interestingly, more than half of the current massive central galaxies near clusters (bottom right panel, solid purple lines) have lost more than half of their gas due to stripping beyond $\sim 1.5 R_{200}$.

\begin{figure*}
    \centering
    \includegraphics[width=0.33\textwidth]{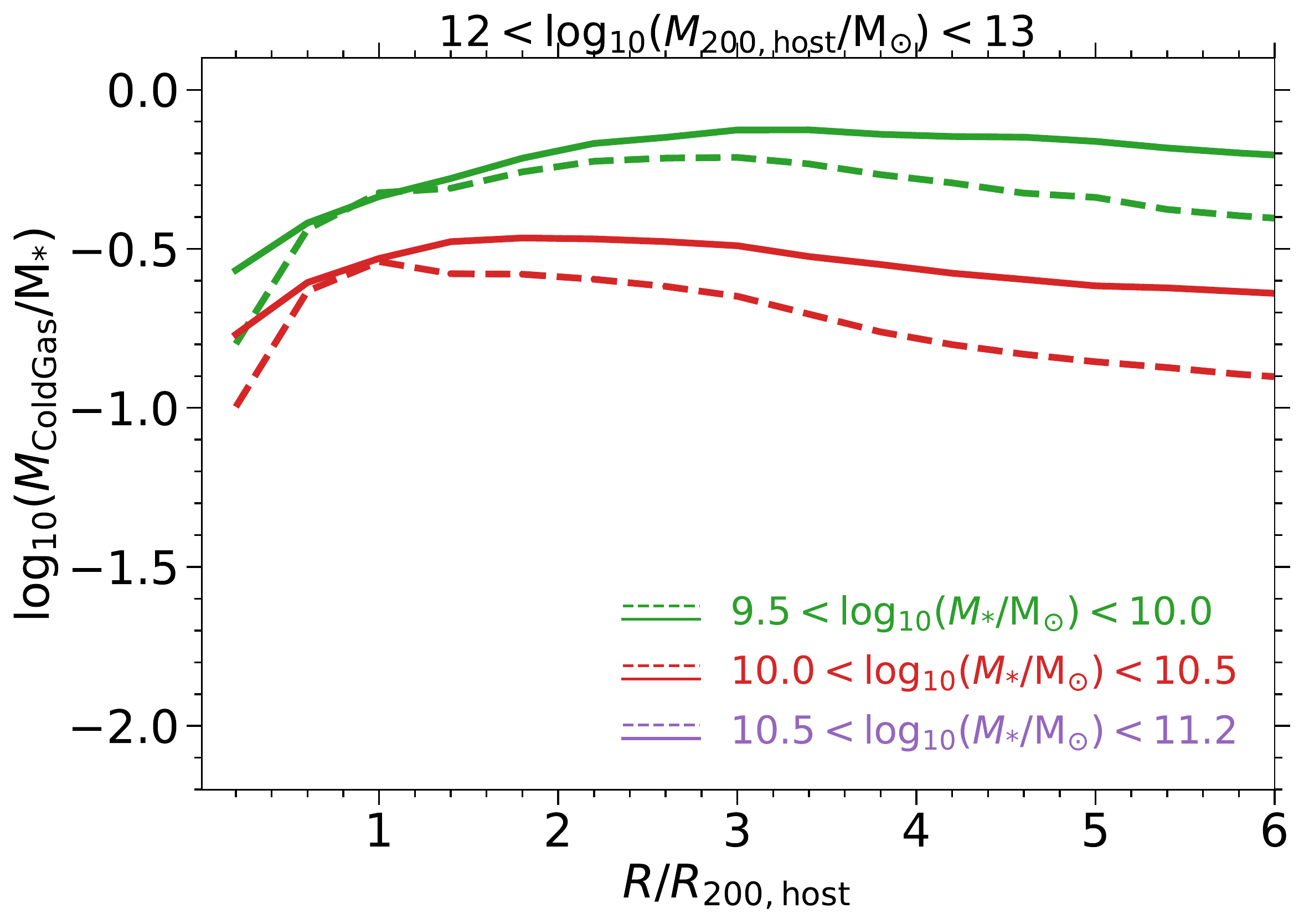}
    \includegraphics[width=0.33\textwidth]{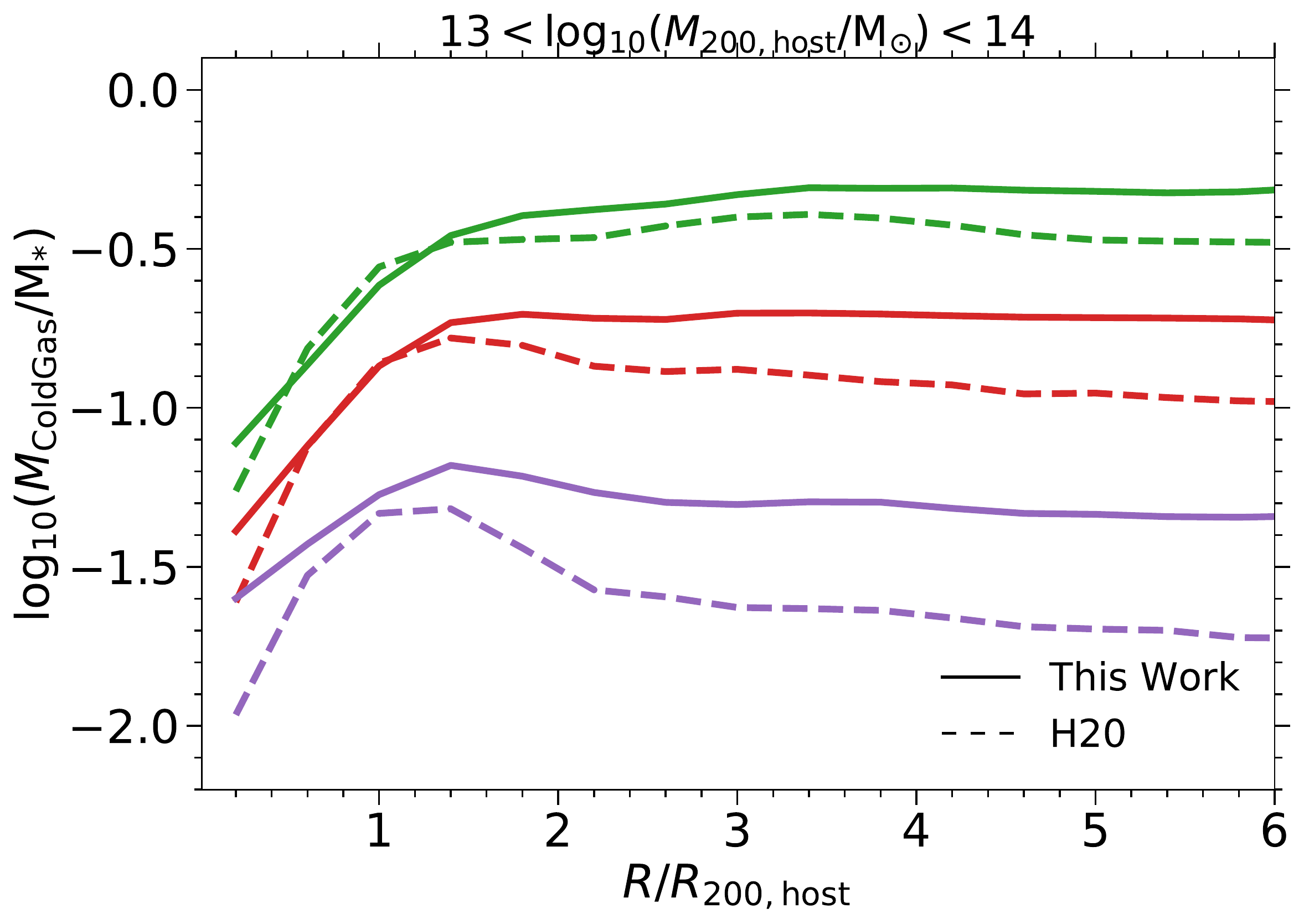}
    \includegraphics[width=0.33\textwidth]{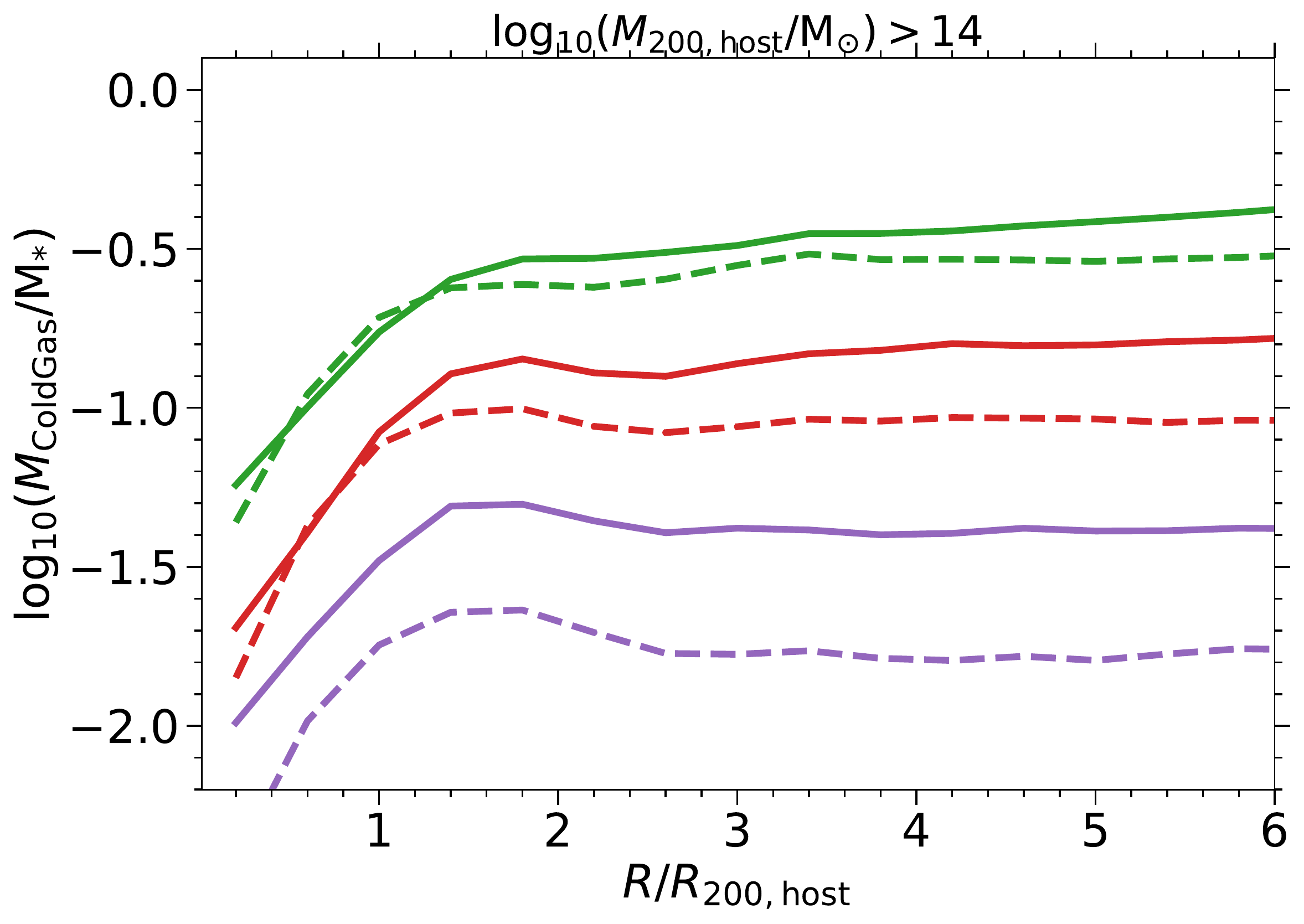}
    \caption{Cold gas to stellar mass ratio as a function of halocentric distance at $z=0$. We show the dependence on host halo mass (three columns), stellar mass (colours), and contrast our updated model with H20 (linestyles).}
\label{Fig: coldgas_comb_plot}
\end{figure*}
\begin{figure}
    \centering
    \includegraphics[width=0.9\columnwidth]{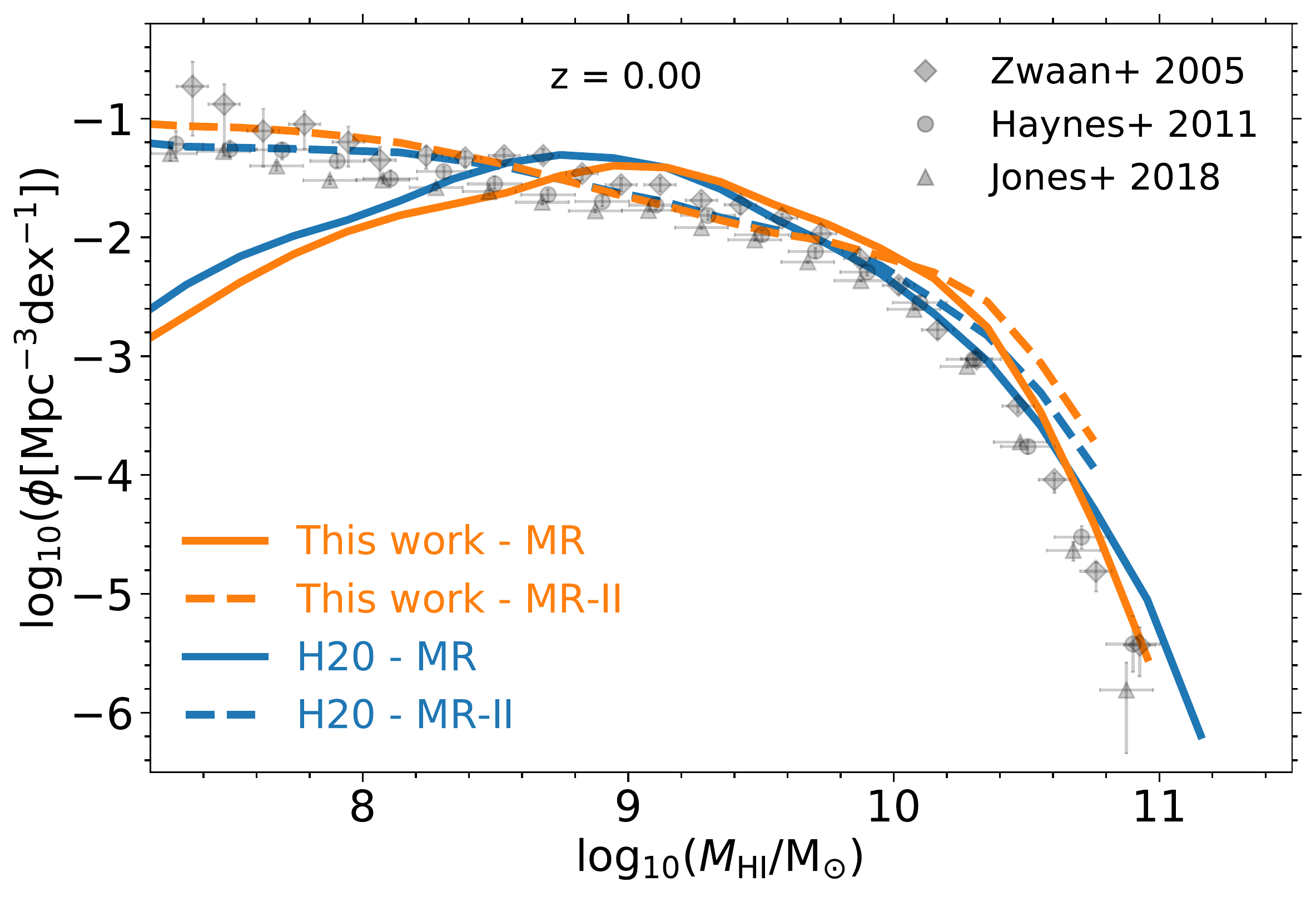}

    \caption{HI mass function from this work in comparison with H20 and with observations from \protect\cite{Zwann2005HIPASS,Haynes2011Arecibo,Jones2018ALFALFA}.}
\label{Fig: HI_MF}
\end{figure}

Our model predicts that $z=0$ satellite galaxies (dotted lines in Fig. \ref{Fig: totGasStripped_dis_z0}) lose a large fraction of their hot gas prior to infall, when they are the central galaxy of their own halo. This is illustrated in the bottom panel of Fig. \ref{Fig: totGasStripped_dis_z0} (dotted lines). There is a local maximum at $R/R_{200}\sim 1-2$, corresponding to where these galaxies change type from central to satellite. At this scale, the overall stripped fraction is very similar for central versus satellite galaxies, reflecting the uniformity of our gas stripping approach in treating different types of galaxies. For satellite galaxies with $R/R_{200}\lesssim1-2$, the fraction of the gas they have lost as a central decreases towards the centre of the halo: galaxies close to the halo centre have earlier infall times, and have spent a correspondingly longer time as satellites than more distant galaxies.

There is an overall trend with stellar mass. The top panel of Fig. \ref{Fig: totGasStripped_dis_z0} shows that the total stripped fraction decreases with stellar mass (green, to red, to purple). The reason is the weaker gravitational binding energy of low-mass systems. The trend is reversed for the stripped fraction \textit{as satellites} (middle panels), because low-mass galaxies lose a larger percentage of their gas prior to infall when they were still the central galaxy of their own halo (bottom panel). Therefore, there is not much residual gas available to be stripped once they later become satellites.

\subsection{Gas content of galaxies and subhaloes}
\label{subsec: gas_to_stellarMass_ratio}
Gas content is the quantity most affected by our new stripping method. In this subsection, we analyse the predictions of our model for hot (non-star-forming) and cold (star-forming) gas, comparing to observations and the H20 model.

\subsubsection{Hot (non-star-forming) gas}
\label{subsubsec: hot_gas_galaxies}

In the top panel of Fig. \ref{Fig: hotgas_comb_plot} we show the hot gas to stellar mass ratio, $M_{\rm hotgas}/M_{\star}$, as a function of halocentric distance for galaxies in the vicinity of clusters (top right), groups (top middle), and lower mass haloes (top left). For both models (different line styles) and at all stellar mass ranges (different colours), the $M_{\rm hotgas}/M_{\star}$ ratio decreases with decreasing halocentric distance from a constant global value. The radius at which the ratio starts to decrease in our model is larger than in H20, showing the impact of more spatially extended gas stripping in the new model.

The hot gas to stellar mass ratio decreases with the halo mass, reflecting the fact that gas stripping processes are more efficient in the dense regions surrounding the most massive halos. Interestingly, the difference between our model and H20 is more significant near massive haloes (right panel), showing the importance of modelling ram-pressure self-consistently in all environments. Moreover, the hot gas to stellar mass ratio increases with galaxy stellar mass. This is also a direct impact of stripping as discussed in \S \ref{subsec: total_stripped_gas} (see Fig. \ref{Fig: totGasStripped_dis_z0}).

The dependence of this ratio on the stellar mass of satellites, as well as their host halo mass, is illustrated in the bottom panels of Fig. \ref{Fig: hotgas_comb_plot}. The 2D histogram is coloured by the median value of $\log_{10}(M_{\rm hotgas}/M_{\star})$. In H20 (right panel), there is an abrupt transition at $\log_{10}(M_{200} / \rm M_{\odot}) \sim 14.7$. This results from H20 only applying ram-pressure stripping to satellites within $R_{200}$ of haloes with $\log_{10}(M_{200} / \rm M_{\odot}) \gtrsim 14.7$ (see \S \ref{subsubsec: L-Galaxies}). On the other hand, such a sharp transition is not present in our model, and low-mass satellite galaxies are more gas-poor than in H20, due to both pre-infall and post-infall stripping.

\subsubsection{Cold star-forming gas in galaxies}
\label{subsubsec: cold_gas_galaxies}

\begin{figure*}
    \centering
    \includegraphics[width=0.33\textwidth]{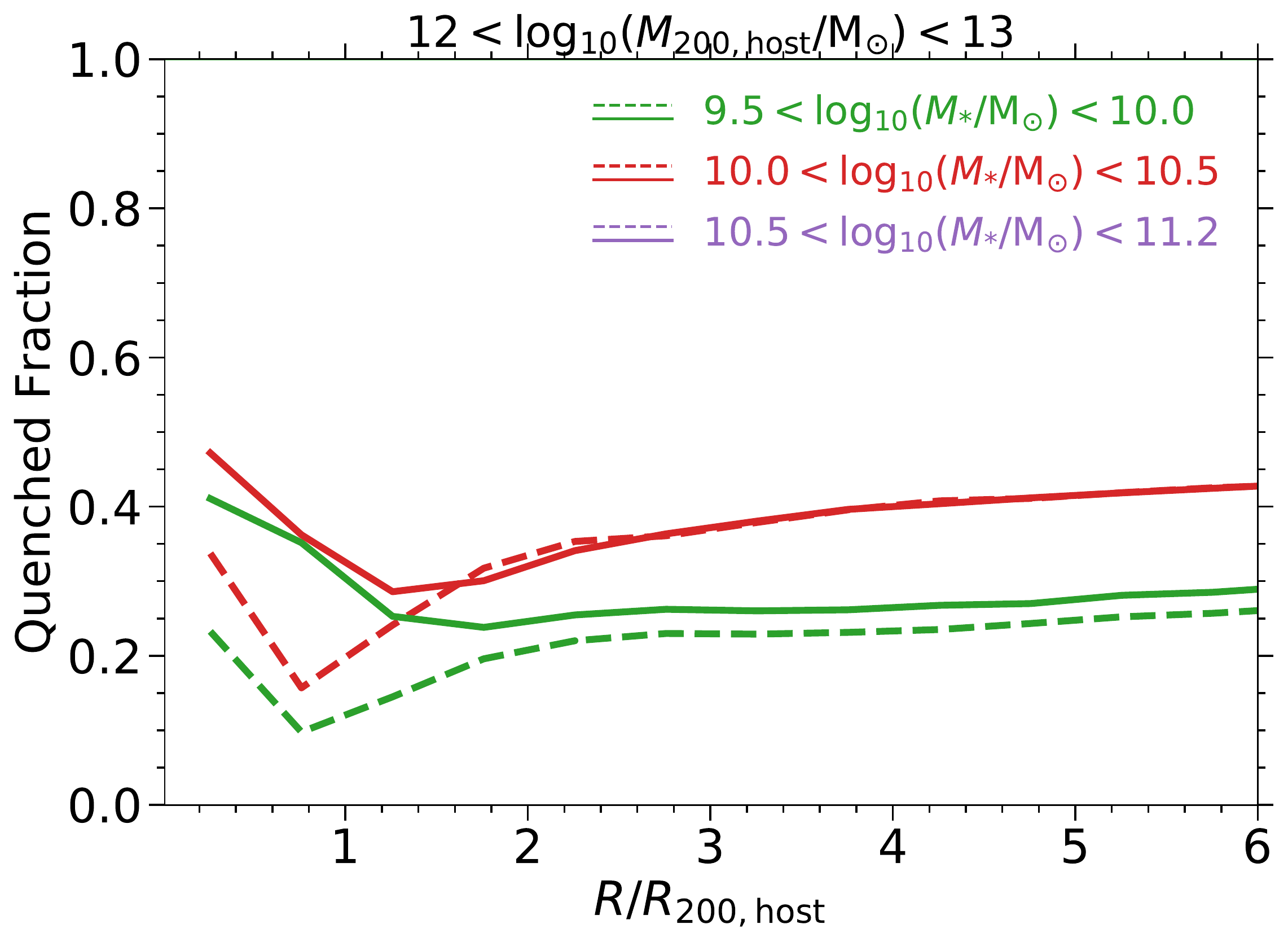}
    \includegraphics[width=0.33\textwidth]{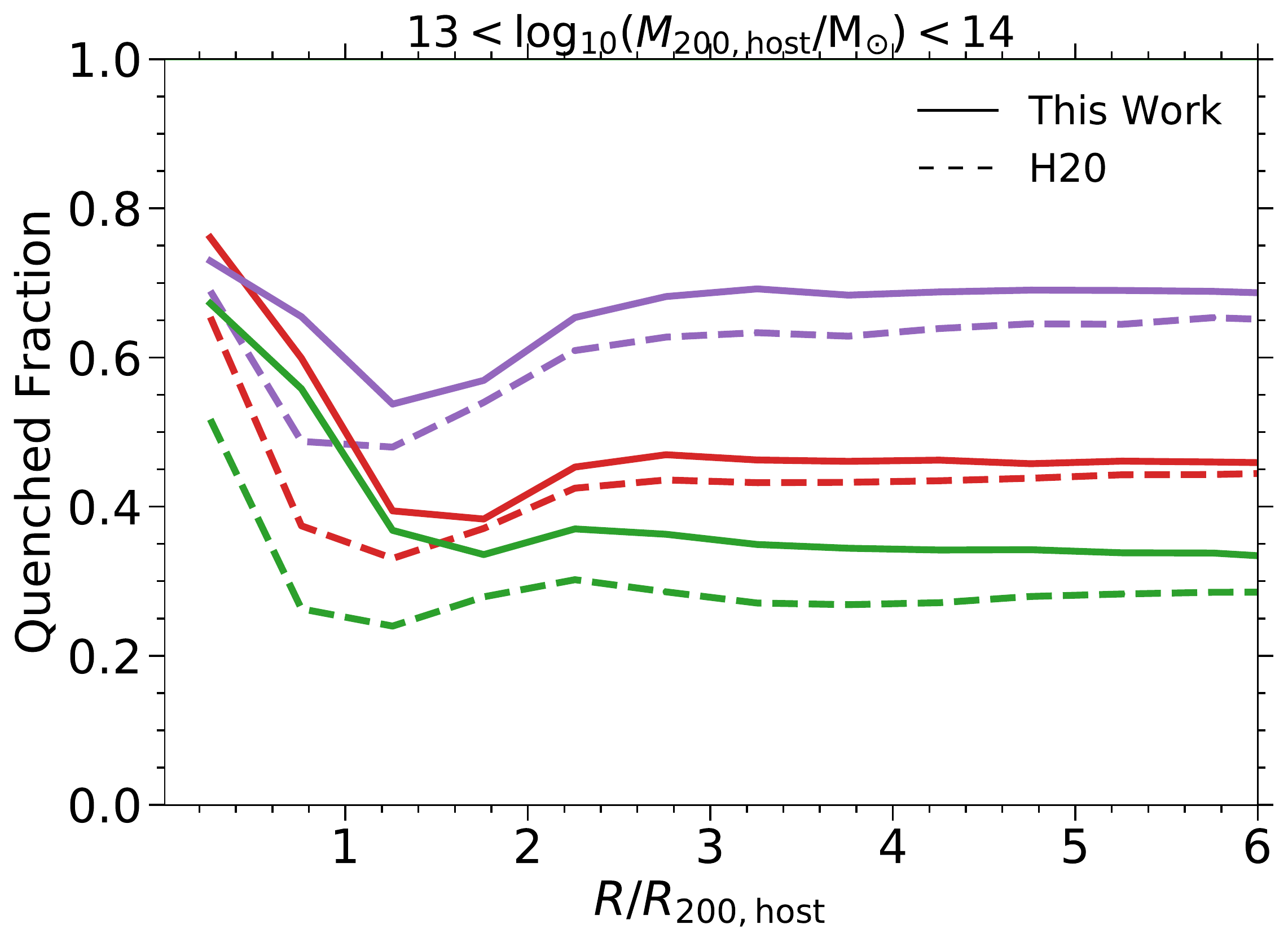}
    \includegraphics[width=0.33\textwidth]{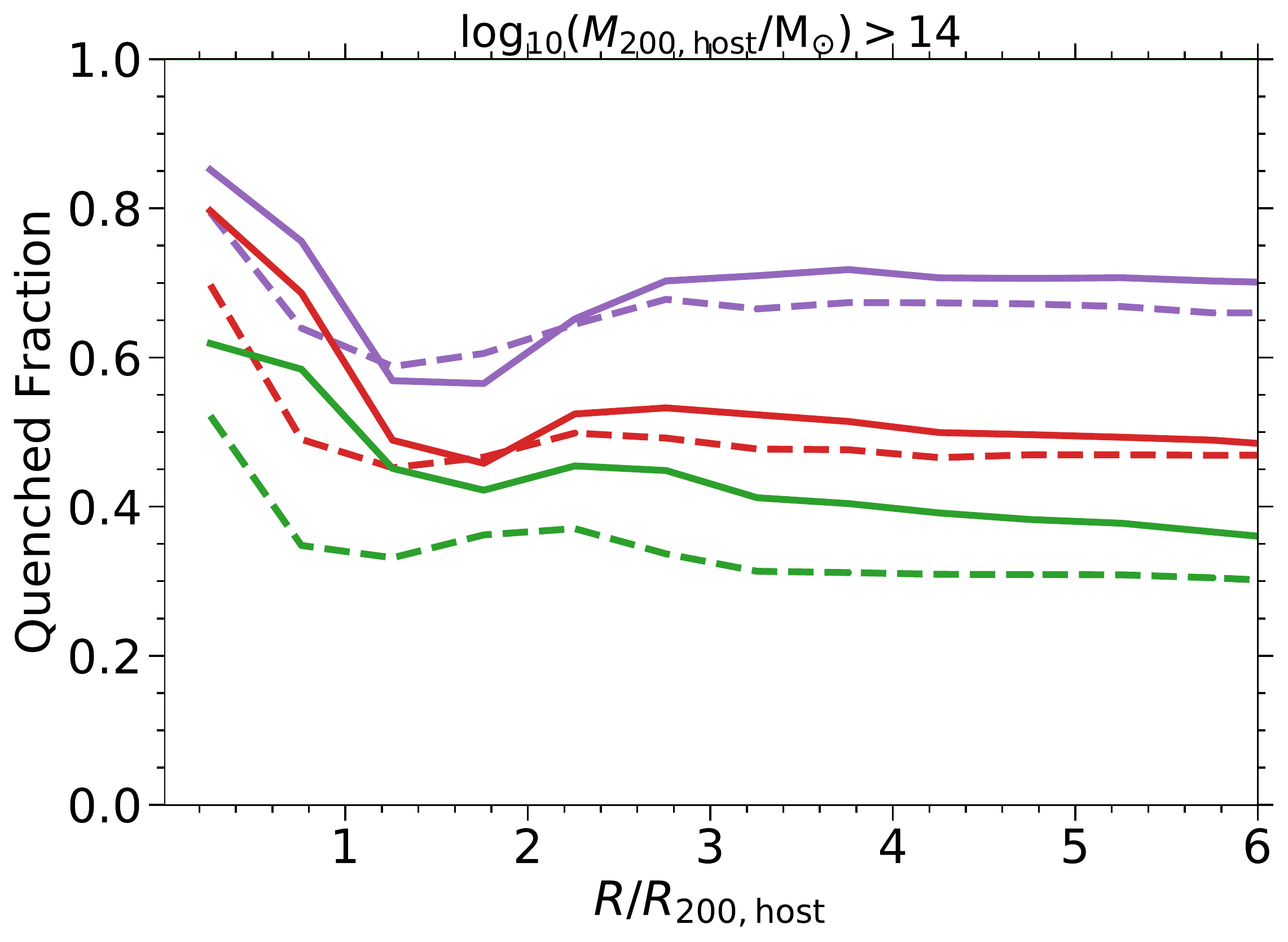}
    
    \includegraphics[width=0.4\textwidth]{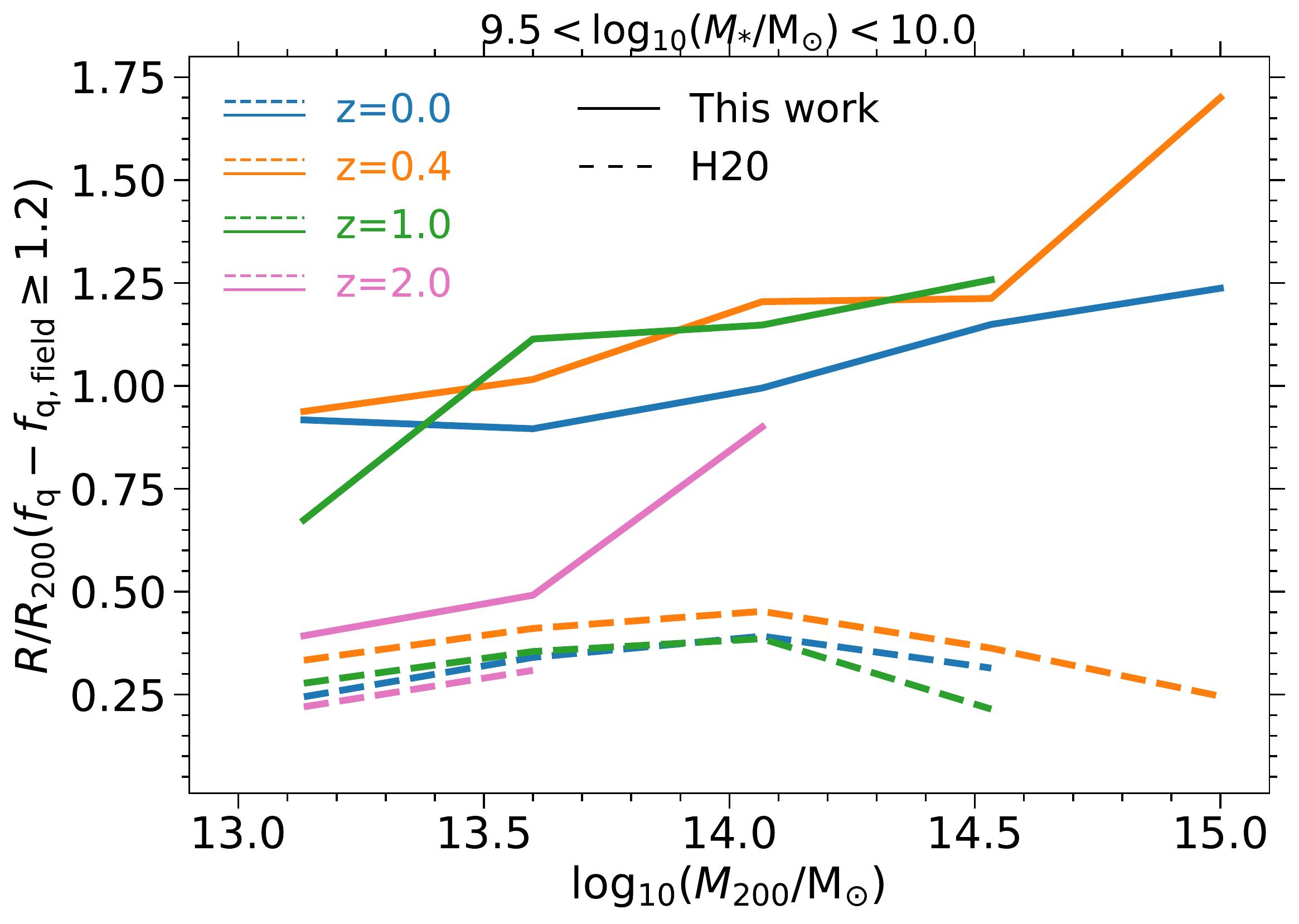}
    \includegraphics[width=0.4\textwidth]{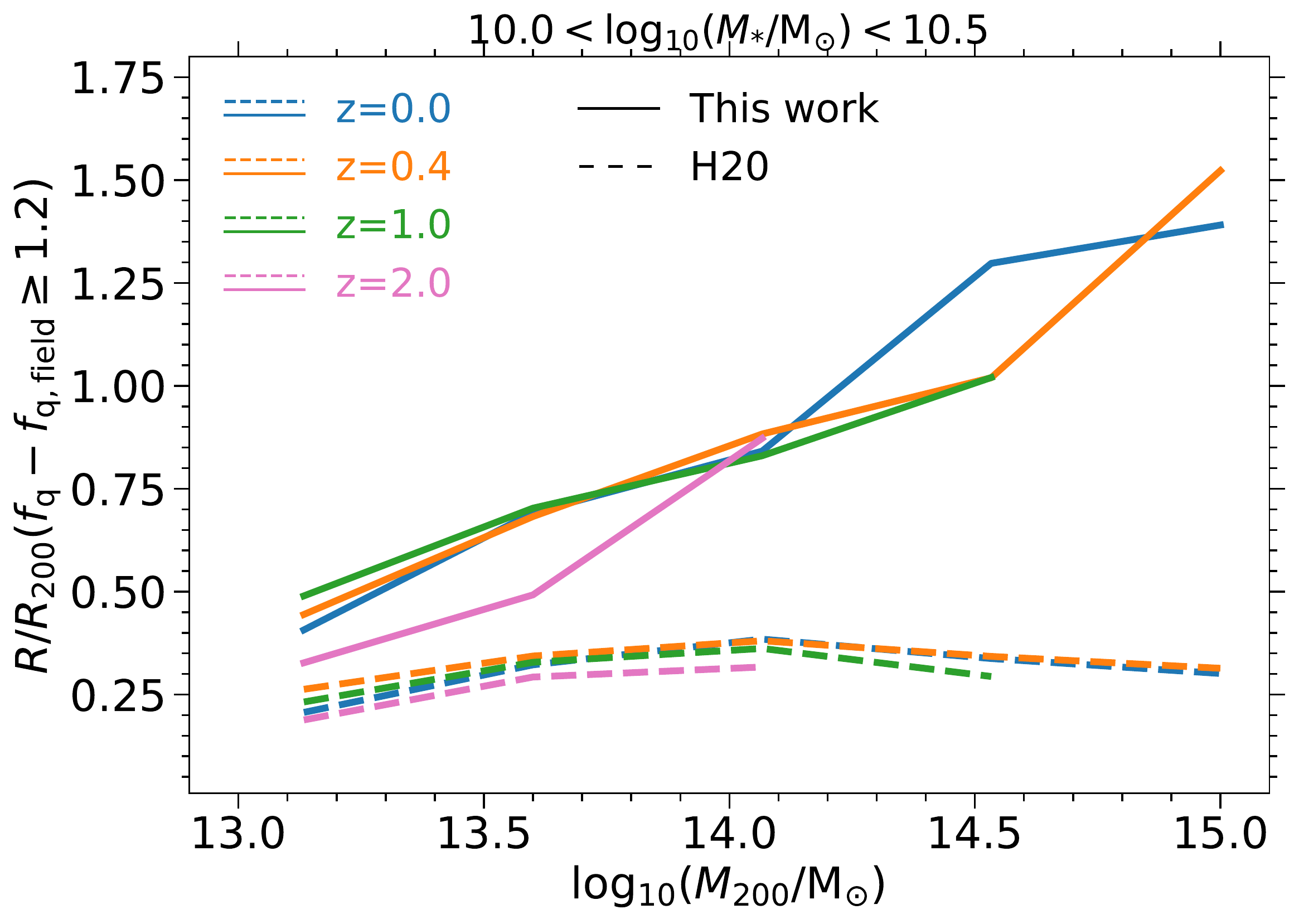}
    
    \includegraphics[width=0.45\textwidth]{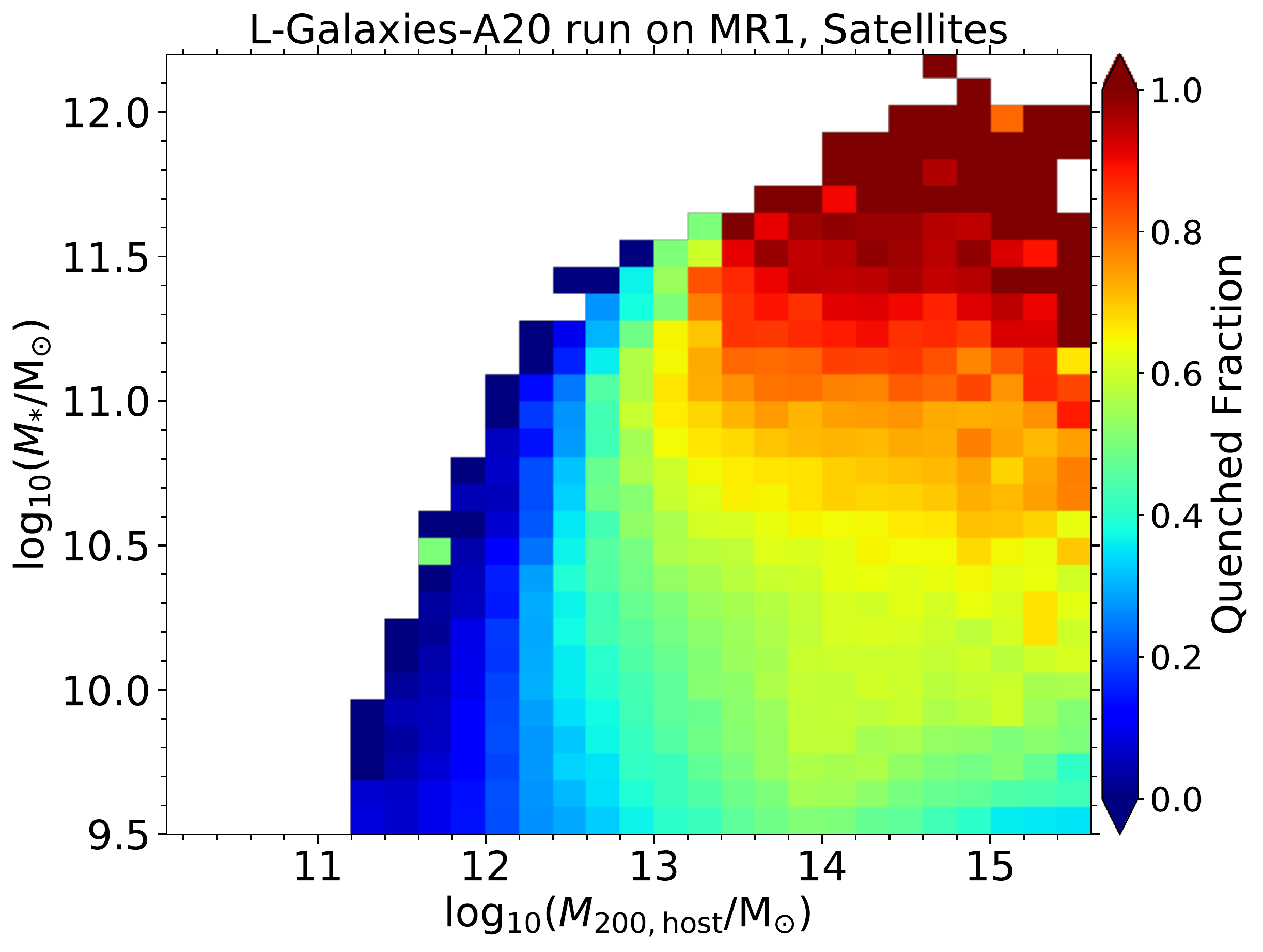}
    \includegraphics[width=0.45\textwidth]{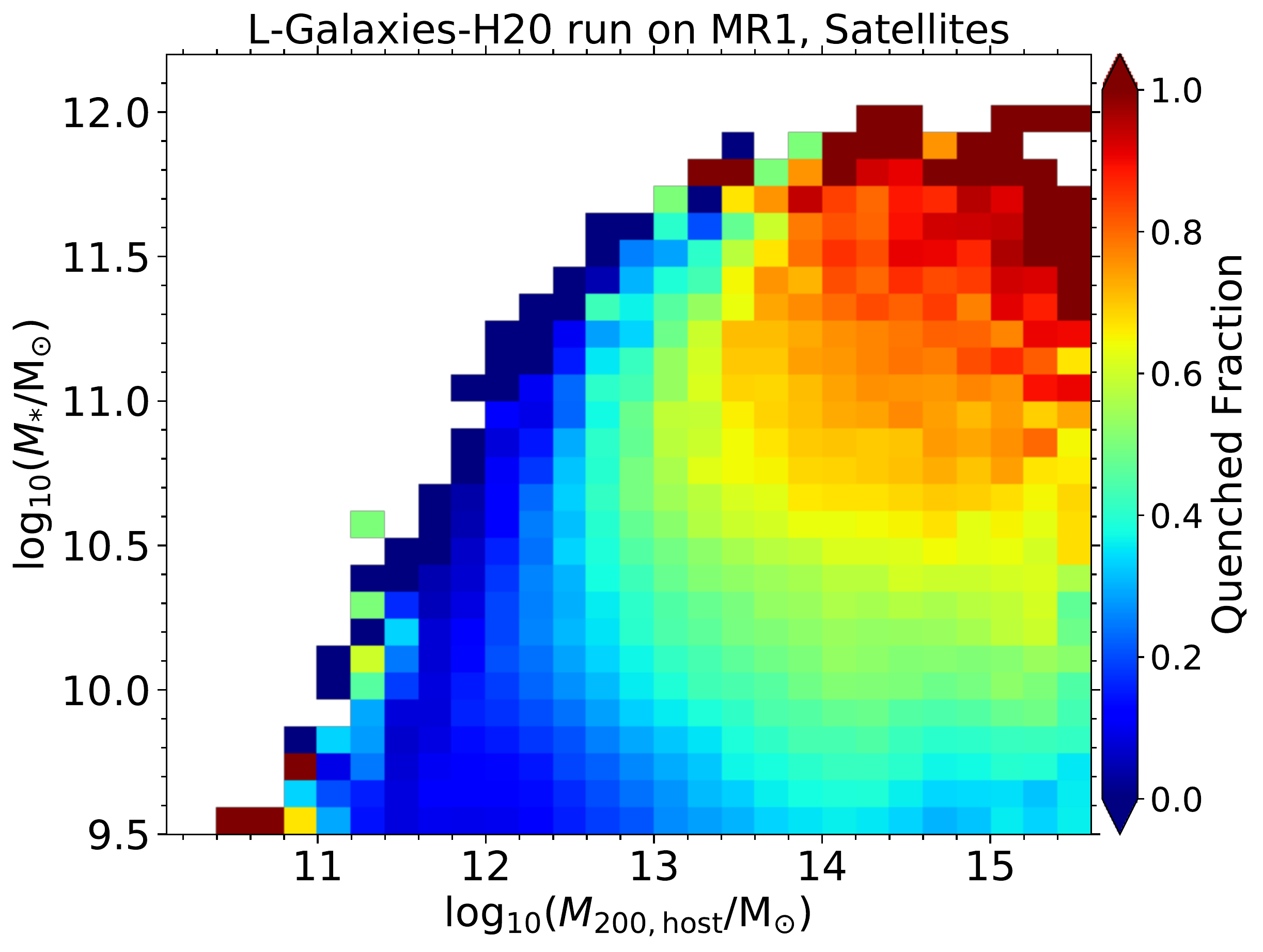}

    \caption{Top panel: Fraction of quenched galaxies as a function of halocentric distance at $z=0$. Middle panel: Halocentric radius at which the quenched fraction is 20\% above the field value vs. the halo mass both for this work and for H20 at different redshifts. Bottom panel: 2D histograms of the quenched fraction of satellite galaxies as a function of their host halo masses (x-axis) and their stellar masses (y-axis) in this work (left panels) and in H20 (right panels) at $z=0$. At $z=0$, galaxies with $\log_{10}(\rm SSFR/yr^{-1})<-11$ are considered as quenched and at $z>0$ galaxies with ${\rm SSFR/yr^{-1}}<(1+z)/(2\times10^{10})$ are considered as quenched (following the quenching criteria in H20).}
\label{Fig: quenchedFrac_comb_plot}
\end{figure*}

The cold gas component is also affected by environmental processes. However, as we do not implement cold gas stripping directly, we do not expect our model to differ significantly from H20 in this regard. Fig. \ref{Fig: coldgas_comb_plot} shows the median cold gas to stellar mass ratio as a function of halocentric distance. As before, galaxies in the vicinity of haloes are stacked and the lines represent the median values at each radial distance bin.

In both models, $M_{\rm coldgas}/M_{\star}$ increases with distance and reaches a constant global value at $R/R_{200}\sim1-2$. The trend with distance roughly follows the trend of $\log_{10}(M_{\rm hotgas}/M_{\star})$ (Fig. \ref{Fig: hotgas_comb_plot}, top panel). However, the influence of environment on the cold gas extends out to smaller radii than is seen for the hot gas. Only when galaxies run out of hot gas does cold gas component cease to grow. Subsequently, as star formation continues, the cold gas mass decreases further until either it is exhausted, or the galaxy merges.

Fig. \ref{Fig: HI_MF} shows the HI mass function for galaxies in our model (orange lines), H20 (blue lines), and observations (grey points, \citealt{Zwann2005HIPASS,Haynes2011Arecibo,Jones2018ALFALFA}). Two different runs are shown for each model: \textsc{L-Galaxies} run on the Millennium and Millennium-II simulations. Due to the better resolution of the Millennium-II simulation we can extend the model results down to lower HI masses. On the other hand, the Millennium simulation provides better statistics for larger $M_{\rm HI}$ values. Both our model and H20 are in relatively good agreement with observations. However, we note that H20 uses the HI mass function as an observational constraint for their model calibration, while we do not.

\subsection{Star formation and galaxy quenching}
\label{subsec: quenchedFraction}

\subsubsection{Direct model predictions}
\label{subsubsec: quenchedFrac_theory}

We move on to study the star-formation activity, and quenching, of galaxies. The top panel of Fig. \ref{Fig: quenchedFrac_comb_plot} shows the fraction of quenched galaxies, $f_{\rm q}$, as a function of halocentric distance at $z=0$. For each halo mass bin (panel) all galaxies in the vicinity of haloes are stacked. The results are shown for different stellar mass bins (different colours) for our model (solid lines) and for H20 (dashed lines). In the vicinity of clusters (top right), $f_{\rm q}$ decreases with distance in both models and flattens at a distance that is always larger than the halo boundary, $R_{200}$. There is a clear trend with halo mass: galaxies in the vicinity of more massive haloes are more quenched. This follows directly from what we showed in \S \ref{subsec: gas_to_stellarMass_ratio} (Figs. \ref{Fig: hotgas_comb_plot}, \ref{Fig: coldgas_comb_plot}) regarding the gas content of galaxies.

The top panel of Fig. \ref{Fig: quenchedFrac_comb_plot} often reveals a minimum in the quenched fraction at $R/R_{200}\sim 1-2$. This is where the most distant satellites of FOF haloes reside, i.e. where they are the least influenced by their environment. At halocentric radii smaller than this scale, satellite galaxies are strongly influenced by their host halo. Beyond this scale, satellites belong to other FOF haloes and could be close to their own centrals and, therefore, more strongly influenced by environmental processes. The contribution of those satellites at $R/R_{200}>2$ in the total quenched fraction (i.e. centrals and satellites together), is sufficient to cause a quenched fraction larger than the value at $R/R_{200}\sim 1-2$, causing the minimum. We note that this directly follows the minimum of the cumulative stripped fraction as shown by dotted lines in the top panel of Fig. \ref{Fig: totGasStripped_dis_z0}.

Comparing with H20, gas stripping in our model affects $f_{\rm q}$ to larger halocentric distances. In order to quantify this, we define a characteristic radius called $R_{\rm c,20\%}$ at which the quenched fraction is 20\% above its field value,
\begin{equation}
    f_{\rm q}(R_{\rm c,20\%}) - f_{\rm q,field} = 0.2\,,
\end{equation}
where $f_{\rm q,field}$ is the fraction of quenched field galaxies, calculated for galaxies with halocentric distance in the range of $5<R/R_{200}<10$. The middle row of Fig. \ref{Fig: quenchedFrac_comb_plot} shows $R_{\rm c,20\%}/R_{200}$ (normalised to the host halo $R_{200}$) for our model (solid lines) and H20 (dashed lines) as a function of halo mass. The results are shown for two different stellar mass ranges (middle left and middle right panels) at different redshifts (different colours).

Our model quenches galaxies in the vicinity of massive haloes to much larger halocentric distances than H20 at all redshifts. At $z>2$, there is no distance within which galaxies become 20\% more quenched than field galaxies. This indicates that the influence of environmental processes on galaxy quenching is non-negligibly visible only since $z\sim2$. At $z\lesssim1$, $R_{\rm c,20\%}/R_{200}$ in our model is on average four times larger than in H20. We also observe a trend with halo mass: galaxies in the vicinity of more massive haloes are quenched up to larger fractions of their R200. The same trend is not present in H20 or is very weak. Moreover, there is no significant correlation with redshift, except for $z\gtrsim2$ where $R_{\rm c,20\%}/R_{200}$ is either small (pink lines, middle row of Fig. \ref{Fig: quenchedFrac_comb_plot}) or is not present. Comparing the two panels in the middle row of Fig. \ref{Fig: quenchedFrac_comb_plot}, $R_{\rm c,20\%}/R_{200}$ decreases with stellar mass, i.e. low-mass galaxies are strongly influenced by their environment out to larger distances.

Environment can also influence intrinsic physical processes such as stellar and black hole feedback. As an example, consider galaxies with $10.5<\log_{10}(M_{\rm star}/{\rm M_{\odot}})<11.2$ that reside in the vicinity of clusters in our model (solid purple line, top right panel). When the gas is removed, there are two main consequences: (i) their star formation will decrease due to hot gas stripping and a lack of cold gas replenishment, and (ii) the efficiency of their black hole feedback will decrease because it is a function of hot halo gas (see Table \ref{tab: MCMC_free_params} and the supplementary material). The decrease in black hole feedback allows the hot gas to cool faster, which increases star-formation. Such non-trivial couplings between environmental effects and feedback processes will be explored in more detail in future work.

Finally, the bottom panel of Fig. \ref{Fig: quenchedFrac_comb_plot} shows 2D histograms of the quenched fraction of satellite galaxies as a function of stellar mass and host halo mass. In both our model (bottom left) and H20 (bottom right), there is a strong trend with halo mass, and in general, satellites of more massive hosts are more quenched. In addition, the quenched fraction increases with stellar mass. Comparing the two models, our model has more quenched satellites than H20 at all stellar and host halo masses. Significantly, satellites with $\log_{10}(M_{\star}/{\rm M_{\odot}})<10$ residing in groups with $13<\log_{10}(M_{200}/{\rm M_{\odot}})<14$ are more quenched in our model than in H20. This is due to gas stripping both within groups and prior to infall. 

\subsubsection{Comparing with observations at $z=0$}

\begin{figure}
    \centering
    \includegraphics[width=0.93\columnwidth]{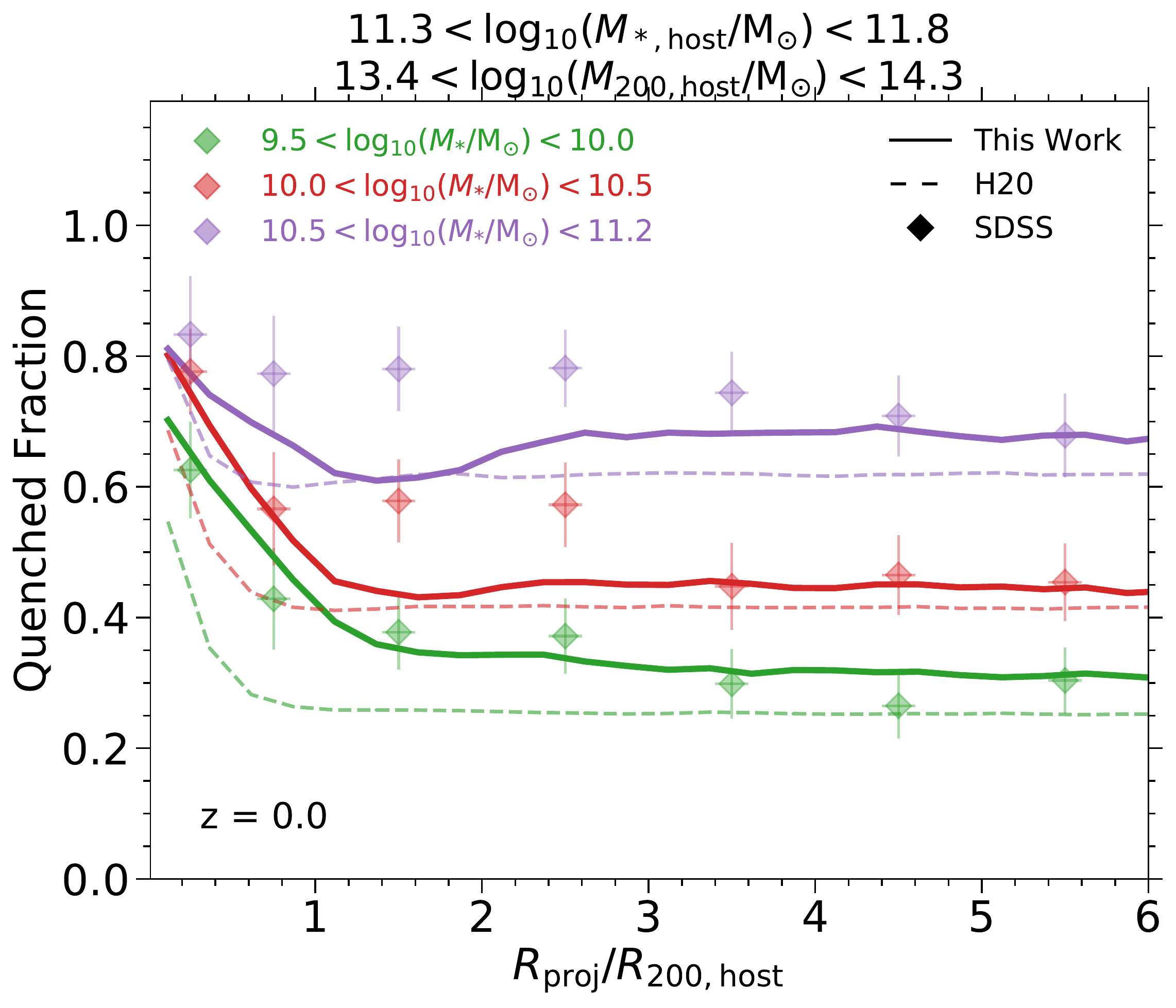}
    \includegraphics[width=0.93\columnwidth]{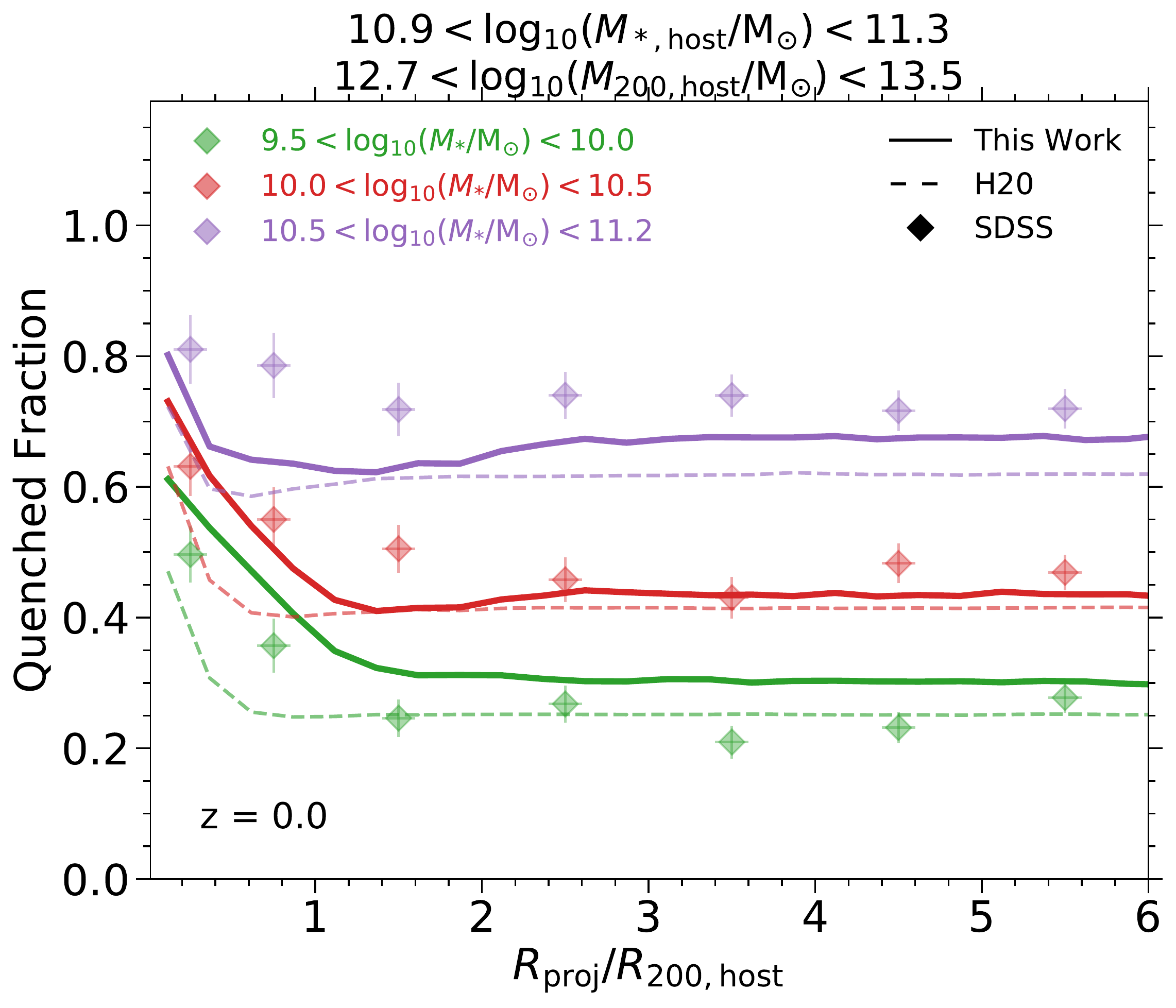}
    \includegraphics[width=0.93\columnwidth]{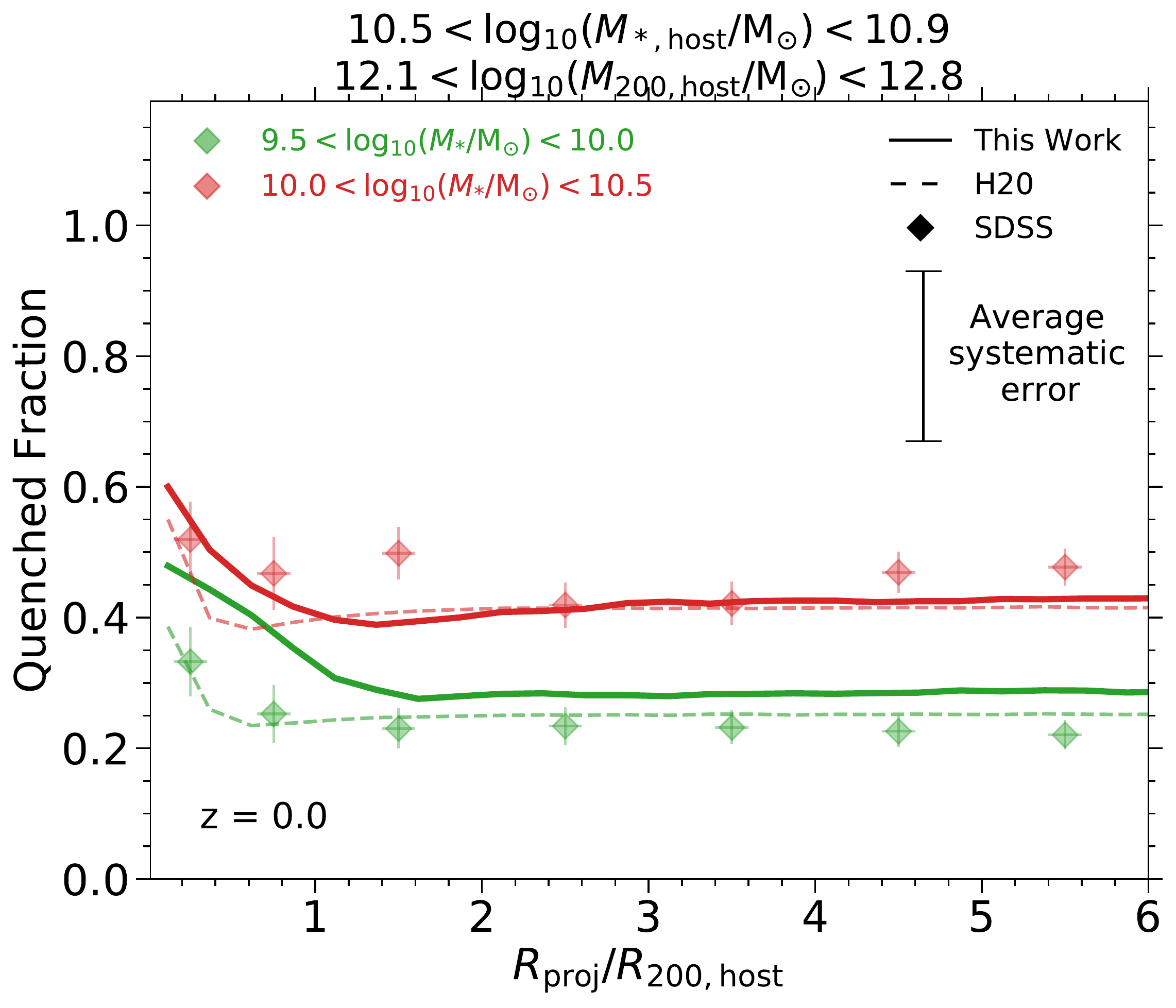}
    \caption{Fraction of quenched galaxies as a function of projected distance to the centre of haloes at $z=0$. For both models and observations, galaxies with $\log_{10}({\rm sSFR/yr^{-1}})<-11$ are considered as quenched. The halo mass in both models and observations is calculated using the stellar masses of the central galaxies, based on Eq. \ref{eq: Mstellar_to_M200_conversion}. The error bars are binomial 95\% confidence intervals based on a Gaussian approximation. The average systematic observational errors of the SDSS data is shown in the figure for $1\sigma$ error of sSFR. The overall trends seen in SDSS are well reproduced by our new model.}
\label{Fig: quenchedFrac_dis_proj_z0}
\end{figure}

\begin{figure}
    \centering
    \includegraphics[width=0.9\columnwidth]{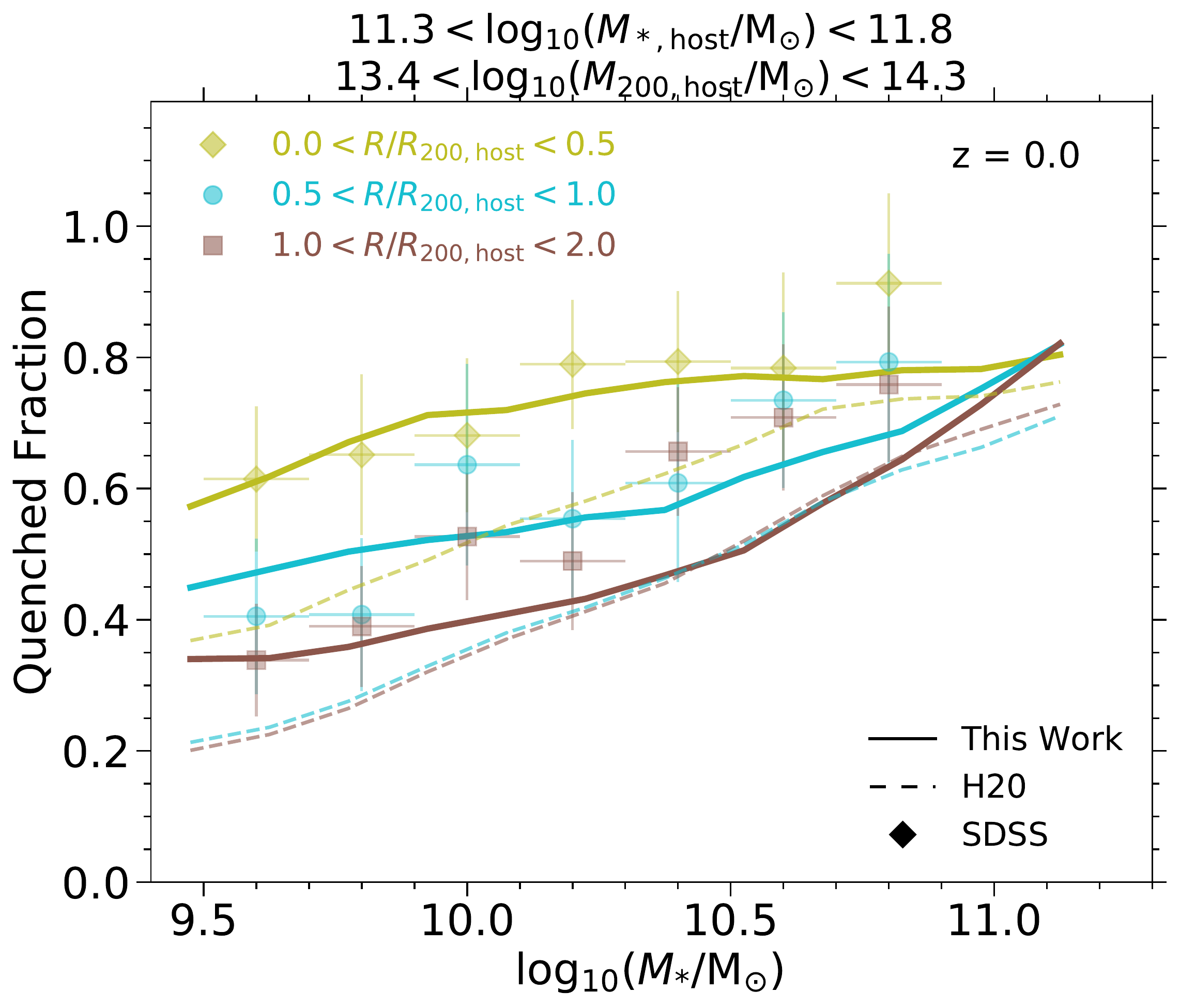}
    \includegraphics[width=0.9\columnwidth]{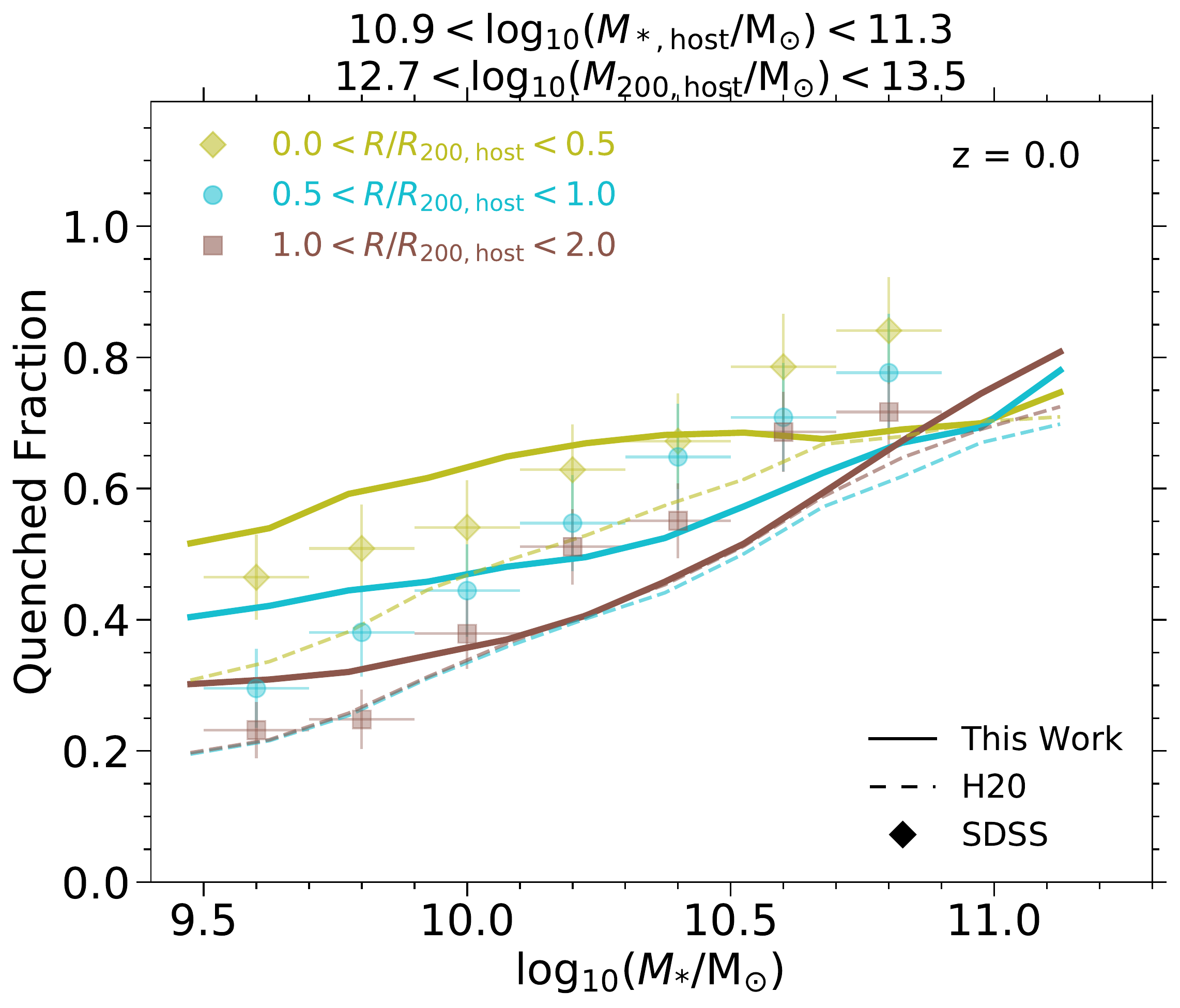}
    \caption{Fraction of quenched galaxies as a function of galaxy stellar mass at $z=0$, comparing the two models to our analysis of SDSS data. For both models and observations, galaxies with $\log_{10}{\rm SSFR/yr}<-11$ are considered as quenched. The halo mass in both models and observations are calculated using the stellar masses of the central galaxies, based on Eq. \ref{eq: Mstellar_to_M200_conversion}. The error bars are binomial 95\% confidence intervals based on a Gaussian approximation.}
\label{Fig: quenchedFrac_M*_proj_z0}
\end{figure}

\begin{figure}
    \centering
    \includegraphics[width=0.8\columnwidth]{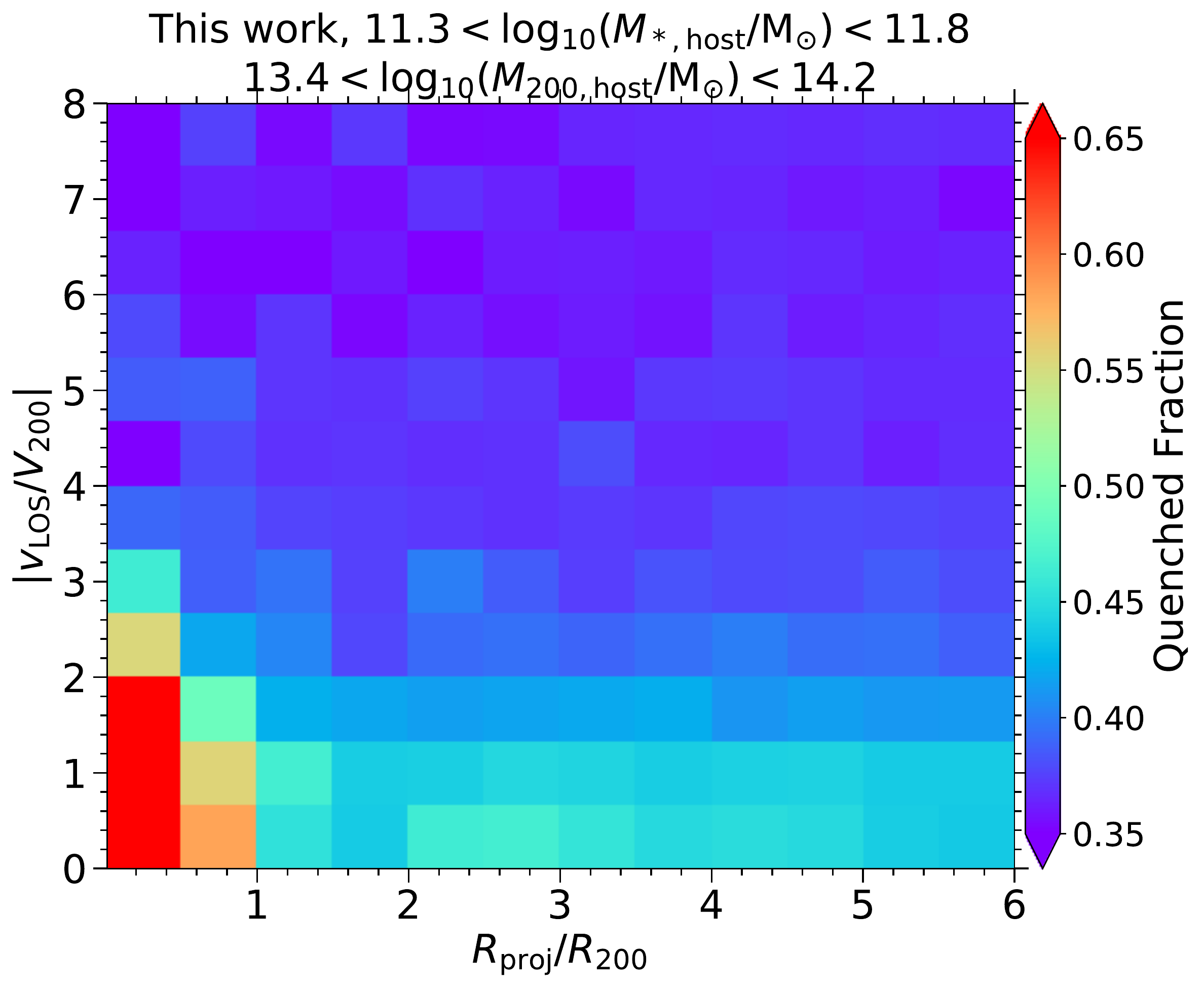}
    \includegraphics[width=0.8\columnwidth]{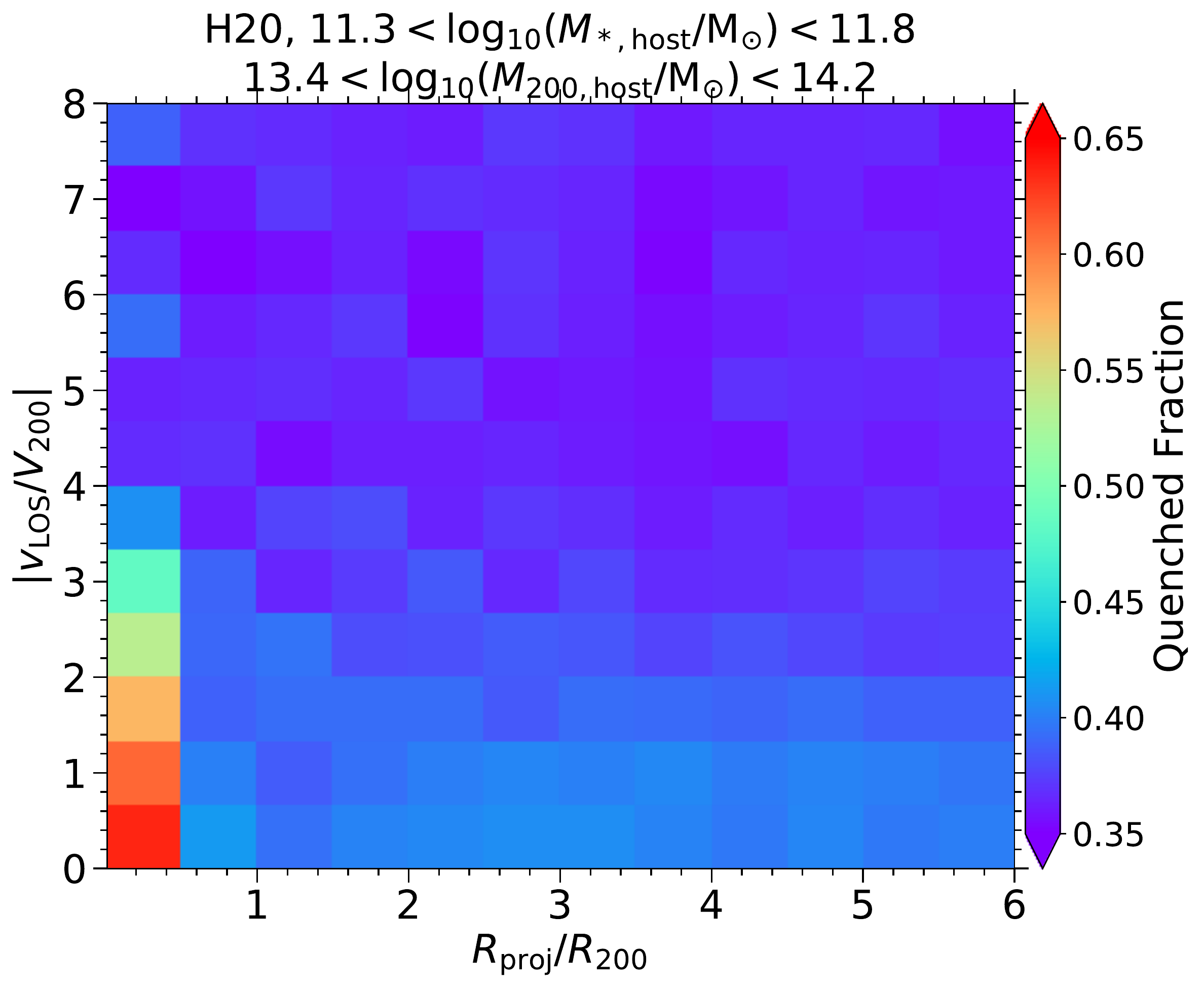} 
    \includegraphics[width=0.8\columnwidth]{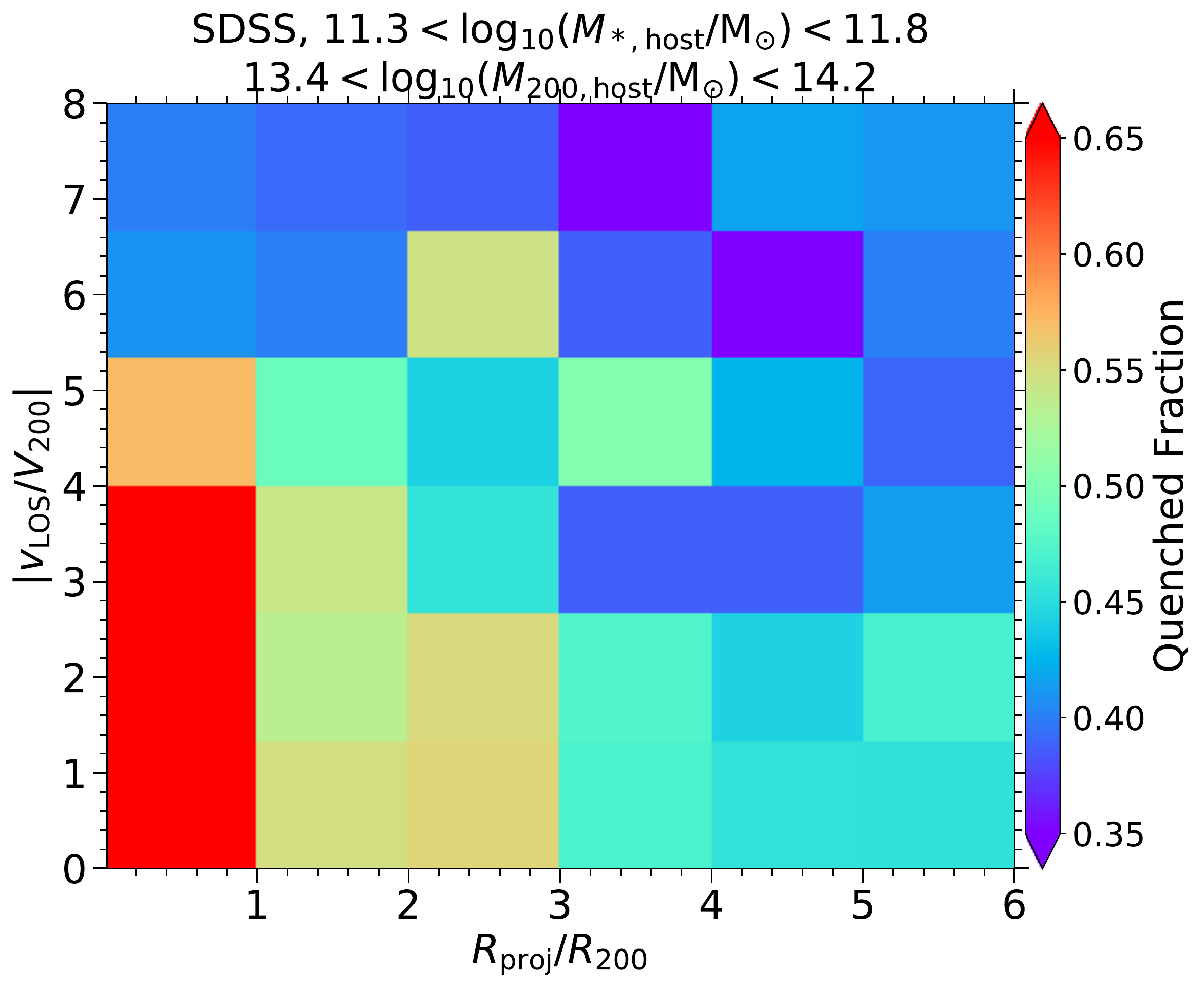}
    \caption{The fraction of quenched galaxies in projected phase space: as a function of projected halocentric distance (x-axis) and line-of-sight velocity (y-axis). We compare our new model (top panel), H20 (middle panel) and SDSS (bottom panel), defining galaxies with $\log_{10} (\rm SSFR/yr^{-1})<-11$ as quenched. All galaxies with $\log_{10}(M_{\star}/{\rm M_{\odot}})>9.5$ are included. Note that the statistics are quite different: the theory panels contain $\sim 18000$ haloes while only $\sim 100$ haloes are available in the data. The new model better reproduces the extended quenched fractions of galaxies out to larger distances and higher line-of-sight velocities.}
\label{Fig: quenchedFrac_dis_vLOS_hist2D}
\end{figure}

\begin{figure}
    \includegraphics[width=0.50\columnwidth]{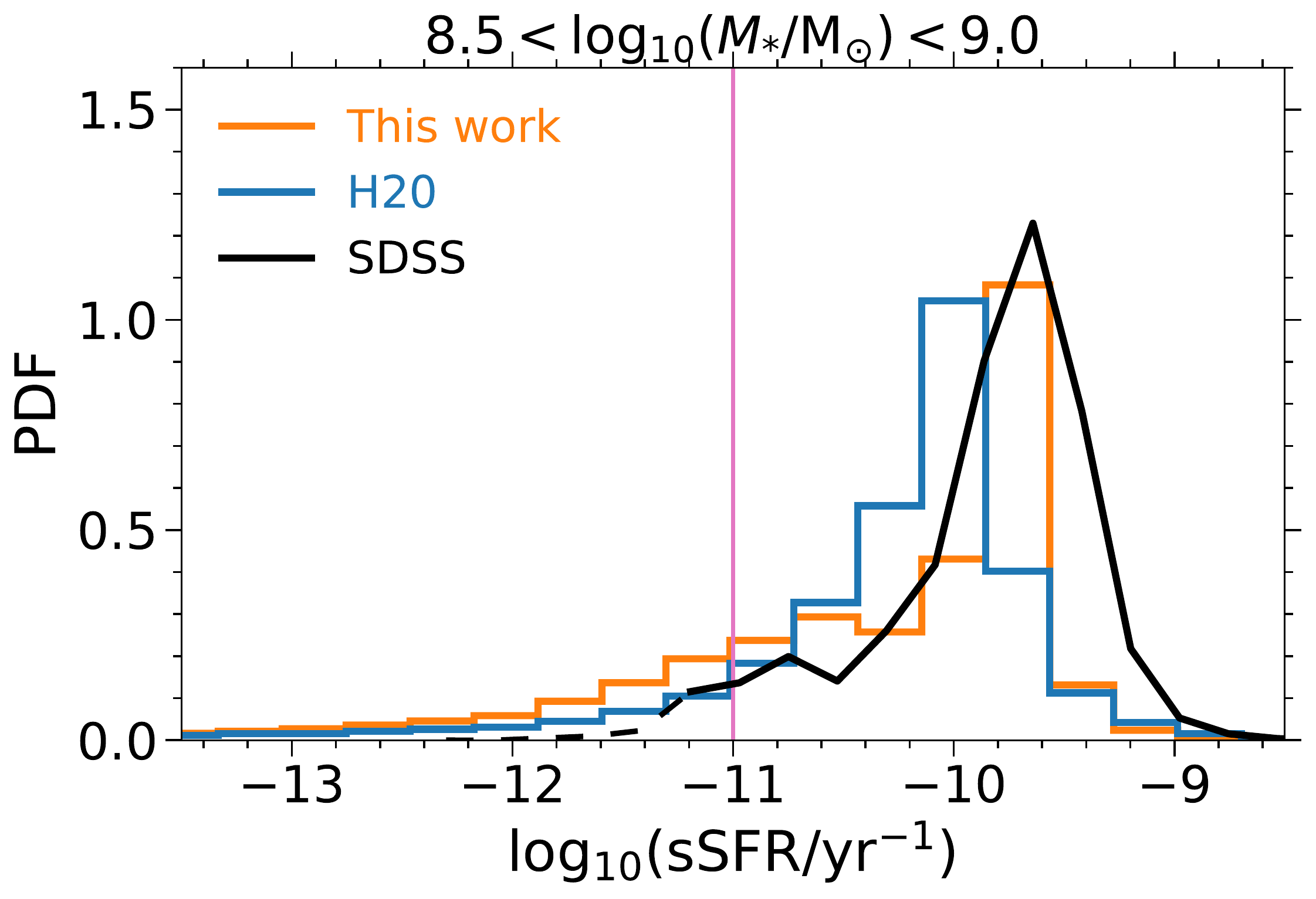}
    \includegraphics[width=0.50\columnwidth]{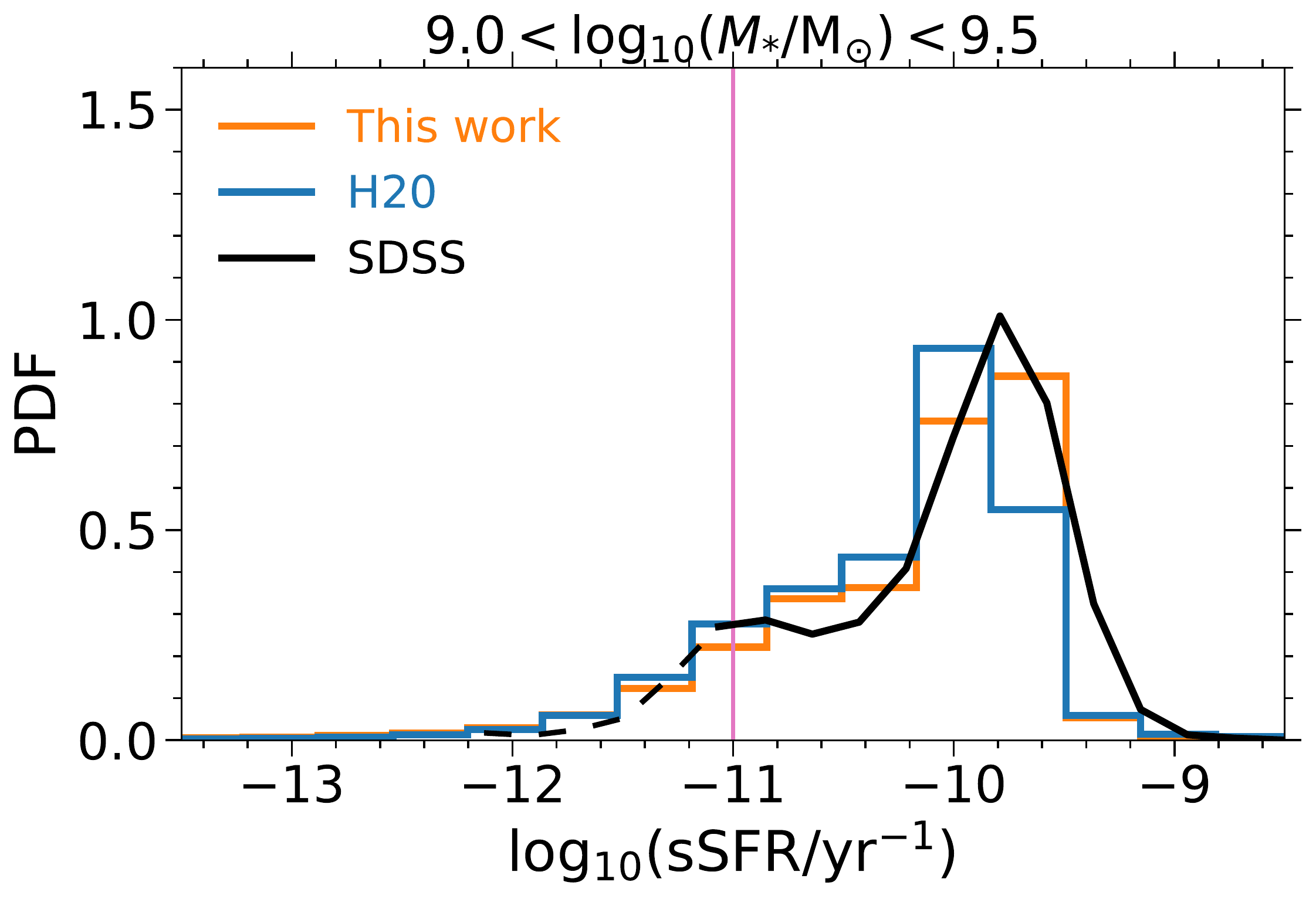}
    \includegraphics[width=0.50\columnwidth]{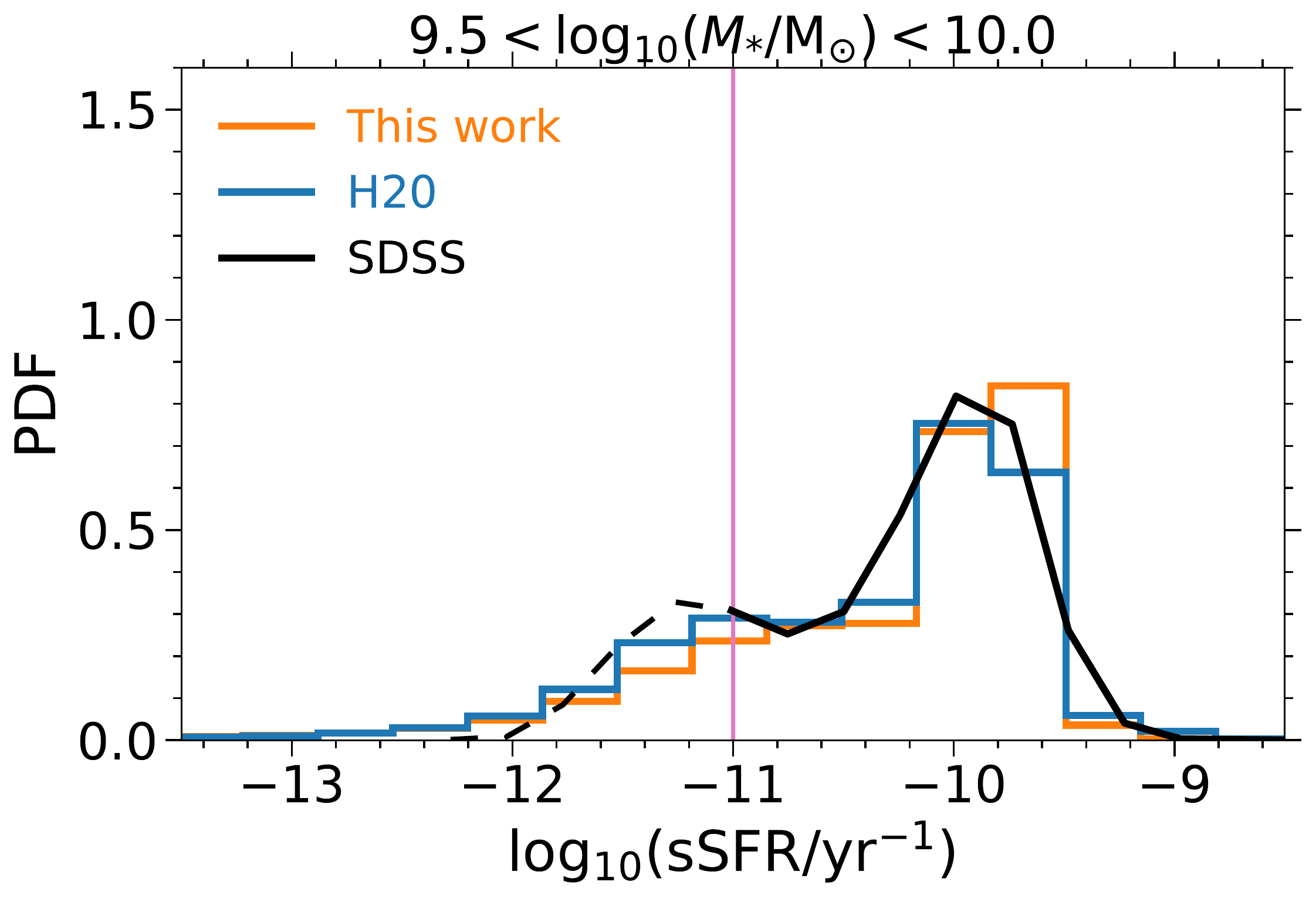}
    \includegraphics[width=0.50\columnwidth]{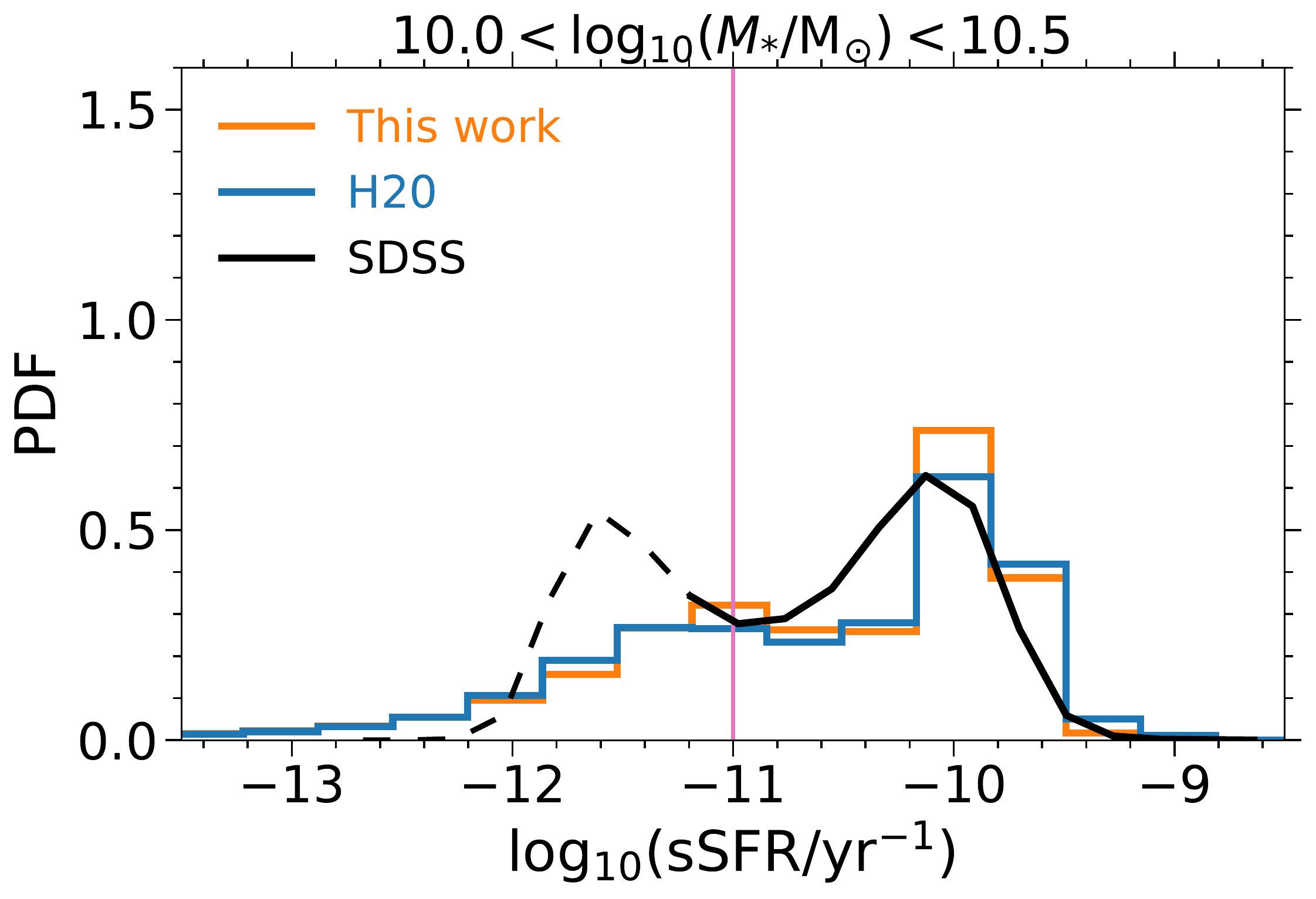}
    \includegraphics[width=0.50\columnwidth]{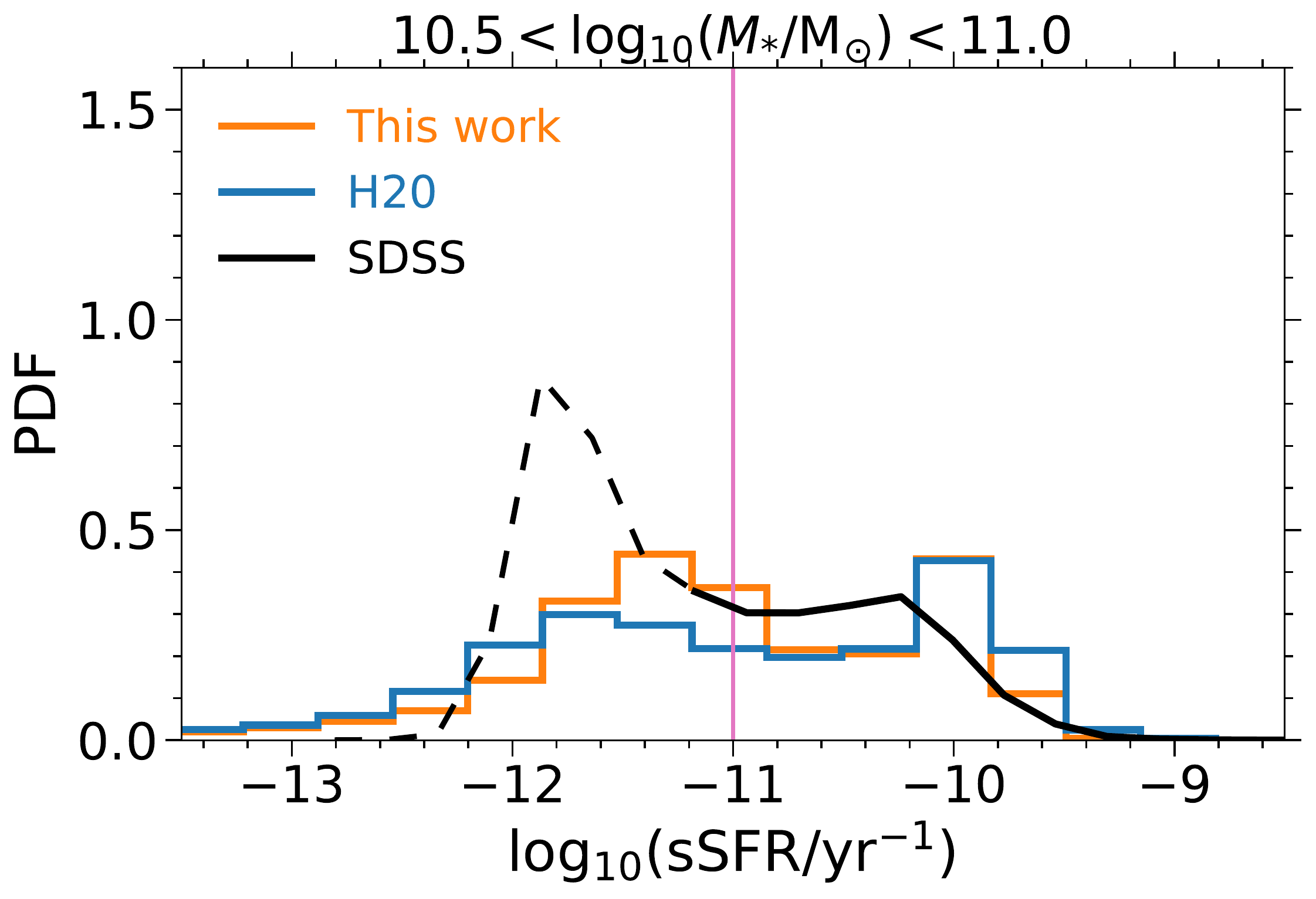}
    \includegraphics[width=0.50\columnwidth]{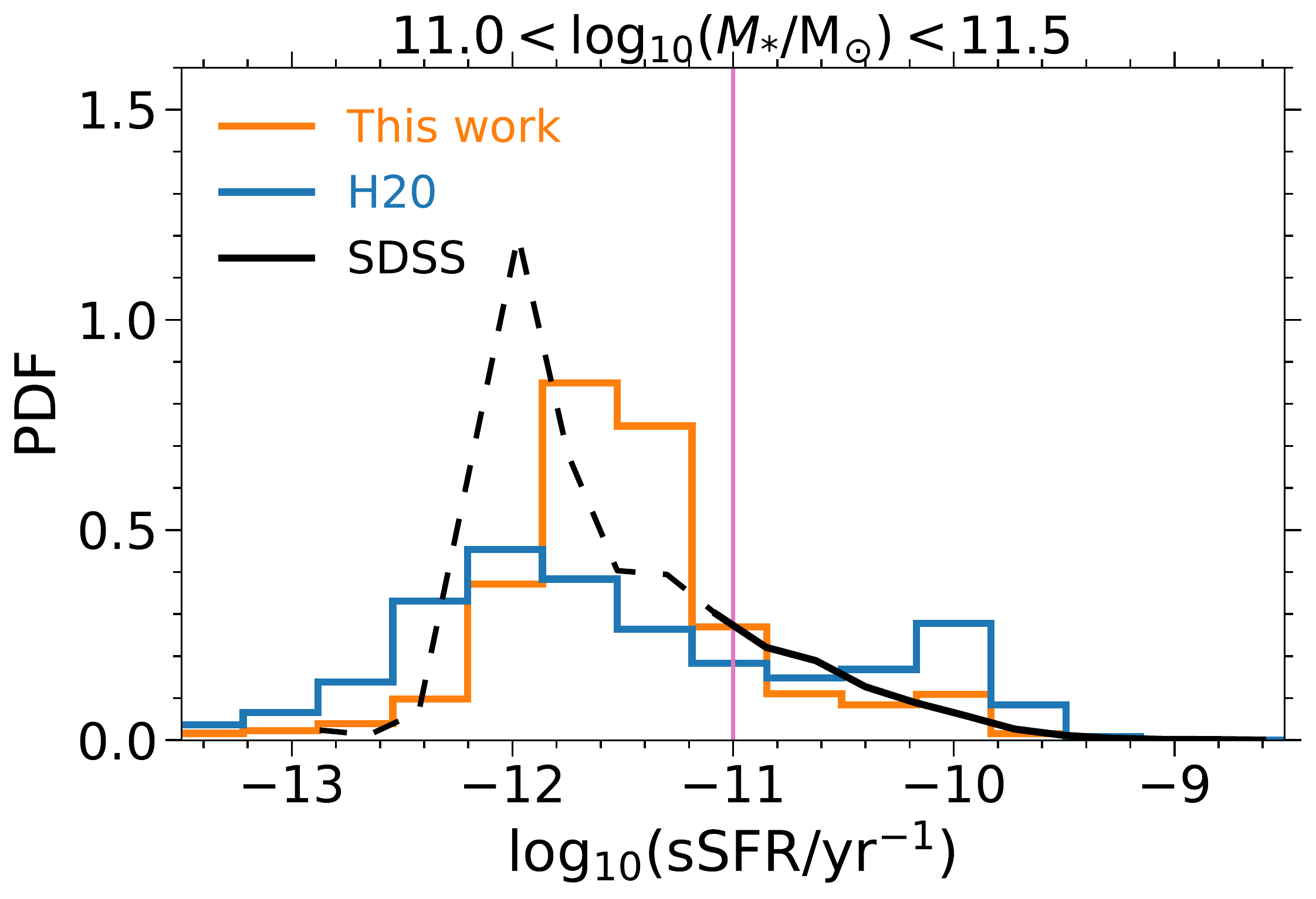}
    \caption{Distributions of sSFR for galaxies in this work, H20 and observations. Each panel corresponds to a particular stellar mass range. The SDSS data are based on \protect\cite{Brinchmann2004Physical} with the corrections of \protect\cite{Salim2007UV}. Galaxies with $\rm \log_{10}(sSFR/yr^{-1})<-11$ are considered as quenched. Observational data are shown with dashed lines in the regions where there is high uncertainty in the observed specific star formation rates. Despite extensive modification to the stripping properties of galaxies, the sSFR distributions are largely unchanged.}
\label{Fig: ssfr_hist_Obs}
\end{figure}

We compare our results for the fraction of quenched galaxies in different environments with SDSS observations. For both the models and observations, the halo mass ($M_{200}$) and radius ($R_{200}$) are estimated from their stellar masses as described in \S \ref{subsubsec: halomass_mock_simulation}. The simulations are transformed into redshift/velocity space based on the method explained in \S \ref{subsubsec: 2d_projection}. For every dark matter halo, we project the outputs of both our model and H20 along the z-axis of the simulation volume in velocity space with the thickness of the projected slice taken as $|v_{\rm gal,LOS}-v_{\rm halo,LOS}|\, \leq \pm 2V_{\rm 200,halo}$. For observations, the velocity separations are calculated along the lone-of-sight using the galaxy redshifts, and the 2D projected distances are calculated from the sky coordinates.

Fig. \ref{Fig: quenchedFrac_dis_proj_z0} shows the fraction of quenched galaxies as a function of projected halocentric distance in our model (solid lines), in H20 (dashed lines), and in the SDSS observations (points with errorbars). Each panel corresponds to a host stellar mass (or equivalently halo mass) bin indicated at the top of each panel. The results are shown for three different stellar mass ranges, specified by different colours. We first note that both models reproduce the field quenched fractions as a function of stellar mass. In all the halo mass ranges, a clear trend with distance is present both in the models and in the observations: the quenched fraction decreases with halocentric distance and reaches a constant, field value at some radius usually larger than the halo boundary, $R_{200}$.

For almost all stellar and host mass regimes our model shows better agreement with observations than H20. In the vicinity of massive haloes (top panel), the environmental dependence of quenched fraction extends to larger radius in our model and in SDSS compared to H20. Looking specifically at low-mass galaxies in the vicinity of clusters (top panel, green lines and points), our model and SDSS show that up to 60-70\% of galaxies near the halo centre are quenched, while this value is 10-20\% lower in H20. While our model is in better agreement for intermediate-mass (red lines and points) and massive (purple lines and points) galaxies, it underpredicts the fraction of quenched massive and intermediate-mass galaxies at $1<R/R_{200}<3$.

We suggest two reasons for this discrepancy which can motivate future model developments. The first is a lack of cold gas stripping in our model, which can be accommodated within our LBE framework, while the second is a lack of ejective AGN feedback. It is possible that strong ejective feedback can push gas outside the halo boundary. Infalling galaxies passing through this ejected gas experience substantial enhancement in ram-pressure and can therefore lose a larger fraction of their halo gas, as we see in the IllustrisTNG simulations (\textcolor{blue}{Ayromlou et al. in prep}). Furthermore, in massive galaxies with $\log_{10}(M_{\star}/{\rm M_{\odot}})>10.5$, AGN feedback makes extended subhalo gas less bound which results in an enhancement of ram-pressure stripping and thus galaxy quenching \citep{Ayromlou2020Comparing}.

Considering galaxies surrounding groups (middle panel) and lower mass haloes (bottom panel; Fig. \ref{Fig: quenchedFrac_dis_proj_z0}), our model is again in reasonably good agreement with observations, while H20 exhibits weaker effects within R200 and the dependence of the quenched fraction of distance flattens at smaller distances comparing to our model and SDSS. The persistence of the variation in quenched fraction out to large halocentric distances is even more marked in the SDSS data, a point we will come back to later.

\begin{figure}
    \centering
    \includegraphics[width=0.85\columnwidth]{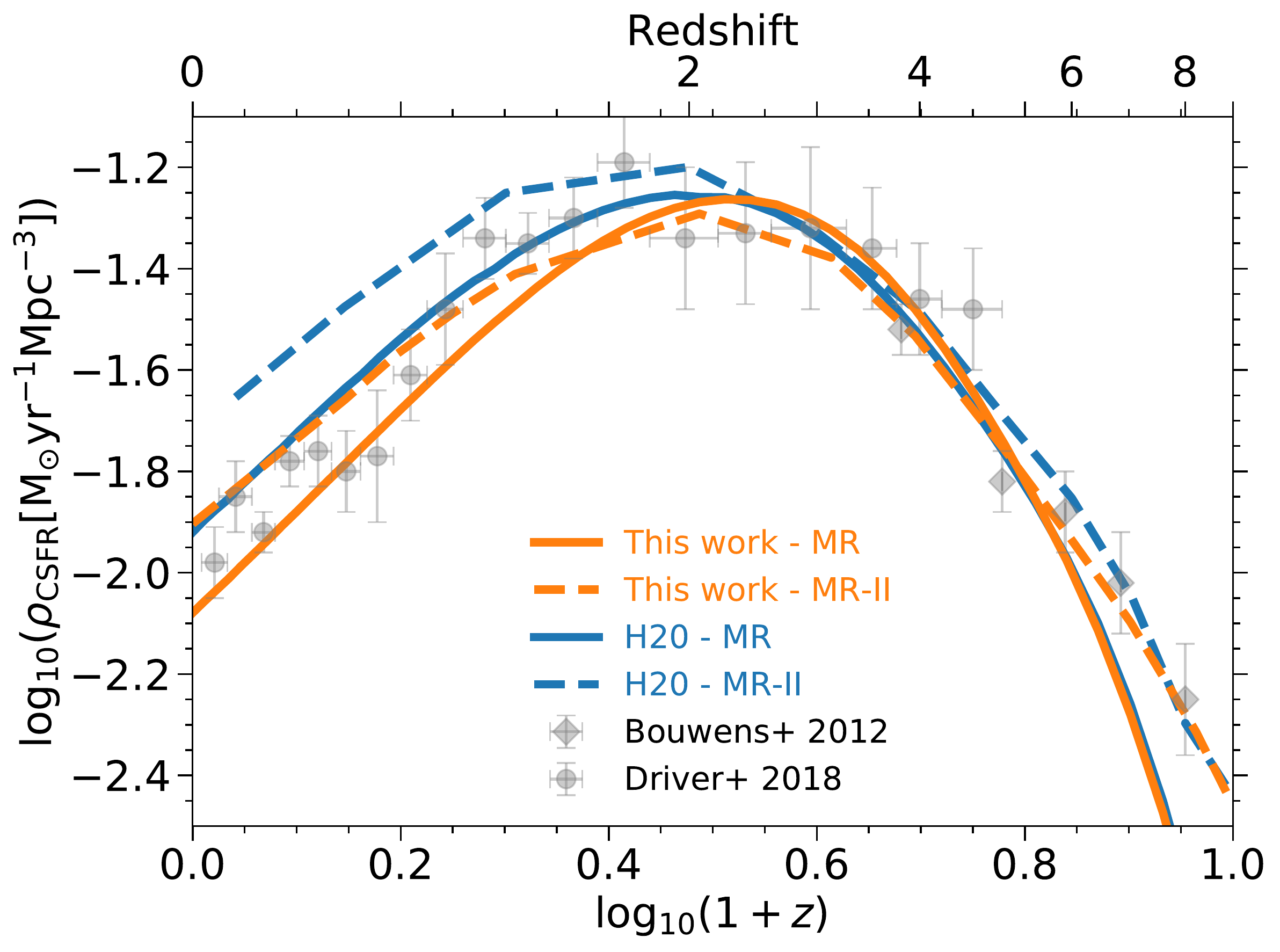}
    \caption{Cosmic star formation rate density as a function of redshift for our model, for H20 and for observations from \protect\cite{Bouwens2012UV-continuum,Driver2018GAMA}.}
\label{Fig: cosmic_SFR_density}
\end{figure}

In Fig. \ref{Fig: quenchedFrac_M*_proj_z0}, we show the fraction of quenched galaxies, now as a function of stellar mass, for three different halocentric distance bins (different colours). Different panels correspond to different host stellar (or halo) masses. In general, the quenched fraction almost always increases monotonically with stellar mass for both the models and the observations. In the vicinity of massive haloes (top panel), our model is in relatively good agreement with observations for all the distance bins, while H20 is off by up to 20\%. The difference between our model and H20 decreases with stellar mass and is the largest for low-mass galaxies.

Similar results are found for galaxies in the vicinity of intermediate mass haloes (bottom panel of Fig. \ref{Fig: quenchedFrac_M*_proj_z0}), although the difference between our model and H20 is smaller. Overall, our model predictions are in better agreement with observations than H20, and low-mass galaxies in cluster environments are the most influenced by our new gas stripping method.

Fig. \ref{Fig: quenchedFrac_dis_vLOS_hist2D} shows the fraction of quenched galaxies as a function of projected halocentric distance (x-axis) and line-of-sight velocity (y-axis), i.e. projected phase space \citep[PPS; see also][]{oman20}. We contrast our model (top panel), H20 (middle panel), and SDSS (bottom panel). In all three cases the quenched fraction decreases both with halocentric distance and with the magnitude of the line-of-sight velocity. As $R_{\rm proj}/R_{200}$ and $|v_{\rm LOS}/v_{200}|$ increase, more field galaxies are included in each bin and the quenched fraction eventually approaches the field value. The trend with distance is stronger in our model than in H20, but is still weaker than in SDSS. In other words, real haloes seem to influence nearby galaxies out to somewhat larger distances than in either model. 

\begin{figure*}
    \centering
    \includegraphics[width=0.8\columnwidth]{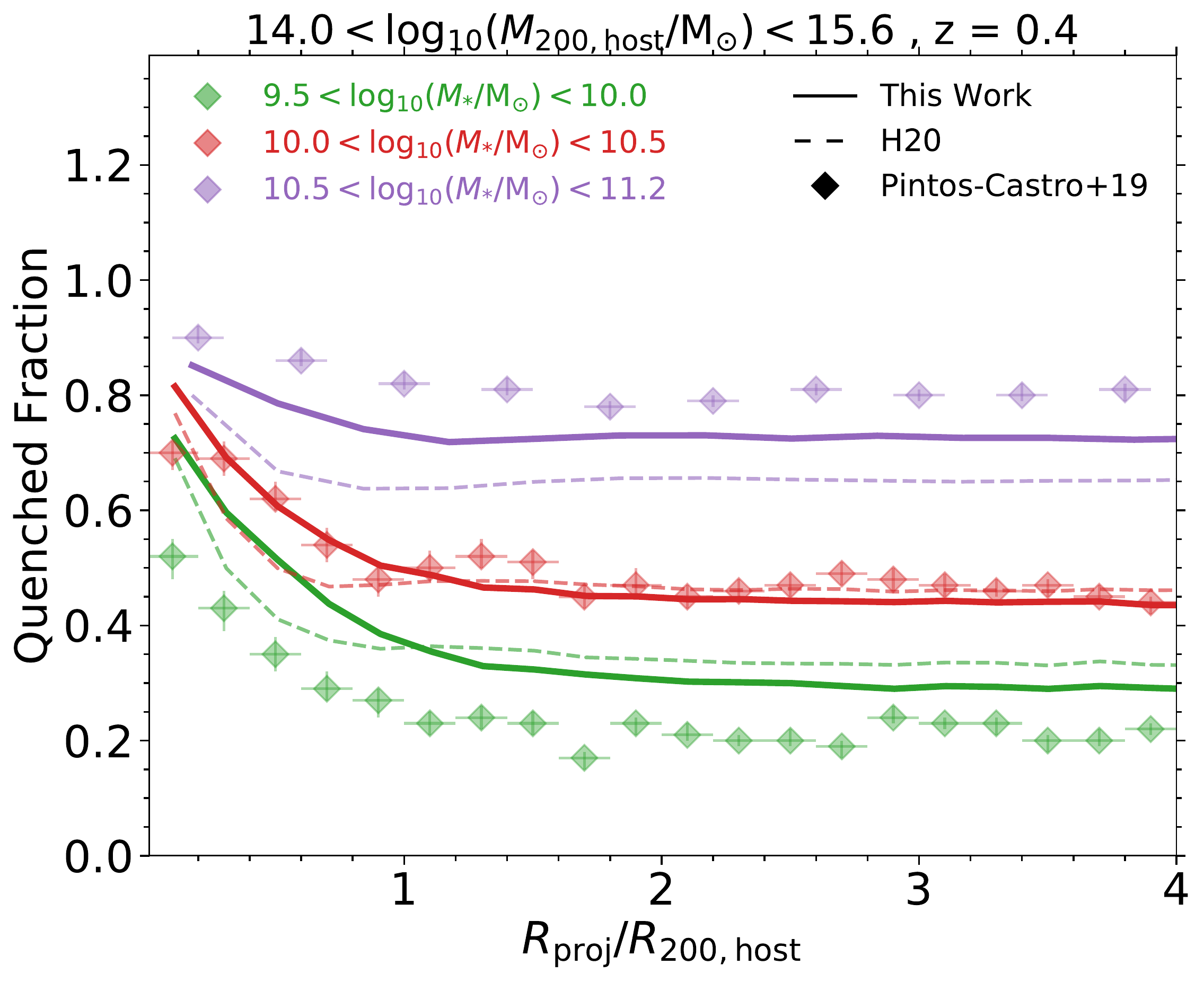}
    \includegraphics[width=0.8\columnwidth]{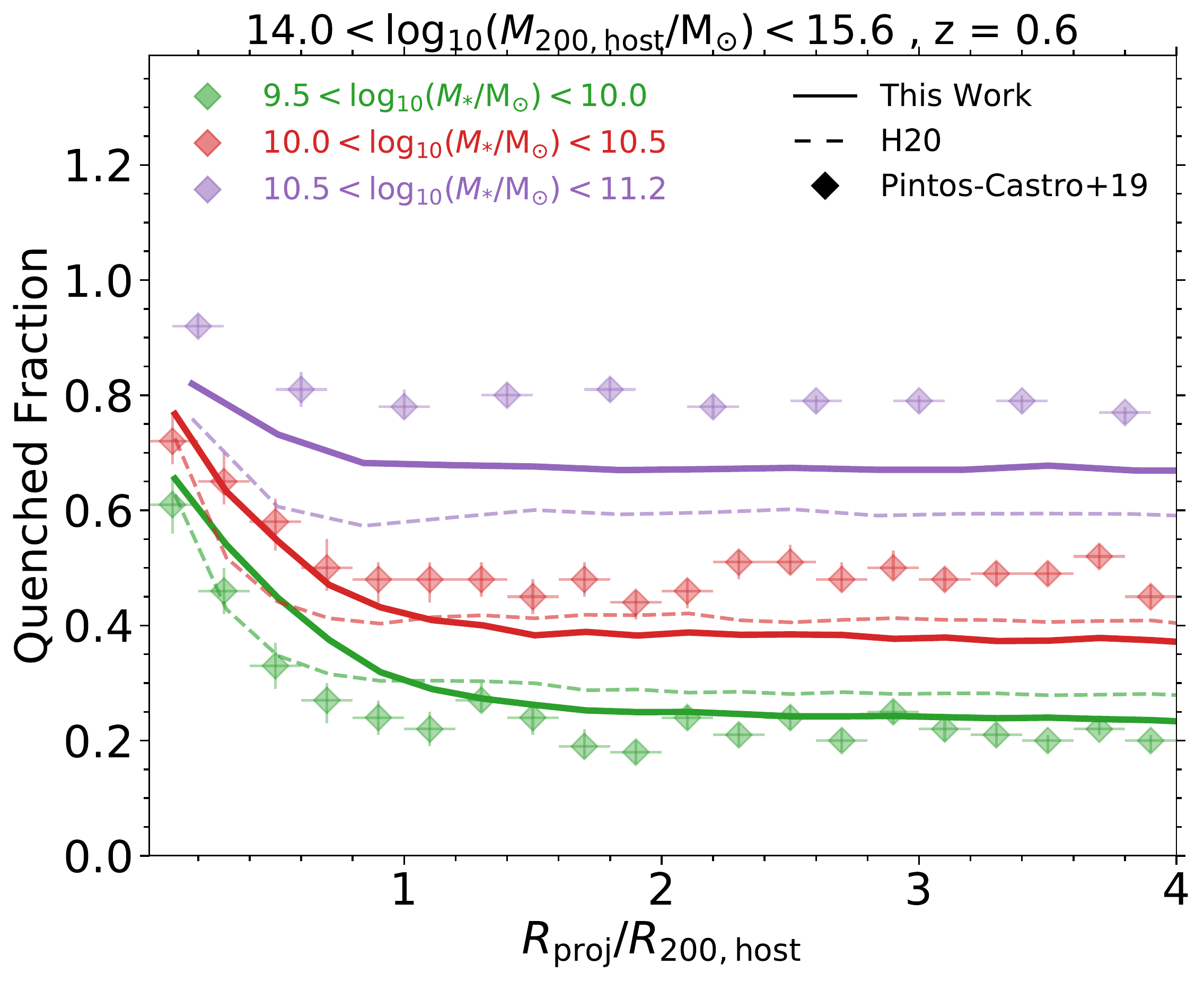}
    \includegraphics[width=0.8\columnwidth]{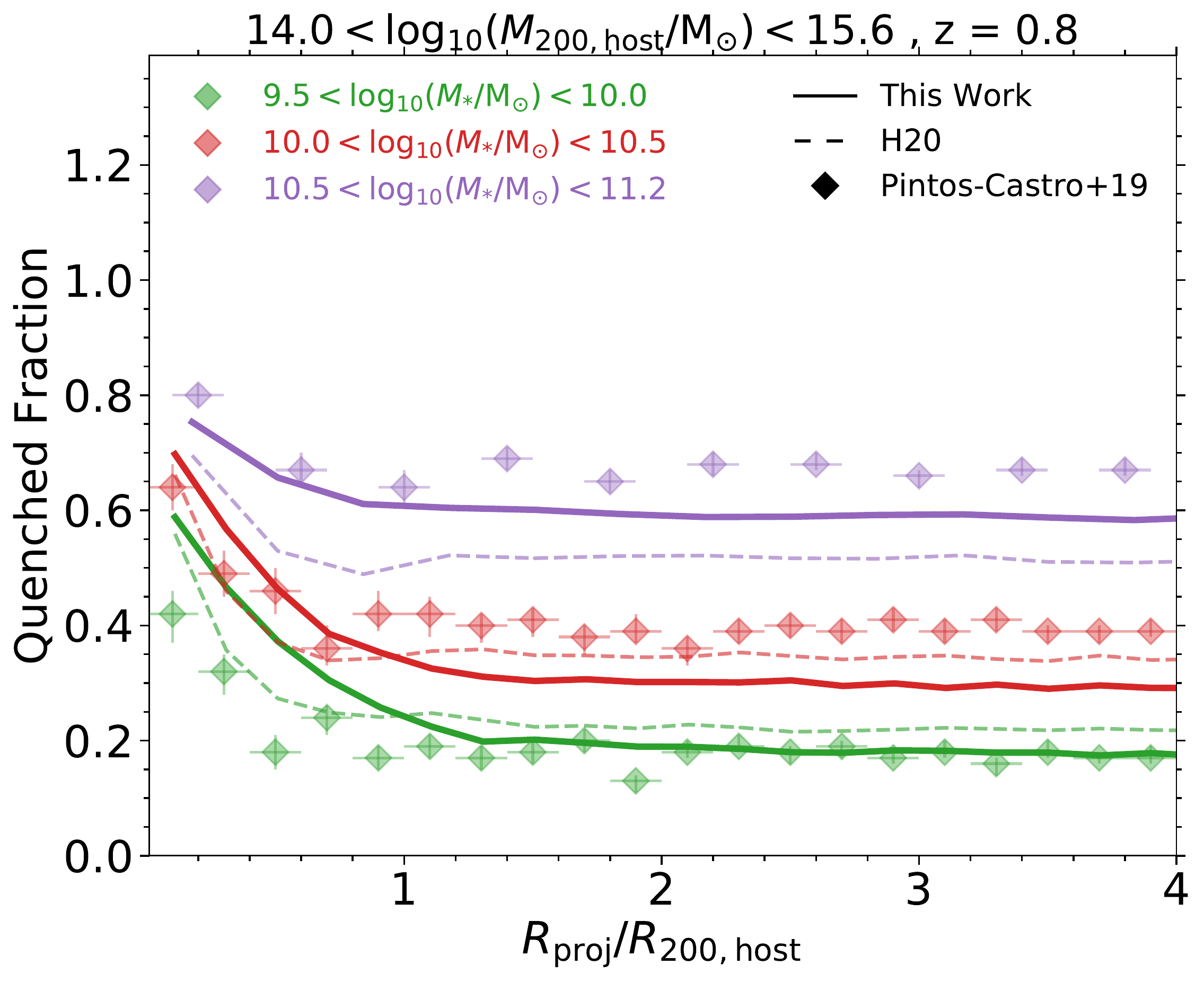}
    \includegraphics[width=0.8\columnwidth]{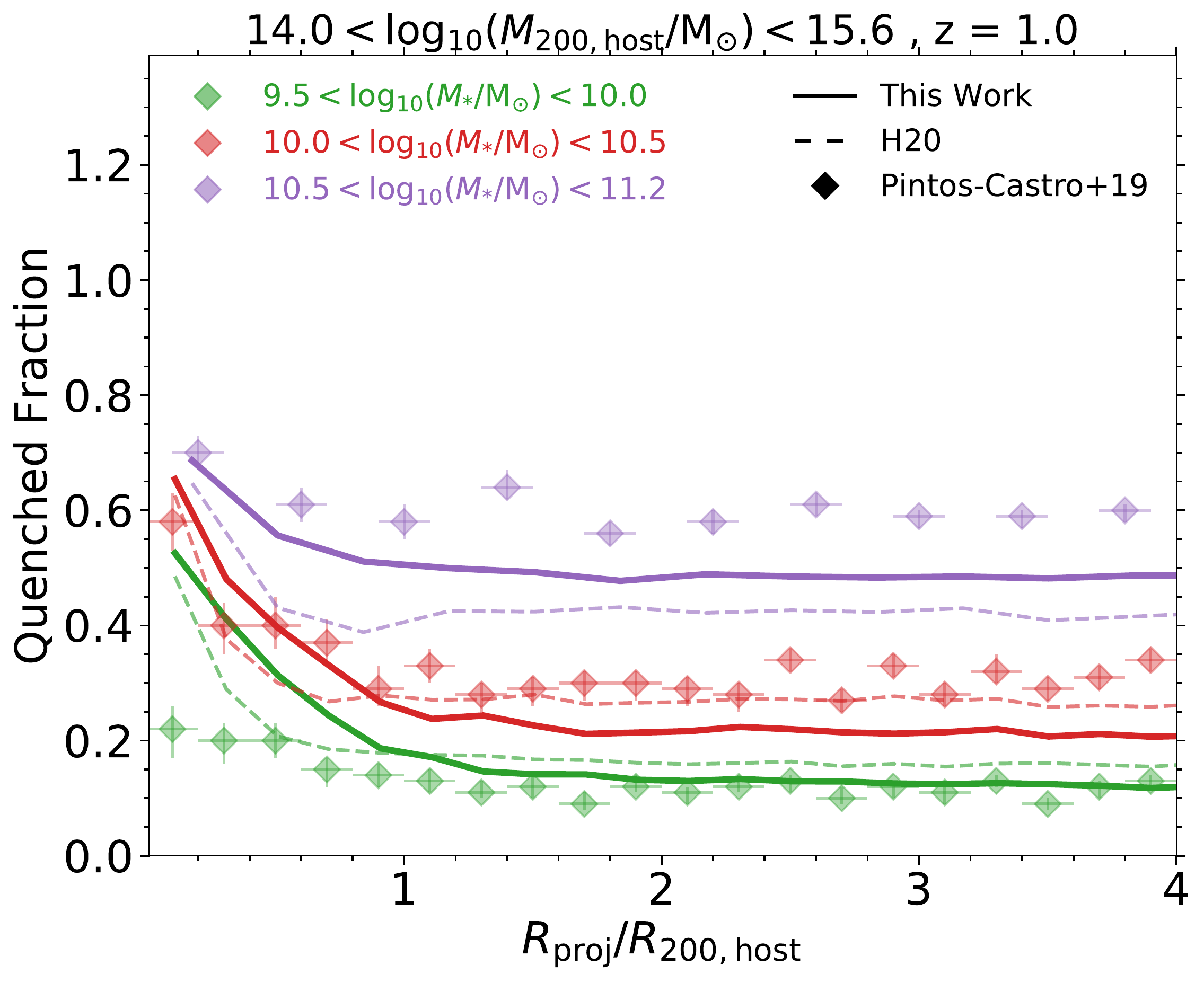}
    \caption{Fraction of quenched galaxies as a function of projected distance to centre of clusters at different redshifts. In the models, galaxies with ${\rm SSFR}<(1+z)/(2\times10^{10})$ are considered as quenched. Here we compare to observational data  at $0.4 \leq z \leq 1.0$ from \protect\cite{Pintos-Castro2019Evolution}. Note that in this data a colour cut is used and galaxies with ${\rm U-J}>0.88\,{\rm (J-V)} + 0.59$ are considered as quenched. In comparison to H20, our updated model shows an environmental effect on the quenched fraction which extends to larger halocentric distance.}
\label{Fig: quenchedFrac_dis_proj_z>0}
\end{figure*}

As a final comparison with SDSS data, Fig. \ref{Fig: ssfr_hist_Obs} shows the distributions of specific star formation rate in this work, in H20, and in SDSS observations at $z\sim0$. Different panels correspond to different stellar mass ranges. SDSS data are shown with black lines, which become dashed where there is high uncertainty in observed star formation rates. The magenta vertical lines at $\rm \log_{10}(sSFR/yr^{-1})=-11$ demarcate the value below which galaxies are considered as quenched. Although we have changed the properties of galaxies near massive hosts significantly, the overall specific star formation rates are rather similar. Both our model and H20 are in relatively good agreement with observation. The noticeable difference is for massive galaxies with $11<\log_{10}(M_{\star}/{\rm M_{\odot})}<11.5$, where our model shows lower specific star formation rates than H20. This is caused by the higher black hole feedback efficiency in our new model.

\subsubsection{Beyond $z=0$: the cosmic star formation rate density}

In Fig. \ref{Fig: cosmic_SFR_density}, we compare the cosmic star formation rate density, $\rho_{\rm CSFR}$, as a function of redshift in the two models to observations taken from \cite{Bouwens2012UV-continuum,Driver2018GAMA}. Overall, both models are in agree quite well with data. The two models are similar at redshifts greater than the peak of cosmic star formation rate density, i.e. $z\gtrsim2$. On the other hand, at lower redshifts, $\rho_{\rm CSFR}$ in our model is smaller than in H20. The maximum difference between our model and H20 occurs at $z=0$ where $\rho_{\rm CSFR}$ is 0.15 dex smaller in our model. This is mainly due to our extended gas stripping implementation but also to our higher AGN feedback efficiency. The former mostly affects low-mass galaxies while the latter influences star formation in more massive objects.

\subsubsection{Quenched fractions versus observations at $z>0$}

Fig. \ref{Fig: quenchedFrac_dis_proj_z>0} shows the fraction of quenched galaxies versus projected halocentric distance in our model, in H20, and as observed in the HSC survey \citep{Pintos-Castro2019Evolution} at four different redshifts. For this comparison, we take the direct output of our simulation for the halo mass and radius (\S \ref{subsubsec: halomass_mock_simulation}) and project the whole simulation depth along the z-axis of the simulation volume as a rough approximation of the survey characteristics (\S \ref{subsec: mock_catalogues}).

At all redshifts the quenched fraction decreases with distance both for models and for observations. The agreement with observations is fair for both models, although the quenched fraction of low-mass galaxies in our model (solid green lines) is higher than in H20 (dashed green lines) and than observed (green points) within the virial radius of clusters. At $z=1$ (bottom right panel), both our model and H20 show a rather strong trend with distance for low-mass galaxies, whereas such a trend is not observed. This could be due to the presence of low-mass quenched galaxies at $z=1$ which fall outside the observed samples. The overall trend with stellar mass is similar for our work, for H20 and for the observational data. At all halocentric distances and redshifts, the quenched fractions increases with stellar mass. 

In both models, the quenched fraction near the cluster centre almost always decreases with redshift, i.e. lower redshift galaxies have higher quenched fractions. On the other hand, comparing observations at $z=0.4$ and $z=0.6$ (top left and top right panels), the quenched fraction is higher at higher redshift, the opposite of the trend predicted by the models, possibly due to different methods or definitions of halo mass. It would be best to compare to high redshift data with spectroscopic redshifts to confirm whether these discrepancies are real or are caused by systematic errors of some kind. Observational constraints beyond the local Universe will undoubtedly pose a challenge to theoretical models, and more sophisticated comparisons and future model improvements will further increase our understanding of the role of environment in galaxy evolution.


\section{Summary and Discussion}
\label{sec: summary}

In this work we study the impact of environment on the formation and evolution of galaxies. We present a variant of the Munich semi-analytical model of galaxy formation, \textsc{L-Galaxies}, with a novel gas-stripping method. Following \cite{ayromlou2019new}, we measure the properties of the local environment of every galaxy and subhalo directly from the particle data of the underlying N-body simulations. This enables us to devise a more accurate treatment of environmental processes, particularly ram-pressure stripping. We re-calibrate the parameters of the new model using an MCMC technique and a set of observational constraints, namely the stellar mass function and quenched fraction at $z=0,1,2$. Due to this re-calibration, global properties of galaxies such as the stellar mass function, quenched fractions versus galaxy mass and HI mass function are all largely unchanged from H20 and remain consistent with observations. Analysing the results of our new model and the standard model on which it is based (H20), our main results are as follows:

\begin{itemize}
    \item Measuring the total amount of stripped gas in our model, we find that galaxies in the vicinity of haloes with $M_{200}/\rm M_{\odot}>10^{12}$ lose a large fraction (median $\sim80\%$ in low-mass galaxies with $9.5<\log_{10}(M_{\star}/{\rm M_{\odot}})<10$) of their gas due to ram-pressure stripping while they were in fact central galaxies (Fig. \ref{Fig: totGasStripped_dis_z0}).
    \item At high redshifts, $z\gtrsim 1$, most stripping is due to gas removal of previously `ejected' material. At $z\lesssim 1$, in contrast, stripping of the hot halo gas is the dominant contributor (Fig. \ref{Fig: Stripped_density_vs_redshift}).
    \item The ratio of hot gas to stellar mass, $M_{\rm hotgas}/M_{*}$, decreases with decreasing halocentric distance towards the centres of haloes, due to stripping. In our model, more than half of all galaxies in the vicinity of clusters and groups, up to several virial radii, are almost devoid of hot gas. The dependence of $M_{\rm hotgas}/M_{*}$ on environment extends to much larger halocentric distances in our model than in H20, with low-mass galaxies being more strongly affected by their environment (Fig. \ref{Fig: hotgas_comb_plot}).
    \item Near clusters and groups in our model, the quenched fraction decreases with halocentric distance, and it flattens to the field value only at $R/R_{200}\lesssim 2-3$, far beyond the halo virial radius (Fig. \ref{Fig: quenchedFrac_comb_plot}).
    \item The characteristic halocentric distance at which the fraction of quenched galaxies is 20\% larger than its field value ($R_{\rm c,20\%}$) is, on average, four times larger in our model than in H20. In our model, $R_{\rm c,20\%}/R_{200}$ increases monotonically with halo mass, whereas such a trend is not present in H20 (Fig. \ref{Fig: quenchedFrac_comb_plot}).
\end{itemize}

We undertake a new analysis of SDSS galaxy data \citep{Abazajian2009SDSS} combined with the \cite{Yang2005halo-based,Yang2007Galaxy} halo catalogues, inferring the quenched fraction versus halocentric distance out to $R_{\rm proj} = 10R_{200}$ with a methodology consistent between simulations and data. Together with observational results from the HSC survey, we compare against our model predictions. Our principal results are:

\begin{itemize}
    \item The $z=0$ observed trend of the quenched fraction of galaxies ($f_{\rm q}$) versus halocentric distance (SDSS) is well reproduced in our model up to several halo virial radii, a noticeable improvement over H20. Nevertheless, the observed environmental dependency of galaxies extends to slightly larger distances (Fig. \ref{Fig: quenchedFrac_dis_proj_z0}).
    \item In our model, as well as in SDSS observations, the quenched fraction near haloes increases with halo mass (Fig. \ref{Fig: quenchedFrac_dis_proj_z0}). The strength of environmental quenching in the vicinity of haloes also decreases with increasing galaxy stellar mass, which is consistent with the observations (Fig. \ref{Fig: quenchedFrac_M*_proj_z0}).
    \item At higher redshifts, $0.4 \leq z \leq 1$, our model is in relatively good agreement with observations from the HSC survey, while quantitative differences remain, particularly within the $R_{200}$ of clusters (Fig. \ref{Fig: quenchedFrac_dis_proj_z>0}). We note that we have not attempted to model the effect of errors on the photometric redshifts that are used in the data.
    \item Our predicted HI gas mass function is in good agreement both with the previous model and with $z=0$ data (Fig. \ref{Fig: HI_MF}).
\end{itemize}

The remaining tensions with data motivate two possible future model improvements: (i) stripping of the cold, star-forming gas discs in galaxies, and (ii) handling the ejection and re-distribution of gas in the (sub)halo due to baryonic feedback processes. Our method for incorporating the local background environment of galaxies can be naturally extended to handle cold gas stripping at smaller scales. At the same time, we have shown that the impact of environment on galaxy properties extends to much larger scales than the often assumed halo virial radius (also see \citealt{ayromlou2019new}), and that related effects are also present in cosmological hydrodynamical simulations \citep{Ayromlou2020Comparing}. Model improvements, incorporating insights from the use of our local environmental measurements in hydrodynamical simulations including IllustrisTNG and EAGLE (\textcolor{blue}{Ayromlou et al. in prep}) will be ideally suited to reveal the links between the physics of galaxy evolution and large scale correlations (e.g. two point correlation function of galaxies of different colour). Complex questions such as the physics behind the galactic conformity, the observed large-scale correlation between the star formation of neighbouring galaxies  \citep{weinmann2006properties,kauffmann2013re} will be interesting avenues of further investigation using our new model.

\section*{Data Availability}
We have made the full output of our updated \textsc{L-Galaxies} model for all the Millennium and Millennium-II snapshots publicly available at \href{https://lgalaxiespublicrelease.github.io}{lgalaxiespublicrelease.github.io}, where the H20 model is already publicly available.

\section*{Acknowledgements}
MA is grateful to Bruno Henriques for very helpful discussions about the MCMC model calibration. MA also thanks Irene Pintos-Castro for kindly sharing her observational data and for fruitful discussions. MA finally thanks Abhijeet Anand, Alireza Vafaei Sadr, Hasti Khoraminezhad, Wolfgang Enzi, Richard D'Souza, and John Helly for useful discussions and assistance. The analysis herein was carried out on the compute cluster of the Max Planck Institute for Astrophysics (MPA) at the Max Planck Computing and Data Facility (MPCDF).


\vspace{-1em}
\bibliographystyle{mnras}
\bibliography{refbibtex}

\begin{thebibliography}{}
\makeatletter
\relax
\def\mn@urlcharsother{\let\do\@makeother \do\$\do\&\do\#\do\^\do\_\do\%\do\~}
\def\mn@doi{\begingroup\mn@urlcharsother \@ifnextchar [ {\mn@doi@}
  {\mn@doi@[]}}
\def\mn@doi@[#1]#2{\def\@tempa{#1}\ifx\@tempa\@empty \href
  {http://dx.doi.org/#2} {doi:#2}\else \href {http://dx.doi.org/#2} {#1}\fi
  \endgroup}
\def\mn@eprint#1#2{\mn@eprint@#1:#2::\@nil}
\def\mn@eprint@arXiv#1{\href {http://arxiv.org/abs/#1} {{\tt arXiv:#1}}}
\def\mn@eprint@dblp#1{\href {http://dblp.uni-trier.de/rec/bibtex/#1.xml}
  {dblp:#1}}
\def\mn@eprint@#1:#2:#3:#4\@nil{\def\@tempa {#1}\def\@tempb {#2}\def\@tempc
  {#3}\ifx \@tempc \@empty \let \@tempc \@tempb \let \@tempb \@tempa \fi \ifx
  \@tempb \@empty \def\@tempb {arXiv}\fi \@ifundefined
  {mn@eprint@\@tempb}{\@tempb:\@tempc}{\expandafter \expandafter \csname
  mn@eprint@\@tempb\endcsname \expandafter{\@tempc}}}

\bibitem[\protect\citeauthoryear{{Abazajian} et~al.,}{{Abazajian}
  et~al.}{2009}]{Abazajian2009SDSS}
{Abazajian} K.~N.,  et~al., 2009, \mn@doi [\apjs]
  {10.1088/0067-0049/182/2/543}, \href
  {https://ui.adsabs.harvard.edu/abs/2009ApJS..182..543A} {182, 543}

\bibitem[\protect\citeauthoryear{{Aihara} et~al.,}{{Aihara}
  et~al.}{2018}]{Aihara2018HSC}
{Aihara} H.,  et~al., 2018, \mn@doi [\pasj] {10.1093/pasj/psx066}, \href
  {https://ui.adsabs.harvard.edu/abs/2018PASJ...70S...4A} {70, S4}

\bibitem[\protect\citeauthoryear{Angulo \& Hilbert}{Angulo \&
  Hilbert}{2015}]{angulo2015cosmological}
Angulo R.~E.,  Hilbert S.,  2015, Monthly Notices of the Royal Astronomical
  Society, 448, 364

\bibitem[\protect\citeauthoryear{Angulo \& White}{Angulo \&
  White}{2010}]{angulo2010one}
Angulo R.~E.,  White S.~D.,  2010, Monthly Notices of the Royal Astronomical
  Society, 405, 143

\bibitem[\protect\citeauthoryear{{Angulo}, {Springel}, {White}, {Jenkins},
  {Baugh}  \& {Frenk}}{{Angulo} et~al.}{2012}]{Angulo2012Scaling}
{Angulo} R.~E.,  {Springel} V.,  {White} S.~D.~M.,  {Jenkins} A.,  {Baugh}
  C.~M.,   {Frenk} C.~S.,  2012, \mn@doi [\mnras]
  {10.1111/j.1365-2966.2012.21830.x}, \href
  {https://ui.adsabs.harvard.edu/abs/2012MNRAS.426.2046A} {426, 2046}

\bibitem[\protect\citeauthoryear{{Ayromlou}, {Nelson}, {Yates}, {Kauffmann}  \&
  {White}}{{Ayromlou} et~al.}{2019}]{ayromlou2019new}
{Ayromlou} M.,  {Nelson} D.,  {Yates} R.~M.,  {Kauffmann} G.,   {White} S.
  D.~M.,  2019, \mn@doi [\mnras] {10.1093/mnras/stz1549}, \href
  {https://ui.adsabs.harvard.edu/abs/2019MNRAS.487.4313A} {487, 4313}

\bibitem[\protect\citeauthoryear{{Ayromlou}, {Nelson}, {Yates}, {Kauffmann},
  {Renneby}  \& {White}}{{Ayromlou} et~al.}{2020}]{Ayromlou2020Comparing}
{Ayromlou} M.,  {Nelson} D.,  {Yates} R.~M.,  {Kauffmann} G.,  {Renneby} M.,
  {White} S. D.~M.,  2020, arXiv e-prints, \href
  {https://ui.adsabs.harvard.edu/abs/2020arXiv200414390A} {p. arXiv:2004.14390}
  (\mn@eprint {arXiv} {2004.14390})

\bibitem[\protect\citeauthoryear{Bah{\'e}, McCarthy, Crain  \& Theuns}{Bah{\'e}
  et~al.}{2012}]{bahe2012competition}
Bah{\'e} Y.~M.,  McCarthy I.~G.,  Crain R.~A.,   Theuns T.,  2012, Monthly
  Notices of the Royal Astronomical Society, 424, 1179

\bibitem[\protect\citeauthoryear{{Baldry}, {Glazebrook}, {Brinkmann},
  {Ivezi{\'c}}, {Lupton}, {Nichol}  \& {Szalay}}{{Baldry}
  et~al.}{2004}]{Baldry2004Quantifying}
{Baldry} I.~K.,  {Glazebrook} K.,  {Brinkmann} J.,  {Ivezi{\'c}} {\v{Z}}.,
  {Lupton} R.~H.,  {Nichol} R.~C.,   {Szalay} A.~S.,  2004, \mn@doi [\apj]
  {10.1086/380092}, \href
  {https://ui.adsabs.harvard.edu/abs/2004ApJ...600..681B} {600, 681}

\bibitem[\protect\citeauthoryear{Baldry, Glazebrook  \& Driver}{Baldry
  et~al.}{2008}]{baldry2008galaxy}
Baldry I.,  Glazebrook K.,   Driver S.,  2008, Monthly Notices of the Royal
  Astronomical Society, 388, 945

\bibitem[\protect\citeauthoryear{Baldry et~al.,}{Baldry
  et~al.}{2012}]{baldry2012galaxy}
Baldry I.~K.,  et~al., 2012, Monthly Notices of the Royal Astronomical Society,
  421, 621

\bibitem[\protect\citeauthoryear{Balogh, Morris, Yee, Carlberg  \&
  Ellingson}{Balogh et~al.}{1999}]{balogh1999differential}
Balogh M.~L.,  Morris S.~L.,  Yee H.,  Carlberg R.,   Ellingson E.,  1999, The
  Astrophysical Journal, 527, 54

\bibitem[\protect\citeauthoryear{{Bell}, {McIntosh}, {Katz}  \&
  {Weinberg}}{{Bell} et~al.}{2003}]{Bell2003Optical}
{Bell} E.~F.,  {McIntosh} D.~H.,  {Katz} N.,   {Weinberg} M.~D.,  2003, \mn@doi
  [\apjs] {10.1086/378847}, \href
  {https://ui.adsabs.harvard.edu/abs/2003ApJS..149..289B} {149, 289}

\bibitem[\protect\citeauthoryear{Bertone, De~Lucia  \& Thomas}{Bertone
  et~al.}{2007}]{bertone2007recycling}
Bertone S.,  De~Lucia G.,   Thomas P.~A.,  2007, Monthly Notices of the Royal
  Astronomical Society, 379, 1143

\bibitem[\protect\citeauthoryear{Binney \& Tremaine}{Binney \&
  Tremaine}{1987}]{binney1987galactic}
Binney J.,  Tremaine S.,  1987, Galactic Dynamics, Princeton Univ

\bibitem[\protect\citeauthoryear{{Boselli} \& {Gavazzi}}{{Boselli} \&
  {Gavazzi}}{2006}]{boselli06}
{Boselli} A.,  {Gavazzi} G.,  2006, \mn@doi [\pasp] {10.1086/500691}, \href
  {http://adsabs.harvard.edu/abs/2006PASP..118..517B} {118, 517}

\bibitem[\protect\citeauthoryear{{Boselli} et~al.,}{{Boselli}
  et~al.}{2016}]{Boselli2016Quenching}
{Boselli} A.,  et~al., 2016, \mn@doi [\aap] {10.1051/0004-6361/201629221},
  \href {https://ui.adsabs.harvard.edu/abs/2016A&A...596A..11B} {596, A11}

\bibitem[\protect\citeauthoryear{{Bouwens} et~al.,}{{Bouwens}
  et~al.}{2012}]{Bouwens2012UV-continuum}
{Bouwens} R.~J.,  et~al., 2012, \mn@doi [\apj] {10.1088/0004-637X/754/2/83},
  \href {https://ui.adsabs.harvard.edu/abs/2012ApJ...754...83B} {754, 83}

\bibitem[\protect\citeauthoryear{Boylan-Kolchin, Springel, White, Jenkins  \&
  Lemson}{Boylan-Kolchin et~al.}{2009}]{boylan2009resolving}
Boylan-Kolchin M.,  Springel V.,  White S.~D.,  Jenkins A.,   Lemson G.,  2009,
  Monthly Notices of the Royal Astronomical Society, 398, 1150

\bibitem[\protect\citeauthoryear{{Bravo}, {Lagos}, {Robotham}, {Bellstedt}  \&
  {Obreschkow}}{{Bravo} et~al.}{2020}]{Bravo2020From}
{Bravo} M.,  {Lagos} C. d.~P.,  {Robotham} A. S.~G.,  {Bellstedt} S.,
  {Obreschkow} D.,  2020, \mn@doi [\mnras] {10.1093/mnras/staa2027}, \href
  {https://ui.adsabs.harvard.edu/abs/2020MNRAS.497.3026B} {497, 3026}

\bibitem[\protect\citeauthoryear{{Brinchmann}, {Charlot}, {White}, {Tremonti},
  {Kauffmann}, {Heckman}  \& {Brinkmann}}{{Brinchmann}
  et~al.}{2004}]{Brinchmann2004Physical}
{Brinchmann} J.,  {Charlot} S.,  {White} S.~D.~M.,  {Tremonti} C.,  {Kauffmann}
  G.,  {Heckman} T.,   {Brinkmann} J.,  2004, \mn@doi [\mnras]
  {10.1111/j.1365-2966.2004.07881.x}, \href
  {https://ui.adsabs.harvard.edu/abs/2004MNRAS.351.1151B} {351, 1151}

\bibitem[\protect\citeauthoryear{Cole, Lacey, Baugh  \& Frenk}{Cole
  et~al.}{2000}]{cole2000hierarchical}
Cole S.,  Lacey C.~G.,  Baugh C.~M.,   Frenk C.~S.,  2000, Monthly Notices of
  the Royal Astronomical Society, 319, 168

\bibitem[\protect\citeauthoryear{Cora et~al.,}{Cora
  et~al.}{2018}]{cora2018semi}
Cora S.~A.,  et~al., 2018, Monthly Notices of the Royal Astronomical Society

\bibitem[\protect\citeauthoryear{{Crain} et~al.,}{{Crain}
  et~al.}{2015}]{Crain2015TheEagle}
{Crain} R.~A.,  et~al., 2015, \mn@doi [\mnras] {10.1093/mnras/stv725}, \href
  {https://ui.adsabs.harvard.edu/abs/2015MNRAS.450.1937C} {450, 1937}

\bibitem[\protect\citeauthoryear{Croton et~al.,}{Croton
  et~al.}{2006}]{croton2006many}
Croton D.~J.,  et~al., 2006, Monthly Notices of the Royal Astronomical Society,
  365, 11

\bibitem[\protect\citeauthoryear{Croton et~al.,}{Croton
  et~al.}{2016}]{croton2016semi}
Croton D.~J.,  et~al., 2016, The Astrophysical Journal Supplement Series, 222,
  22

\bibitem[\protect\citeauthoryear{{Davies} et~al.,}{{Davies}
  et~al.}{2019}]{Davies2020Galaxy}
{Davies} L.~J.~M.,  et~al., 2019, \mn@doi [\mnras] {10.1093/mnras/sty3393},
  \href {https://ui.adsabs.harvard.edu/abs/2019MNRAS.483.5444D} {483, 5444}

\bibitem[\protect\citeauthoryear{{Davis}, {Efstathiou}, {Frenk}  \&
  {White}}{{Davis} et~al.}{1985}]{Davis1985TheEvolution}
{Davis} M.,  {Efstathiou} G.,  {Frenk} C.~S.,   {White} S.~D.~M.,  1985,
  \mn@doi [\apj] {10.1086/163168}, \href
  {https://ui.adsabs.harvard.edu/abs/1985ApJ...292..371D} {292, 371}

\bibitem[\protect\citeauthoryear{{De Lucia}, {Kauffmann}, {Springel}, {White},
  {Lanzoni}, {Stoehr}, {Tormen}  \& {Yoshida}}{{De Lucia}
  et~al.}{2004}]{DeLucia2004Substructures}
{De Lucia} G.,  {Kauffmann} G.,  {Springel} V.,  {White} S.~D.~M.,  {Lanzoni}
  B.,  {Stoehr} F.,  {Tormen} G.,   {Yoshida} N.,  2004, \mn@doi [\mnras]
  {10.1111/j.1365-2966.2004.07372.x}, \href
  {https://ui.adsabs.harvard.edu/abs/2004MNRAS.348..333D} {348, 333}

\bibitem[\protect\citeauthoryear{De~Lucia, Springel, White, Croton  \&
  Kauffmann}{De~Lucia et~al.}{2006}]{de2006formation}
De~Lucia G.,  Springel V.,  White S.~D.,  Croton D.,   Kauffmann G.,  2006,
  Monthly Notices of the Royal Astronomical Society, 366, 499

\bibitem[\protect\citeauthoryear{Dom{\'\i}nguez~S{\'a}nchez
  et~al.,}{Dom{\'\i}nguez~S{\'a}nchez et~al.}{2011}]{dominguez2011evolution}
Dom{\'\i}nguez~S{\'a}nchez H.,  et~al., 2011, Monthly Notices of the Royal
  Astronomical Society, 417, 900

\bibitem[\protect\citeauthoryear{{Donnari} et~al.,}{{Donnari}
  et~al.}{2020a}]{donnari2020}
{Donnari} M.,  et~al., 2020a, \mn@doi [\mnras] {10.1093/mnras/staa3006}, \href
  {https://ui.adsabs.harvard.edu/abs/2020MNRAS.tmp.2921D} {}

\bibitem[\protect\citeauthoryear{{Donnari}, {Pillepich}, {Nelson}, {Marinacci},
  {Vogelsberger}  \& {Hernquist}}{{Donnari} et~al.}{2020b}]{donnari2020b}
{Donnari} M.,  {Pillepich} A.,  {Nelson} D.,  {Marinacci} F.,  {Vogelsberger}
  M.,   {Hernquist} L.,  2020b, arXiv e-prints, \href
  {https://ui.adsabs.harvard.edu/abs/2020arXiv200800004D} {p. arXiv:2008.00004}
  (\mn@eprint {arXiv} {2008.00004})

\bibitem[\protect\citeauthoryear{Dressler}{Dressler}{1980}]{dressler1980galaxy}
Dressler A.,  1980, The Astrophysical Journal, 236, 351

\bibitem[\protect\citeauthoryear{{Driver} et~al.,}{{Driver}
  et~al.}{2018}]{Driver2018GAMA}
{Driver} S.~P.,  et~al., 2018, \mn@doi [\mnras] {10.1093/mnras/stx2728}, \href
  {https://ui.adsabs.harvard.edu/abs/2018MNRAS.475.2891D} {475, 2891}

\bibitem[\protect\citeauthoryear{{Dubois} et~al.,}{{Dubois}
  et~al.}{2020}]{dubois2020}
{Dubois} Y.,  et~al., 2020, arXiv e-prints, \href
  {https://ui.adsabs.harvard.edu/abs/2020arXiv200910578D} {p. arXiv:2009.10578}
  (\mn@eprint {arXiv} {2009.10578})

\bibitem[\protect\citeauthoryear{Font et~al.,}{Font
  et~al.}{2008}]{font2008colours}
Font A.~S.,  et~al., 2008, Monthly Notices of the Royal Astronomical Society,
  389, 1619

\bibitem[\protect\citeauthoryear{{Gladders} \& {Yee}}{{Gladders} \&
  {Yee}}{2000}]{Gladders2000New}
{Gladders} M.~D.,  {Yee} H.~K.~C.,  2000, \mn@doi [\aj] {10.1086/301557}, \href
  {https://ui.adsabs.harvard.edu/abs/2000AJ....120.2148G} {120, 2148}

\bibitem[\protect\citeauthoryear{{Gladders} \& {Yee}}{{Gladders} \&
  {Yee}}{2005}]{Gladders2005Red-Sequence}
{Gladders} M.~D.,  {Yee} H.~K.~C.,  2005, \mn@doi [\apjs] {10.1086/427327},
  \href {https://ui.adsabs.harvard.edu/abs/2005ApJS..157....1G} {157, 1}

\bibitem[\protect\citeauthoryear{{Grand} et~al.,}{{Grand}
  et~al.}{2017}]{Grand2017Auriga}
{Grand} R. J.~J.,  et~al., 2017, \mn@doi [\mnras] {10.1093/mnras/stx071}, \href
  {https://ui.adsabs.harvard.edu/abs/2017MNRAS.467..179G} {467, 179}

\bibitem[\protect\citeauthoryear{{Gunn} \& {Gott}}{{Gunn} \&
  {Gott}}{1972}]{Gunn_Gott1972}
{Gunn} J.~E.,  {Gott} J.~Richard I.,  1972, \mn@doi [\apj] {10.1086/151605},
  \href {https://ui.adsabs.harvard.edu/abs/1972ApJ...176....1G} {176, 1}

\bibitem[\protect\citeauthoryear{Guo et~al.,}{Guo et~al.}{2011}]{guo2011dwarf}
Guo Q.,  et~al., 2011, Monthly Notices of the Royal Astronomical Society, 413,
  101

\bibitem[\protect\citeauthoryear{{Guo}, {White}, {Angulo}, {Henriques},
  {Lemson}, {Boylan-Kolchin}, {Thomas}  \& {Short}}{{Guo}
  et~al.}{2013}]{Guo2013Galaxy}
{Guo} Q.,  {White} S.,  {Angulo} R.~E.,  {Henriques} B.,  {Lemson} G.,
  {Boylan-Kolchin} M.,  {Thomas} P.,   {Short} C.,  2013, \mn@doi [\mnras]
  {10.1093/mnras/sts115}, \href
  {https://ui.adsabs.harvard.edu/abs/2013MNRAS.428.1351G} {428, 1351}

\bibitem[\protect\citeauthoryear{Hansen, Sheldon, Wechsler  \& Koester}{Hansen
  et~al.}{2009}]{hansen2009galaxy}
Hansen S.~M.,  Sheldon E.~S.,  Wechsler R.~H.,   Koester B.~P.,  2009, The
  Astrophysical Journal, 699, 1333

\bibitem[\protect\citeauthoryear{{Haynes} et~al.,}{{Haynes}
  et~al.}{2011}]{Haynes2011Arecibo}
{Haynes} M.~P.,  et~al., 2011, \mn@doi [\aj] {10.1088/0004-6256/142/5/170},
  \href {https://ui.adsabs.harvard.edu/abs/2011AJ....142..170H} {142, 170}

\bibitem[\protect\citeauthoryear{{Henriques}, {Thomas}, {Oliver}  \&
  {Roseboom}}{{Henriques} et~al.}{2009}]{henriques2009monte}
{Henriques} B. M.~B.,  {Thomas} P.~A.,  {Oliver} S.,   {Roseboom} I.,  2009,
  \mn@doi [\mnras] {10.1111/j.1365-2966.2009.14730.x}, \href
  {https://ui.adsabs.harvard.edu/abs/2009MNRAS.396..535H} {396, 535}

\bibitem[\protect\citeauthoryear{{Henriques}, {White}, {Thomas}, {Angulo},
  {Guo}, {Lemson}  \& {Springel}}{{Henriques}
  et~al.}{2013}]{Henriques2013Simulations}
{Henriques} B. M.~B.,  {White} S. D.~M.,  {Thomas} P.~A.,  {Angulo} R.~E.,
  {Guo} Q.,  {Lemson} G.,   {Springel} V.,  2013, \mn@doi [\mnras]
  {10.1093/mnras/stt415}, \href
  {https://ui.adsabs.harvard.edu/abs/2013MNRAS.431.3373H} {431, 3373}

\bibitem[\protect\citeauthoryear{Henriques, White, Thomas, Angulo, Guo, Lemson,
  Springel  \& Overzier}{Henriques et~al.}{2015}]{henriques2015galaxy}
Henriques B.~M.,  White S.~D.,  Thomas P.~A.,  Angulo R.,  Guo Q.,  Lemson G.,
  Springel V.,   Overzier R.,  2015, Monthly Notices of the Royal Astronomical
  Society, 451, 2663

\bibitem[\protect\citeauthoryear{{Henriques}, {Yates}, {Fu}, {Guo},
  {Kauffmann}, {Srisawat}, {Thomas}  \& {White}}{{Henriques}
  et~al.}{2020}]{henriques2020galaxies}
{Henriques} B. M.~B.,  {Yates} R.~M.,  {Fu} J.,  {Guo} Q.,  {Kauffmann} G.,
  {Srisawat} C.,  {Thomas} P.~A.,   {White} S. D.~M.,  2020, \mn@doi [\mnras]
  {10.1093/mnras/stz3233}, \href
  {https://ui.adsabs.harvard.edu/abs/2020MNRAS.491.5795H} {491, 5795}

\bibitem[\protect\citeauthoryear{{Hernquist} \& {Katz}}{{Hernquist} \&
  {Katz}}{1989}]{hernquist89}
{Hernquist} L.,  {Katz} N.,  1989, \mn@doi [\apjs] {10.1086/191344}, \href
  {https://ui.adsabs.harvard.edu/abs/1989ApJS...70..419H} {70, 419}

\bibitem[\protect\citeauthoryear{{Hopkins} et~al.,}{{Hopkins}
  et~al.}{2018}]{hopkins18}
{Hopkins} P.~F.,  et~al., 2018, \mn@doi [\mnras] {10.1093/mnras/sty1690}, \href
  {https://ui.adsabs.harvard.edu/abs/2018MNRAS.480..800H} {480, 800}

\bibitem[\protect\citeauthoryear{Hubble \& Humason}{Hubble \&
  Humason}{1931}]{hubble1931velocity}
Hubble E.,  Humason M.~L.,  1931, The Astrophysical Journal, 74, 43

\bibitem[\protect\citeauthoryear{Ilbert et~al.,}{Ilbert
  et~al.}{2010}]{ilbert2010galaxy}
Ilbert O.,  et~al., 2010, The Astrophysical Journal, 709, 644

\bibitem[\protect\citeauthoryear{Ilbert et~al.,}{Ilbert
  et~al.}{2013}]{ilbert2013mass}
Ilbert O.,  et~al., 2013, Astronomy \& Astrophysics, 556, A55

\bibitem[\protect\citeauthoryear{{Jones}, {Haynes}, {Giovanelli}  \&
  {Moorman}}{{Jones} et~al.}{2018}]{Jones2018ALFALFA}
{Jones} M.~G.,  {Haynes} M.~P.,  {Giovanelli} R.,   {Moorman} C.,  2018,
  \mn@doi [\mnras] {10.1093/mnras/sty521}, \href
  {https://ui.adsabs.harvard.edu/abs/2018MNRAS.477....2J} {477, 2}

\bibitem[\protect\citeauthoryear{{Joshi}, {Pillepich}, {Nelson}, {Marinacci},
  {Springel}, {Rodriguez-Gomez}, {Vogelsberger}  \& {Hernquist}}{{Joshi}
  et~al.}{2020}]{joshi2020}
{Joshi} G.~D.,  {Pillepich} A.,  {Nelson} D.,  {Marinacci} F.,  {Springel} V.,
  {Rodriguez-Gomez} V.,  {Vogelsberger} M.,   {Hernquist} L.,  2020, \mn@doi
  [\mnras] {10.1093/mnras/staa1668}, \href
  {https://ui.adsabs.harvard.edu/abs/2020MNRAS.496.2673J} {496, 2673}

\bibitem[\protect\citeauthoryear{Kauffmann, White  \& Guiderdoni}{Kauffmann
  et~al.}{1993}]{kauffmann1993formation}
Kauffmann G.,  White S.~D.,   Guiderdoni B.,  1993, Monthly Notices of the
  Royal Astronomical Society, 264, 201

\bibitem[\protect\citeauthoryear{Kauffmann, Colberg, Diaferio  \&
  White}{Kauffmann et~al.}{1999}]{kauffmann1999clustering}
Kauffmann G.,  Colberg J.~M.,  Diaferio A.,   White S.~D.,  1999, Monthly
  Notices of the Royal Astronomical Society, 303, 188

\bibitem[\protect\citeauthoryear{{Kauffmann} et~al.,}{{Kauffmann}
  et~al.}{2003}]{Kauffmann2003Stellar}
{Kauffmann} G.,  et~al., 2003, \mn@doi [\mnras]
  {10.1046/j.1365-8711.2003.06291.x}, \href
  {https://ui.adsabs.harvard.edu/abs/2003MNRAS.341...33K} {341, 33}

\bibitem[\protect\citeauthoryear{Kauffmann, White, Heckman, M{\'e}nard,
  Brinchmann, Charlot, Tremonti  \& Brinkmann}{Kauffmann
  et~al.}{2004}]{kauffmann2004environmental}
Kauffmann G.,  White S.~D.,  Heckman T.~M.,  M{\'e}nard B.,  Brinchmann J.,
  Charlot S.,  Tremonti C.,   Brinkmann J.,  2004, Monthly Notices of the Royal
  Astronomical Society, 353, 713

\bibitem[\protect\citeauthoryear{Kauffmann, Li, Zhang  \& Weinmann}{Kauffmann
  et~al.}{2013}]{kauffmann2013re}
Kauffmann G.,  Li C.,  Zhang W.,   Weinmann S.,  2013, Monthly Notices of the
  Royal Astronomical Society, 430, 1447

\bibitem[\protect\citeauthoryear{{Klypin}, {Trujillo-Gomez}  \&
  {Primack}}{{Klypin} et~al.}{2011}]{Klypin2011Dark}
{Klypin} A.~A.,  {Trujillo-Gomez} S.,   {Primack} J.,  2011, \mn@doi [\apj]
  {10.1088/0004-637X/740/2/102}, \href
  {https://ui.adsabs.harvard.edu/abs/2011ApJ...740..102K} {740, 102}

\bibitem[\protect\citeauthoryear{Lacey et~al.,}{Lacey
  et~al.}{2016}]{lacey2016unified}
Lacey C.~G.,  et~al., 2016, Monthly Notices of the Royal Astronomical Society,
  462, 3854

\bibitem[\protect\citeauthoryear{Lagos, Tobar, Robotham, Obreschkow, Mitchell,
  Power  \& Elahi}{Lagos et~al.}{2018}]{lagos2018shark}
Lagos C. d.~P.,  Tobar R.~J.,  Robotham A.~S.,  Obreschkow D.,  Mitchell P.~D.,
   Power C.,   Elahi P.~J.,  2018, Monthly Notices of the Royal Astronomical
  Society, 481, 3573

\bibitem[\protect\citeauthoryear{{Lee}, {Kimm}, {Katz}, {Rosdahl}, {Devriendt}
  \& {Slyz}}{{Lee} et~al.}{2020}]{lee2020}
{Lee} J.,  {Kimm} T.,  {Katz} H.,  {Rosdahl} J.,  {Devriendt} J.,   {Slyz} A.,
  2020, arXiv e-prints, \href
  {https://ui.adsabs.harvard.edu/abs/2020arXiv201011028L} {p. arXiv:2010.11028}
  (\mn@eprint {arXiv} {2010.11028})

\bibitem[\protect\citeauthoryear{Li \& White}{Li \&
  White}{2009}]{li2009distribution}
Li C.,  White S.~D.,  2009, Monthly Notices of the Royal Astronomical Society,
  398, 2177

\bibitem[\protect\citeauthoryear{{Libeskind} et~al.,}{{Libeskind}
  et~al.}{2020}]{Libeskind2020Hestia}
{Libeskind} N.~I.,  et~al., 2020, \mn@doi [\mnras] {10.1093/mnras/staa2541},
  \href {https://ui.adsabs.harvard.edu/abs/2020MNRAS.tmp.2509L} {}

\bibitem[\protect\citeauthoryear{Lu, Gilbank, McGee, Balogh  \& Gallagher}{Lu
  et~al.}{2012}]{lu2012cfht}
Lu T.,  Gilbank D.~G.,  McGee S.~L.,  Balogh M.~L.,   Gallagher S.,  2012,
  Monthly Notices of the Royal Astronomical Society, 420, 126

\bibitem[\protect\citeauthoryear{{Luo}, {Kang}, {Kauffmann}  \& {Fu}}{{Luo}
  et~al.}{2016}]{Luo2016Resolution}
{Luo} Y.,  {Kang} X.,  {Kauffmann} G.,   {Fu} J.,  2016, \mn@doi [\mnras]
  {10.1093/mnras/stw268}, \href
  {http://adsabs.harvard.edu/abs/2016MNRAS.458..366L} {458, 366}

\bibitem[\protect\citeauthoryear{Marchesini, Van~Dokkum, Schreiber, Franx,
  Labb{\'e}  \& Wuyts}{Marchesini et~al.}{2009}]{marchesini2009evolution}
Marchesini D.,  Van~Dokkum P.~G.,  Schreiber N. M.~F.,  Franx M.,  Labb{\'e}
  I.,   Wuyts S.,  2009, The Astrophysical Journal, 701, 1765

\bibitem[\protect\citeauthoryear{Marchesini et~al.,}{Marchesini
  et~al.}{2010}]{marchesini2010most}
Marchesini D.,  et~al., 2010, The Astrophysical Journal, 725, 1277

\bibitem[\protect\citeauthoryear{{McCarthy}, {Frenk}, {Font}, {Lacey}, {Bower},
  {Mitchell}, {Balogh}  \& {Theuns}}{{McCarthy} et~al.}{2008}]{mccarthy2007ram}
{McCarthy} I.~G.,  {Frenk} C.~S.,  {Font} A.~S.,  {Lacey} C.~G.,  {Bower}
  R.~G.,  {Mitchell} N.~L.,  {Balogh} M.~L.,   {Theuns} T.,  2008, \mn@doi
  [\mnras] {10.1111/j.1365-2966.2007.12577.x}, \href
  {https://ui.adsabs.harvard.edu/abs/2008MNRAS.383..593M} {383, 593}

\bibitem[\protect\citeauthoryear{Mo, Van~den Bosch  \& White}{Mo
  et~al.}{2010}]{mo2010galaxy}
Mo H.,  Van~den Bosch F.,   White S.,  2010, Galaxy formation and evolution.
Cambridge University Press

\bibitem[\protect\citeauthoryear{{Mosleh}, {Tavasoli}  \& {Tacchella}}{{Mosleh}
  et~al.}{2018}]{Mosleh2018Stellar}
{Mosleh} M.,  {Tavasoli} S.,   {Tacchella} S.,  2018, \mn@doi [\apj]
  {10.3847/1538-4357/aac5e6}, \href
  {https://ui.adsabs.harvard.edu/abs/2018ApJ...861..101M} {861, 101}

\bibitem[\protect\citeauthoryear{{Moster}, {Somerville}, {Maulbetsch}, {van den
  Bosch}, {Macci{\`o}}, {Naab}  \& {Oser}}{{Moster}
  et~al.}{2010}]{Moster2010Constraints}
{Moster} B.~P.,  {Somerville} R.~S.,  {Maulbetsch} C.,  {van den Bosch} F.~C.,
  {Macci{\`o}} A.~V.,  {Naab} T.,   {Oser} L.,  2010, \mn@doi [\apj]
  {10.1088/0004-637X/710/2/903}, \href
  {https://ui.adsabs.harvard.edu/abs/2010ApJ...710..903M} {710, 903}

\bibitem[\protect\citeauthoryear{{Muzzin}, {Wilson}, {Lacy}, {Yee}  \&
  {Stanford}}{{Muzzin} et~al.}{2008}]{Muzzin2008Evolution}
{Muzzin} A.,  {Wilson} G.,  {Lacy} M.,  {Yee} H.~K.~C.,   {Stanford} S.~A.,
  2008, \mn@doi [\apj] {10.1086/591542}, \href
  {https://ui.adsabs.harvard.edu/abs/2008ApJ...686..966M} {686, 966}

\bibitem[\protect\citeauthoryear{{Muzzin} et~al.,}{{Muzzin}
  et~al.}{2009}]{Muzzin2009Spectroscopic}
{Muzzin} A.,  et~al., 2009, \mn@doi [\apj] {10.1088/0004-637X/698/2/1934},
  \href {https://ui.adsabs.harvard.edu/abs/2009ApJ...698.1934M} {698, 1934}

\bibitem[\protect\citeauthoryear{Muzzin et~al.,}{Muzzin
  et~al.}{2013}]{muzzin2013evolution}
Muzzin A.,  et~al., 2013, The Astrophysical Journal, 777, 18

\bibitem[\protect\citeauthoryear{{Nelson} et~al.,}{{Nelson}
  et~al.}{2019}]{nelson2019First}
{Nelson} D.,  et~al., 2019, \mn@doi [\mnras] {10.1093/mnras/stz2306}, \href
  {https://ui.adsabs.harvard.edu/abs/2019MNRAS.490.3234N} {490, 3234}

\bibitem[\protect\citeauthoryear{Oemler}{Oemler}{1974}]{oemler1974systematic}
Oemler A.,  1974, PhD thesis, California Institute of Technology

\bibitem[\protect\citeauthoryear{{Oman}, {Bah{\'e}}, {Healy}, {Hess}, {Hudson}
  \& {Verheijen}}{{Oman} et~al.}{2020}]{oman20}
{Oman} K.~A.,  {Bah{\'e}} Y.~M.,  {Healy} J.,  {Hess} K.~M.,  {Hudson} M.~J.,
  {Verheijen} M. A.~W.,  2020, arXiv e-prints, \href
  {https://ui.adsabs.harvard.edu/abs/2020arXiv200900667O} {p. arXiv:2009.00667}
  (\mn@eprint {arXiv} {2009.00667})

\bibitem[\protect\citeauthoryear{{Pallero}, {G{\'o}mez}, {Padilla},
  {Torres-Flores}, {Demarco}, {Cerulo}  \& {Olave-Rojas}}{{Pallero}
  et~al.}{2019}]{Pallero2019Tracing}
{Pallero} D.,  {G{\'o}mez} F.~A.,  {Padilla} N.~D.,  {Torres-Flores} S.,
  {Demarco} R.,  {Cerulo} P.,   {Olave-Rojas} D.,  2019, \mn@doi [\mnras]
  {10.1093/mnras/stz1745}, \href
  {https://ui.adsabs.harvard.edu/abs/2019MNRAS.488..847P} {488, 847}

\bibitem[\protect\citeauthoryear{{Peng} et~al.,}{{Peng} et~al.}{2010}]{peng10}
{Peng} Y.-j.,  et~al., 2010, \mn@doi [\apj] {10.1088/0004-637X/721/1/193},
  \href {http://adsabs.harvard.edu/abs/2010ApJ...721..193P} {721, 193}

\bibitem[\protect\citeauthoryear{Pillepich et~al.,}{Pillepich
  et~al.}{2018}]{pillepich2018First}
Pillepich A.,  et~al., 2018, Monthly Notices of the Royal Astronomical Society,
  475, 648

\bibitem[\protect\citeauthoryear{Pillepich et~al.,}{Pillepich
  et~al.}{2019}]{pillepich19}
Pillepich A.,  et~al., 2019, Monthly Notices of the Royal Astronomical Society,
  490, 3196

\bibitem[\protect\citeauthoryear{{Pintos-Castro}, {Yee}, {Muzzin}, {Old}  \&
  {Wilson}}{{Pintos-Castro} et~al.}{2019}]{Pintos-Castro2019Evolution}
{Pintos-Castro} I.,  {Yee} H.~K.~C.,  {Muzzin} A.,  {Old} L.,   {Wilson} G.,
  2019, \mn@doi [\apj] {10.3847/1538-4357/ab14ee}, \href
  {https://ui.adsabs.harvard.edu/abs/2019ApJ...876...40P} {876, 40}

\bibitem[\protect\citeauthoryear{{Planck Collaboration}}{{Planck
  Collaboration}}{2016}]{planck2015_xiii}
{Planck Collaboration} 2016, \mn@doi [\aap] {10.1051/0004-6361/201525830},
  \href {http://adsabs.harvard.edu/abs/2016A%26A...594A..13P} {594, A13}

\bibitem[\protect\citeauthoryear{{Riebe} et~al.,}{{Riebe}
  et~al.}{2011}]{Riebe2011MultiDark}
{Riebe} K.,  et~al., 2011, arXiv e-prints, \href
  {https://ui.adsabs.harvard.edu/abs/2011arXiv1109.0003R} {p. arXiv:1109.0003}
  (\mn@eprint {arXiv} {1109.0003})

\bibitem[\protect\citeauthoryear{{Roediger} \& {Br{\"u}ggen}}{{Roediger} \&
  {Br{\"u}ggen}}{2007}]{roediger07}
{Roediger} E.,  {Br{\"u}ggen} M.,  2007, \mn@doi [\mnras]
  {10.1111/j.1365-2966.2007.12241.x}, \href
  {http://adsabs.harvard.edu/abs/2007MNRAS.380.1399R} {380, 1399}

\bibitem[\protect\citeauthoryear{{Salim} et~al.,}{{Salim}
  et~al.}{2007}]{Salim2007UV}
{Salim} S.,  et~al., 2007, \mn@doi [\apjs] {10.1086/519218}, \href
  {https://ui.adsabs.harvard.edu/abs/2007ApJS..173..267S} {173, 267}

\bibitem[\protect\citeauthoryear{{Sarron}, {Adami}, {Durret}  \&
  {Laigle}}{{Sarron} et~al.}{2019}]{Sarron2017filaments}
{Sarron} F.,  {Adami} C.,  {Durret} F.,   {Laigle} C.,  2019, \mn@doi [\aap]
  {10.1051/0004-6361/201935394}, \href
  {https://ui.adsabs.harvard.edu/abs/2019A&A...632A..49S} {632, A49}

\bibitem[\protect\citeauthoryear{{Schaye} et~al.,}{{Schaye}
  et~al.}{2015}]{Schaye2015eagle}
{Schaye} J.,  et~al., 2015, \mn@doi [\mnras] {10.1093/mnras/stu2058}, \href
  {https://ui.adsabs.harvard.edu/abs/2015MNRAS.446..521S} {446, 521}

\bibitem[\protect\citeauthoryear{{Skillman}, {Warren}, {Turk}, {Wechsler},
  {Holz}  \& {Sutter}}{{Skillman} et~al.}{2014}]{Skillman2014Dark}
{Skillman} S.~W.,  {Warren} M.~S.,  {Turk} M.~J.,  {Wechsler} R.~H.,  {Holz}
  D.~E.,   {Sutter} P.~M.,  2014, arXiv e-prints, \href
  {https://ui.adsabs.harvard.edu/abs/2014arXiv1407.2600S} {p. arXiv:1407.2600}
  (\mn@eprint {arXiv} {1407.2600})

\bibitem[\protect\citeauthoryear{Somerville \& Primack}{Somerville \&
  Primack}{1999}]{somerville1999semi}
Somerville R.~S.,  Primack J.~R.,  1999, Monthly Notices of the Royal
  Astronomical Society, 310, 1087

\bibitem[\protect\citeauthoryear{Spergel et~al.,}{Spergel
  et~al.}{2003}]{spergel2003first}
Spergel D.~N.,  et~al., 2003, The Astrophysical Journal Supplement Series, 148,
  175

\bibitem[\protect\citeauthoryear{Springel}{Springel}{2010}]{springel2010pur}
Springel V.,  2010, Monthly Notices of the Royal Astronomical Society, 401, 791

\bibitem[\protect\citeauthoryear{Springel, White, Tormen  \&
  Kauffmann}{Springel et~al.}{2001}]{springel2001populating}
Springel V.,  White S.~D.,  Tormen G.,   Kauffmann G.,  2001, Monthly Notices
  of the Royal Astronomical Society, 328, 726

\bibitem[\protect\citeauthoryear{Springel et~al.,}{Springel
  et~al.}{2005}]{springel2005simulations}
Springel V.,  et~al., 2005, nature, 435, 629

\bibitem[\protect\citeauthoryear{Stevens, Croton  \& Mutch}{Stevens
  et~al.}{2016}]{stevens2016building}
Stevens A.~R.,  Croton D.~J.,   Mutch S.~J.,  2016, Monthly Notices of the
  Royal Astronomical Society, 461, 859

\bibitem[\protect\citeauthoryear{{Tavasoli}, {Vasei}  \& {Mohayaee}}{{Tavasoli}
  et~al.}{2013}]{Tavasoli2013Challenge}
{Tavasoli} S.,  {Vasei} K.,   {Mohayaee} R.,  2013, \mn@doi [\aap]
  {10.1051/0004-6361/201220774}, \href
  {https://ui.adsabs.harvard.edu/abs/2013A&A...553A..15T} {553, A15}

\bibitem[\protect\citeauthoryear{Tecce, Cora, Tissera, Abadi  \& Lagos}{Tecce
  et~al.}{2010}]{tecce2010ram}
Tecce T.~E.,  Cora S.~A.,  Tissera P.~B.,  Abadi M.~G.,   Lagos C. d.~P.,
  2010, Monthly Notices of the Royal Astronomical Society, 408, 2008

\bibitem[\protect\citeauthoryear{{Tinker}}{{Tinker}}{2020}]{Tinker2020Self-calibrating}
{Tinker} J.~L.,  2020, arXiv e-prints, \href
  {https://ui.adsabs.harvard.edu/abs/2020arXiv200712200T} {p. arXiv:2007.12200}
  (\mn@eprint {arXiv} {2007.12200})

\bibitem[\protect\citeauthoryear{Tomczak et~al.,}{Tomczak
  et~al.}{2014}]{tomczak2014galaxy}
Tomczak A.~R.,  et~al., 2014, The Astrophysical Journal, 783, 85

\bibitem[\protect\citeauthoryear{{Vijayaraghavan} \& {Ricker}}{{Vijayaraghavan}
  \& {Ricker}}{2017}]{vijayaraghavan17}
{Vijayaraghavan} R.,  {Ricker} P.~M.,  2017, \mn@doi [\apj]
  {10.3847/1538-4357/aa6eac}, \href
  {http://adsabs.harvard.edu/abs/2017ApJ...841...38V} {841, 38}

\bibitem[\protect\citeauthoryear{Vogelsberger et~al.,}{Vogelsberger
  et~al.}{2014}]{vogelsberger2014Introducing}
Vogelsberger M.,  et~al., 2014, Monthly Notices of the Royal Astronomical
  Society, 444, 1518

\bibitem[\protect\citeauthoryear{{Wang}, {Xu}, {Lee}, {Du}, {Overzier}  \&
  {Shao}}{{Wang} et~al.}{2020a}]{wang2020}
{Wang} J.,  {Xu} W.,  {Lee} B.,  {Du} M.,  {Overzier} R.,   {Shao} L.,  2020a,
  arXiv e-prints, \href {https://ui.adsabs.harvard.edu/abs/2020arXiv200908159W}
  {p. arXiv:2009.08159} (\mn@eprint {arXiv} {2009.08159})

\bibitem[\protect\citeauthoryear{{Wang}, {Bose}, {Frenk}, {Gao}, {Jenkins},
  {Springel}  \& {White}}{{Wang} et~al.}{2020b}]{Wang2020Universal}
{Wang} J.,  {Bose} S.,  {Frenk} C.~S.,  {Gao} L.,  {Jenkins} A.,  {Springel}
  V.,   {White} S.~D.~M.,  2020b, \mn@doi [\nat] {10.1038/s41586-020-2642-9},
  \href {https://ui.adsabs.harvard.edu/abs/2020Natur.585...39W} {585, 39}

\bibitem[\protect\citeauthoryear{{Webb} et~al.,}{{Webb}
  et~al.}{2020}]{webb2020}
{Webb} K.,  et~al., 2020, \mn@doi [\mnras] {10.1093/mnras/staa2752}, \href
  {https://ui.adsabs.harvard.edu/abs/2020MNRAS.498.5317W} {498, 5317}

\bibitem[\protect\citeauthoryear{Weinmann, Van Den~Bosch, Yang  \& Mo}{Weinmann
  et~al.}{2006}]{weinmann2006properties}
Weinmann S.~M.,  Van Den~Bosch F.~C.,  Yang X.,   Mo H.,  2006, Monthly Notices
  of the Royal Astronomical Society, 366, 2

\bibitem[\protect\citeauthoryear{{Wetzel}, {Tinker}  \& {Conroy}}{{Wetzel}
  et~al.}{2012}]{Wetzel2012Galaxy}
{Wetzel} A.~R.,  {Tinker} J.~L.,   {Conroy} C.,  2012, \mn@doi [\mnras]
  {10.1111/j.1365-2966.2012.21188.x}, \href
  {https://ui.adsabs.harvard.edu/abs/2012MNRAS.424..232W} {424, 232}

\bibitem[\protect\citeauthoryear{{Wetzel}, {Tinker}, {Conroy}  \& {van den
  Bosch}}{{Wetzel} et~al.}{2014}]{Wetzel2014Galaxy}
{Wetzel} A.~R.,  {Tinker} J.~L.,  {Conroy} C.,   {van den Bosch} F.~C.,  2014,
  \mn@doi [\mnras] {10.1093/mnras/stu122}, \href
  {https://ui.adsabs.harvard.edu/abs/2014MNRAS.439.2687W} {439, 2687}

\bibitem[\protect\citeauthoryear{{White} \& {Rees}}{{White} \&
  {Rees}}{1978}]{white1978core}
{White} S.~D.~M.,  {Rees} M.~J.,  1978, \mn@doi [\mnras]
  {10.1093/mnras/183.3.341}, \href
  {https://ui.adsabs.harvard.edu/abs/1978MNRAS.183..341W} {183, 341}

\bibitem[\protect\citeauthoryear{{Wilson} et~al.,}{{Wilson}
  et~al.}{2009}]{Wilson2009Spectroscopic}
{Wilson} G.,  et~al., 2009, \mn@doi [\apj] {10.1088/0004-637X/698/2/1943},
  \href {https://ui.adsabs.harvard.edu/abs/2009ApJ...698.1943W} {698, 1943}

\bibitem[\protect\citeauthoryear{{Yan}, {Fan}  \& {White}}{{Yan}
  et~al.}{2013}]{Yan2013Dependence}
{Yan} H.,  {Fan} Z.,   {White} S. D.~M.,  2013, \mn@doi [\mnras]
  {10.1093/mnras/stt141}, \href
  {https://ui.adsabs.harvard.edu/abs/2013MNRAS.430.3432Y} {430, 3432}

\bibitem[\protect\citeauthoryear{{Yang}, {Mo}  \& {van den Bosch}}{{Yang}
  et~al.}{2003}]{Yang2003Costraining}
{Yang} X.,  {Mo} H.~J.,   {van den Bosch} F.~C.,  2003, \mn@doi [\mnras]
  {10.1046/j.1365-8711.2003.06254.x}, \href
  {https://ui.adsabs.harvard.edu/abs/2003MNRAS.339.1057Y} {339, 1057}

\bibitem[\protect\citeauthoryear{{Yang}, {Mo}, {van den Bosch}  \&
  {Jing}}{{Yang} et~al.}{2005}]{Yang2005halo-based}
{Yang} X.,  {Mo} H.~J.,  {van den Bosch} F.~C.,   {Jing} Y.~P.,  2005, \mn@doi
  [\mnras] {10.1111/j.1365-2966.2005.08560.x}, \href
  {https://ui.adsabs.harvard.edu/abs/2005MNRAS.356.1293Y} {356, 1293}

\bibitem[\protect\citeauthoryear{{Yang}, {Mo}, {van den Bosch}, {Pasquali},
  {Li}  \& {Barden}}{{Yang} et~al.}{2007}]{Yang2007Galaxy}
{Yang} X.,  {Mo} H.~J.,  {van den Bosch} F.~C.,  {Pasquali} A.,  {Li} C.,
  {Barden} M.,  2007, \mn@doi [\apj] {10.1086/522027}, \href
  {https://ui.adsabs.harvard.edu/abs/2007ApJ...671..153Y} {671, 153}

\bibitem[\protect\citeauthoryear{Yates, Henriques, Thomas, Kauffmann, Johansson
   \& White}{Yates et~al.}{2013}]{yates2013modelling}
Yates R.~M.,  Henriques B.,  Thomas P.~A.,  Kauffmann G.,  Johansson J.,
  White S.~D.,  2013, Monthly Notices of the Royal Astronomical Society, 435,
  3500

\bibitem[\protect\citeauthoryear{{Yates}, {Thomas}  \& {Henriques}}{{Yates}
  et~al.}{2017}]{Yates+17}
{Yates} R.~M.,  {Thomas} P.~A.,   {Henriques} B.~M.~B.,  2017, \mn@doi [\mnras]
  {10.1093/mnras/stw2361}, \href
  {http://adsabs.harvard.edu/abs/2017MNRAS.464.3169Y} {464, 3169}

\bibitem[\protect\citeauthoryear{{Yun} et~al.,}{{Yun} et~al.}{2019}]{yun19}
{Yun} K.,  et~al., 2019, \mn@doi [\mnras] {10.1093/mnras/sty3156}, \href
  {http://adsabs.harvard.edu/abs/2019MNRAS.483.1042Y} {483, 1042}

\bibitem[\protect\citeauthoryear{{Zwaan}, {Meyer}, {Staveley-Smith}  \&
  {Webster}}{{Zwaan} et~al.}{2005}]{Zwann2005HIPASS}
{Zwaan} M.~A.,  {Meyer} M.~J.,  {Staveley-Smith} L.,   {Webster} R.~L.,  2005,
  \mn@doi [\mnras] {10.1111/j.1745-3933.2005.00029.x}, \href
  {https://ui.adsabs.harvard.edu/abs/2005MNRAS.359L..30Z} {359, L30}

\bibitem[\protect\citeauthoryear{von~der Linden, Wild, Kauffmann, White  \&
  Weinmann}{von~der Linden et~al.}{2010}]{von2010star}
von~der Linden A.,  Wild V.,  Kauffmann G.,  White S.~D.,   Weinmann S.,  2010,
  Monthly Notices of the Royal Astronomical Society, 404, 1231

\makeatother
\end{thebibliography}


\appendix

\section{Deriving the halo mass from stellar mass}
\label{app: halomass_stellarmass_fit}

The total halo mass, $M_{200}$, is not observable and is usually estimated using other well-defined observables. As explained in \S \ref{subsec: mock_catalogues}, we derive the halo mass from the stellar mass of each halo's central galaxy to make mock catalogues comparable with observations. Comparing these mock halo masses with their actual values from the simulation, we here explore the resulting scatter.

\begin{figure}
    \centering
    \includegraphics[width=0.8\columnwidth]{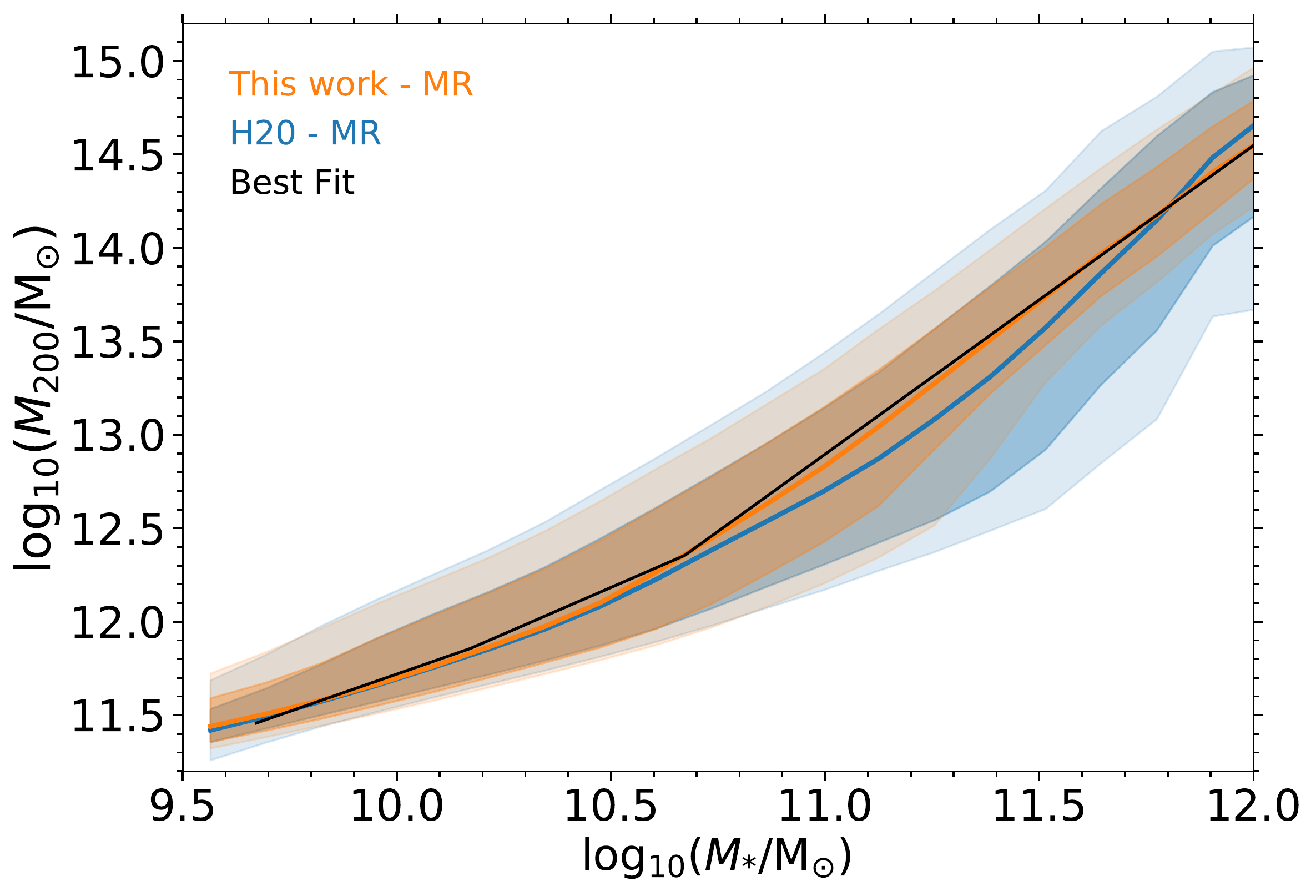}
    \caption{Halo mass as a function of stellar mass in this work (orange) and in H20 (blue). The solid black line shows our best fit using L-Galaxies (this work) to the halo mass, as given in Eq. \ref{eq: Mstellar_to_M200_conversion}. The orange and blue solid lines correspond to the median values and the shaded regions illustrate $1\sigma$ and $2\sigma$ of the distribution. Our best fit (black line) is very close to the median value from the simulation (orange line), but slightly differs from the median line of the other model (H20).}
\label{Fig: append_SMHM}
\end{figure}

Fig. \ref{Fig: append_SMHM} shows the direct simulation halo mass from the Millennium simulation as a function of the stellar mass of the central galaxy in our model (orange) and H20 (blue). The solid lines denote the median values and the light and dark shaded regions correspond to the $1\sigma$ and $2\sigma$ scatter of the distribution. The black line shows our best fit, which is in good agreement with our model (orange line) based on which the fitting is performed.

\begin{figure}
    \centering
    \includegraphics[width=0.8\columnwidth]{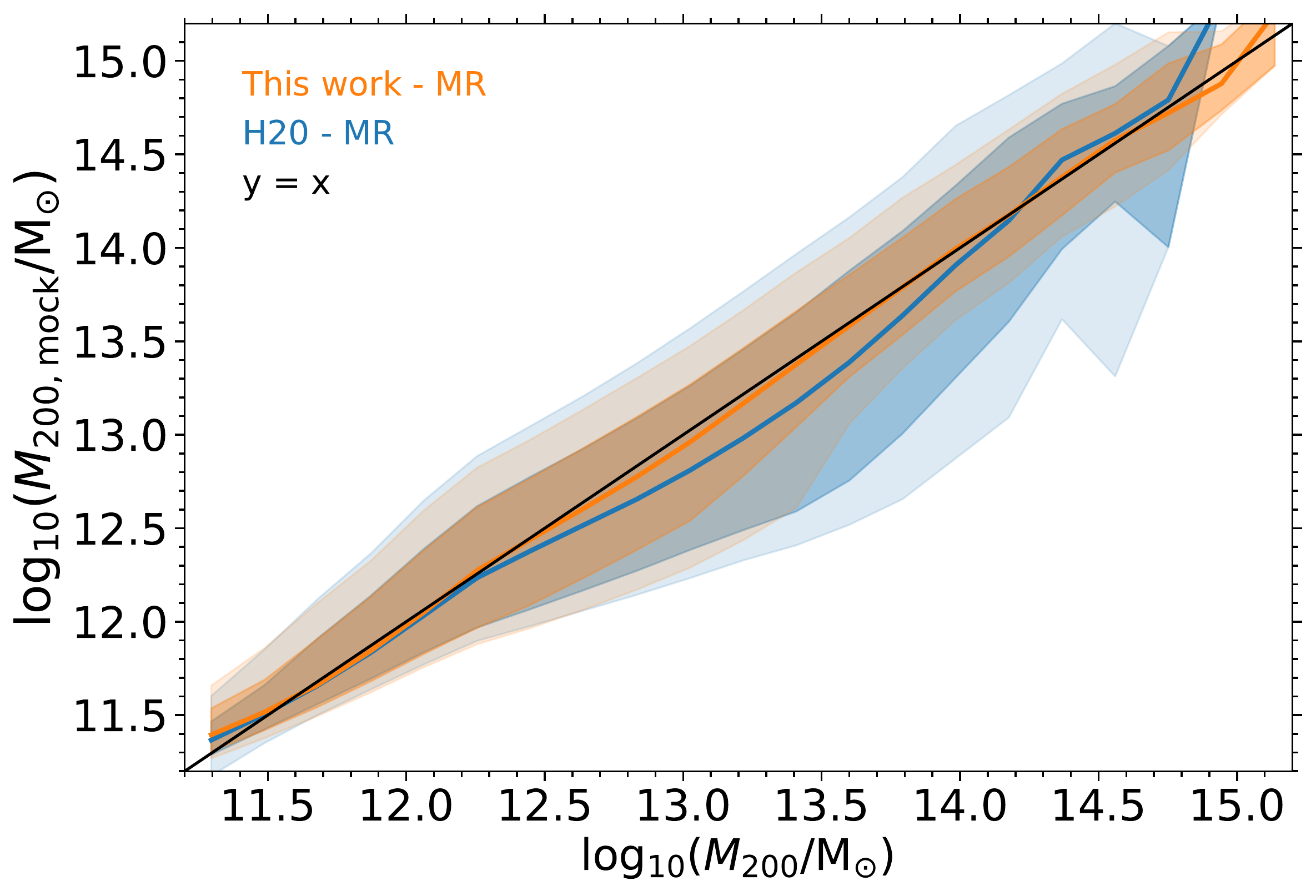}
    \caption{The halo mass calculated using the stellar mass (Eq. \ref{eq: Mstellar_to_M200_conversion}) as a function of the halo mass measured directly from the Millennium simulation. The solid lines correspond to the median values and the shaded regions illustrate $1\sigma$ and $2\sigma$ of the distribution. The $1\sigma$ scatter around $y=x$ line is $\sim$ 0.2-0.3 dex, and the $2\sigma$ scatter is up to 0.5-0.6 dex.}
\label{Fig: append_M200_mock}
\end{figure}

Additionally, Fig. \ref{Fig: append_M200_mock} shows the fit halo mass from Eq. \ref{eq: Mstellar_to_M200_conversion} (y-axis) as a function of the direct simulation halo mass (x-axis) in our model and H20. The two halo masses (fit and model) are in rather close agreement with the $1\sigma$ scatter of 0.2-0.3 dex. The H20 is a bit off near $M_{200}/\rm M_{\odot}\sim 10^{13}$, which reflects different stellar mass to halo mass relationships in our model and H20. In this work, whenever we needed to convert stellar mass to halo mass, we used the fit based on our model (see \S \ref{subsubsec: halomass_mock_simulation}).

Although this scatter is likely smaller than the typical error on stellar mass estimates from observations, its impact on the virial radius, $R_{200}$, could be substantial. For instance, if the mock halo mass is $0.3$ dex larger/smaller than the true halo mass, this leads to the mock virial radius being $\sim 20-25\%$ larger/smaller than the true virial radius. As a result, in our comparison with observations, where we report results as a function of halocentric distance (Figs. \ref{Fig: quenchedFrac_dis_proj_z0} and \ref{Fig: quenchedFrac_dis_proj_z>0}), the trends with distance could contain $20-25 \%$ error.

\label{lastpage}

\end{document}